\documentclass[aps,prc,groupedaddress,showpacs]{revtex4}

\usepackage[dvips]{graphicx}
\usepackage{amssymb,amsmath,amstext,amsthm,amsfonts}

\newcommand{\be}{\begin{equation}}
\newcommand{\ee}{\end{equation}}
\newcommand{\bea}{\begin{eqnarray}}
\newcommand{\eea}{\end{eqnarray}}

\newcommand{\myunit}[1]{\; \text{#1} }

\newcommand{\myvec}[1]{\,\vec{#1}\,}
\newcommand{\reffig}[1]{Fig.~\ref{#1}}
\newcommand{\refeq}[1]{Eq.~(\ref{#1})}
\newcommand{\refch}[1]{Sec.~\ref{#1}}
\newcommand{\refsubch}[1]{Sec.~\ref{#1}}
\newcommand{\refcite}[1]{Ref.~\cite{#1}}
\newcommand{\refscite}[1]{Refs.~\cite{#1}}
\begin{document}

\title{Charged current neutrino-nucleus interactions at intermediate energies}

\author{T.~Leitner}
\email[electronic address: ]{tina.j.leitner@theo.physik.uni-giessen.de}
\author{L.~Alvarez-Ruso}
\author{U.~Mosel}
\affiliation{Institut f\"ur Theoretische Physik, Universit\"at Giessen, Germany}

\date{January 31, 2006}

\begin{abstract}
We have developed a model to describe the interactions of neutrinos with nucleons and nuclei, focusing on the region of the quasielastic and $\Delta (1232)$ peaks. We describe neutrino-nucleon collisions with a fully relativistic formalism which incorporates state-of-the-art parametrizations of the form factors for both the nucleon and the $N-\Delta$ transition. The model has then been extended to finite nuclei, taking into account nuclear effects such as Fermi motion, Pauli blocking (both within the local density approximation), nuclear binding and final-state interactions. The in-medium modification of the $\Delta$ resonance due to Pauli blocking and collisional broadening have also been included. Final-state interactions are implemented by means of the Boltzmann-Uehling-Uhlenbeck (BUU) coupled-channel 
transport model. Results for charged current inclusive cross sections and exclusive channels as pion production and nucleon knockout are presented and discussed.
\end{abstract}


\maketitle

\section{Introduction}

Neutrino experiments now provide conclusive evidence that neutrino oscillations exist~\cite{Fukuda:1998mi,Ahmad:2001an} and, therefore, neutrinos are not massless. 
An extensive experimental program~\cite{minos,k2k,miniboone} aiming at the precise determination of $\nu$ masses, mixing angles and CP parameters is being developed in several labs around the world. The success of these oscillation experiments depends critically on a good knowledge of neutrino-nucleus interactions. In this way it is possible to minimize the systematic uncertainties in neutrino fluxes, backgrounds and detector responses.

The interest in neutrinos goes beyond the study of their intrinsic properties and extends to a variety of
topics in astro-, nuclear and hadronic physics. In particular, they are a valuable tool for nuclear and hadronic physics. 
The availability of a high intensity $\nu$ beam at Fermilab offers a unique opportunity 
to gain new information on the structure of the nucleon and baryonic resonances. Experiments
such as MINER$\nu$A~\cite{minervaprop} and FINeSSE~\cite{finesseprop} shall address relevant problems like the
extraction of the nucleon and $N −- \Delta$ axial form factors, or the measurement of the
strange spin of the nucleon~\cite{Alberico:2001sd}. However, those experiments will be performed mainly on
nuclear targets since they provide relatively large cross sections. 

Particles produced in neutrino interactions can reinteract before leaving the nucleus and can be absorbed, change their kinematics or even charge before being detected. For example, even if an initial interaction might be quasielastic scattering the observed final particles can include pions and multiple nucleons.
Thus, understanding nuclear effects is essential for the interpretation of the data and represents both a challenge and an opportunity~\cite{Cavanna:2005yp}.

There is a rather extensive literature on the theoretical study of neutrino-nucleus collisions. Nuclear shell model~\cite{Haxton:1987kc} and RPA calculations~\cite{Kolbe:1994xb,Volpe:2000zn} address the region of low energy neutrinos. Those reactions are important for the synthesis of heavy elements in stars~\cite{Kolbe:2003ys,Qian:1996db}. 

At intermediate energies, quasielastic collisions have been investigated using a relativistic treatment of the matrix element in Fermi gas models~\cite{Smith:1972xh,Donnelly:1978tz,Horowitz:1993rj} that take into account Fermi motion, Pauli blocking and binding energies. 
The importance of strong renormalization effects on the weak transition strengths in the nuclear medium has also been emphasized. These nuclear correlations are often taken into account as RPA resummation of particle-hole and $\Delta$-hole states~\cite{Singh:1992dc, Nieves:2004wx, Marteau:1999kt}, but realistic nuclear spectral functions have been used as well~\cite{Benhar:2005dj}.

Nucleon knockout has attracted considerable attention, mainly in connection with the possibility of extracting the axial strange content of the nucleon~\cite{Alberico:1997vh, Meucci:2006ir}.
The most straightforward approach to this problem is the plane wave impulse approximation. It neglects all interactions between the outgoing nucleons and the nucleus~\cite{vanderVentel:2003km}, but it has been claimed that the distortions in the observables due to rescattering largely cancel for the ratio of proton to neutron yields~\cite{vanderVentel:2005ke}.
This situation can be improved by means of a distorted wave impulse approximations, where the final nucleon scattering state is computed using optical potentials~\cite{Martinez:2005xe,Meucci:2004ip}.
For energies above 1~GeV the Glauber model, which is based on the eikonal and frozen spectator approximation, is a better alternative~\cite{Martinez:2005xe}. 
However, in those approaches the nucleons that go into unobserved states as a result of an interaction are lost, while, in fact, they are ejected with a different energy, angle and maybe charge. Monte Carlo methods permit to take into account interactions of nucleons leading to energy losses, charge exchange and multiple nucleon emission~\cite{Nieves:2005rq}.

Theoretical studies of neutrino-induced pion production in nuclei are 
considerably more scarce. The main pion production mechanism is via 
the excitation and subsequent decay of the $\Delta$ resonance~\cite{Singh:1998ha, Paschos:2000be,Kim:1996bt}. The contribution of higher order resonances 
at $\nu$ energies $\lesssim$~2~GeV has been found to be small~\cite{Fogli:1979cz,Paschos:2003qr}. 
Singh et al.~\cite{Singh:1998ha} pointed out that the in-medium modification of the 
$\Delta$~width causes a reduction of the $\nu \, A \rightarrow \Delta\, X$
cross section. The final-state interactions of the produced pions 
is an essential ingredient of any realistic model of this reaction. 
However, the available calculations deal with it in an oversimplified way. 
Singh et al. only estimate the pion flux reduction in an eikonal 
approximation, neglecting elastic and charge changing scattering. 
In \refscite{Adler:1974qu, Paschos:2000be}, those processes are considered in a \textit{random walk} approximation where the pion energy stays constant during their propagation out of the nucleus. The nucleons are frozen within the nucleus (no Fermi motion) and
Pauli blocking appears only as a global factor common to all collisions. However, this approach does not take into account the knowledge gathered in extensive research of pion production in nuclei using different probes in the past years~\cite{Ericson:1988gk}. 

Finally, at higher energies, deep inelastic scattering becomes progressively important~\cite{Ashie:2005ik}. The role of nuclear corrections in this case is currently debated in connection with the recent NuTeV findings~\cite{Zeller:2001hh}. The impact of deep inelastic scattering in the energy region under study here is expected to be minor. Therefore, we do not take it into consideration.

Quasielastic scattering, pion production and deep inelastic scattering are put together in event generators (e.~g.~NUANCE~\cite{Casper:2002sd}, NEUGEN~\cite{Gallagher:2002sf} and NEUT~\cite{Hayato:2002sd}) used by the neutrino experiments in their simulations. 
For resonance production, these generators take the model of Rein and Sehgal~\cite{Rein:1980wg} based on old quark model calculations for the form factors.
They differ substantially in the implementation of in-medium effects and final-state interactions. 
For example, NEUT and NUANCE apply a relativistic Fermi gas model with a global Fermi momentum and a constant nuclear binding energy. The in-medium modification of the $\Delta$~width is taken into account only in the NEUT model. The final-state interactions are treated by using a cascade model with rescattering, charge exchange and absorption of pions.

In this work we study neutrino interactions with nuclei at intermediate energies where we have included the two most important contributions to the $\nu N$ collision at neutrino energies up to about 1.5~GeV, namely quasielastic scattering and $\Delta$~excitation. As in photonuclear reactions we assume that the neutrino interacts with a single bound nucleon in the nucleus. Thus, the first essential step is a reliable description of the $\nu N$ collision. For that we use a fully relativistic formalism with state-of-the-art parametrizations of the form factors. Then, the elementary cross section has to be modified in order to account for nuclear effects. We describe the nucleus by a local Fermi gas, leading to density dependent Fermi momenta. We further account for the binding of the nucleons in a density and momentum dependent mean-field. Pauli blocking of the final state nucleons is also considered. 

Once produced, the particles experience final-state interactions when propagating through the nuclear medium. A realistic treatment of the final-state interactions can be achieved in the framework of a coupled-channel transport theory --- the Giessen BUU model --- which allows the investigation of exclusive channels. 
Our results show that the coupled-channel effects influence in particular the low energy part of the hadron spectra by side-feeding --- this is an important feature of our model that cannot be achieved with simple absorption models, as e.~g.~Glauber approaches.
Furthermore, we emphasize that our transport model has been extensively and successfully tested against experimental data for $\gamma A$, $e A$, $\pi A$ and also heavy ion collisions. Thus, the same theory is able to describe many different nuclear reactions with the same set of parameters and physics assumptions. Since there are no free nuclear parameters for neutrino-nucleus reactions, our theory should be able to give reliable predictions for $\nu A$ scattering relevant for future neutrino experiments.

This article is organized as follows. In \refch{sec:model} we first present our model for neutrino-nucleon scattering. Then we discuss the in-medium modifications needed in the case of scattering with a bound nucleon. We give a brief overview of the BUU transport model that is used for the calculation of exclusive channels. 
Our results for inclusive cross sections, pion production and nucleon knockout are presented in \refch{sec:results}. 
Our model is general since it includes all $\nu$ flavor for both neutrino and antineutrino, as well as charged and neutral current in any arbitrary nuclei. Here, we focus on charged current $\nu_{\mu}$ scattering off $^{56}$Fe. First results of our model were already presented in \refcite{Leitner:2005jg, Alvarez-Ruso:2006tv, leitner_diplom}. 
We close with a summary and conclusions in \refch{sec:summary}.

\section{Theory \label{sec:model}}

In the impulse approximation, the neutrino-nucleus interaction can be treated as a two-step process. The first step is the interaction of the neutrino with a single bound nucleon in the target nucleus. Then, as second step, we propagate the produced particles through the nucleus using the BUU transport model including a full coupled-channel treatment of the final-state interactions.
Below we describe those ingredients in some detail.

\subsection{Elementary neutrino-nucleon reaction \label{subsec:nuN}}

The cross section for the elementary interaction of a neutrino with a nucleon,
\be
\nu(k) + N(p) \to l^-(k') + X(p'),
\ee
with the momenta $k_{\alpha}=\left(E_{\nu},\myvec{k}\right)$, $k'_{\alpha}=\left(E_{l},\myvec{k}'\right)$, $p_{\alpha}=\left(E,\myvec{p}\right)$ and $p'_{\alpha}=\left(E',\myvec{p}'\right)$,
is given by
\be
\frac{{\rm d}^2\sigma_{\nu N}}{{\rm d}Q^2{\rm d}E_{l}} = \int {\rm d} \phi \; \frac{1}{64 \pi^2} \frac{1}{| k \cdot p|} \frac{1}{E_{\nu}} \delta\left(p'^2-M^{'\,2}\right) |\bar{\mathcal{M}}|^2  \label{eq:crosssection}
\ee
with $Q^2=-(k-k')^2$ and the azimuthal angle $\phi \in (0,2\pi)$. $M'$ denotes the (invariant) mass of the outgoing baryon.
The matrix element squared $|\bar{\mathcal{M}}|^2$ for charged current (CC) interactions, summed and averaged over spins, reads
\be
|\bar{\mathcal{M}}|^2=\frac{G_F^2 \cos^2 \theta_C}{2} L_{\alpha \beta} H^{\alpha \beta},
\ee
where the leptonic tensor, $L_{\alpha \beta}$, is given by 
\bea
L_{\alpha \beta} 
&=& {\rm Tr}\left[ (k \hspace{-2mm}/ \, + m_{\nu}) \gamma_{\alpha} (1-\gamma_5) (k' \hspace{-3mm}/ \, + m_{l}) \gamma_{\beta} (1-\gamma_5) \right] \nonumber \\
 &=& 8 \left[k_{\alpha}' k_{\beta} + k_{\alpha} k_{\beta}' - g_{\alpha \beta} k \cdot k' + i \epsilon_{\alpha \beta \rho \sigma} k^{\rho} k'^{\sigma} \right]. \label{eq:leptonictensor}
\eea
$m_l$ is the mass of the outgoing lepton. It has been shown, that in particular at low momentum transfer, the lepton mass becomes important~\cite{Alvarez-Ruso:1998hi,Lalakulich:2005cs}. Thus, we do not neglect it like in other works (e.~g.~\refcite{Paschos:2003qr}). The neutrino mass $m_{\nu}$, however, is set to zero.
For the corresponding antineutrino reaction, 
\be
\bar{\nu}(k) + N(p) \to l^+(k') + X(p'),
\ee
the antisymmetric term proportional to the fully antisymmetric tensor $\epsilon_{\alpha \beta \rho \sigma}$ gets a minus sign.

\begin{figure}
\begin{center}
\includegraphics[scale=.75]{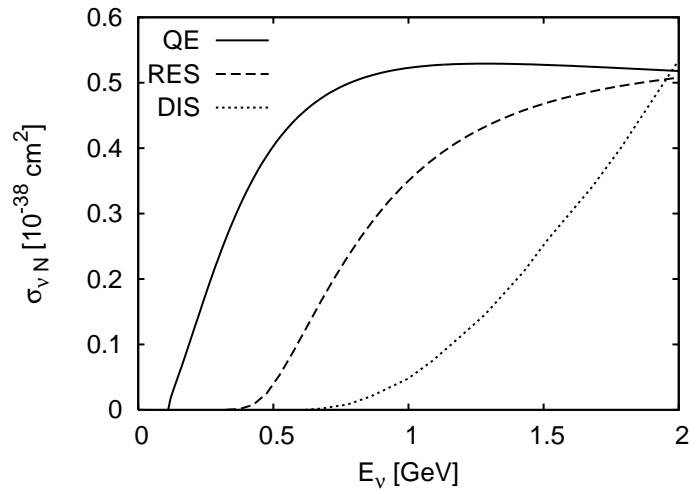}
\end{center}
\caption{Total neutrino cross sections for $\nu_{\mu} N \rightarrow \mu^- X$ for an isoscalar target as a function of the neutrino energy decomposed in the contributions of quasielastic scattering (QE), $\Delta$ excitation (RES) and deep inelastic scattering (DIS). The DIS curve is taken from \refcite{Ashie:2005ik}; the QE and RES curves are calculated with the formalism presented in \refsubch{subsec:nuN} averaging over protons and neutrons: $\sigma_{\nu_{\mu} N}= (\sigma_{\nu_{\mu} p} +\sigma_{\nu_{\mu} n})/2$.
\label{fig:decomposition_QE_RES_DIS}}
\end{figure}

The hadronic tensor $H^{\alpha \beta}$ depends on the specific reaction. In the intermediate energy region, i.~e.~up to neutrino energies of about 1.5 GeV, the reaction is dominated by two processes, namely quasielastic scattering (QE) and $\Delta$ production (cf.~\reffig{fig:decomposition_QE_RES_DIS}). 
Restricting to CC interactions, we consider in this work 
\be
\nu n \to l^- p, 
\ee
(note that the charge changing QE reaction is not possible on protons) and
\be
\nu n \to l^- \Delta^+, \quad \quad \nu p \to l^- \Delta^{++}.\label{eq:deltapl}
\ee
Both contributions to the $\nu N$ cross section are described within a fully relativistic formalism which is presented below.

We want to note that, even though quasielastic reactions and $\Delta$ production are the most important processes in the energy regime under investigation, for a full description one has to include higher-mass resonances, the non-resonant background and deep inelastic scattering (DIS) as well as more exotic channels such as strangeness production. 

\subsubsection{Quasielastic scattering} 

Since quantum chromodynamics (QCD) does not allow a direct calculation of the hadronic vertex, we need to construct an explicit expression for the hadronic current of the QE reaction based on general assumptions.
Following the arguments of \refcite{Nowakowski:2004cv} we construct the most general form from the four-vectors at our disposal: $p_{\alpha}$, $p'_{\alpha}$ and $q_{\alpha}=p'_{\alpha}-p_{\alpha}$. Gordon identities (cf.~e.~g.~Appendix~A.2 in \refcite{Itzykson:1980rhy}) limit the number of terms and we obtain 
\bea
J_{\alpha}^{QE}&=& \langle p |J_{\alpha}^{QE}(0) | n \rangle  \nonumber \\
          &=&  \bar{u}(p') A_{\alpha } u(p)
\eea
with
\be
A_{\alpha } =\left(\gamma_{\alpha} - \frac{q\hspace{-2mm}/ \,q_{\alpha}}{q^2} \right) F_1^V + \frac{i}{2 M_N} \sigma_{\alpha \beta} q^{\beta} F_2^V + \gamma_{\alpha} \gamma_5 F_A +  \frac{q_{\alpha}}{M_N}\gamma_5 F_P, \label{eq:qehadcurrent}
\ee
with $M_N$ being the nucleon mass.
$F_{1,2}^V$ are the vector form factors; $F_A$ and $F_P$ are the axial and pseudoscalar form factor, respectively. 
The term $(q\hspace{-2mm}/ \,q_{\alpha})/q^2$ appears when the masses of in- ($M$) and outgoing nucleons ($M'$) are not equal. It ensures that the vector part of the current is conserved even for nonequal masses.
This is the case in the nuclear medium where a momentum dependent mean-field can cause such a difference in masses (cf.~\refsubch{subsec:inmed}). This extra term vanishes by applying Gordon identities in the limit of $M=M'$, and \refeq{eq:qehadcurrent} reduces to the "standard expression" (see e.~g.~\refcite{LlewellynSmith:1971zm})
\be
A_{\alpha} \to \gamma_{\alpha} F_1^V + \frac{i}{2 M_N} \sigma_{\alpha \beta} q^{\beta} F_2^V + \gamma_{\alpha} \gamma_5 F_A +  \frac{q_{\alpha}}{M_N}\gamma_5 F_P.
\ee

The hadronic tensor $H^{\alpha \beta}$ for QE reactions is given by
\be
H^{\alpha \beta}_{QE}=\frac12 {\rm Tr} \left[ (p \hspace{-2mm}/ \, + M) \tilde{A}^{\alpha} (p' \hspace{-3mm}/ \, + M') A^{\beta}  \right] \label{eq:hadrtensorQE}
\ee
with
\be
\tilde{A}_{\alpha}=\gamma_0 A_{\alpha}^{\dagger} \gamma_0
\ee
and --- in the free nucleon case --- with $M=M'=M_N$.

The vector form factors $F_{1,2}^V$ can be related to electron scattering form factors. The conserved vector current hypothesis (CVC) implies that the vector part of the current in \refeq{eq:qehadcurrent} and the electromagnetic current are components of the same isospin multiplet of conserved currents, and that therefore their form factors are related. We obtain
\bea
F_1^V(Q^2)&=&\left[\left( G_E^p-G_E^n \right)+ \frac{Q^2}{4M_N^2}\left( G_M^p-G_M^n \right)\right] \left[1+\frac{Q^2}{4M_N^2}\right]^{-1}, \label{eq:f1sachs}\\
F_2^V(Q^2)&=&\left[\left( G_M^p-G_M^n\right)- \left( G_E^p-G_E^n \right)\right]\left[1+\frac{Q^2}{4M_N^2}\right]^{-1}.\label{eq:f2sachs} 
\eea
$G_M(Q^2)$ and $G_E(Q^2)$ are the magnetic and the electric form factors of the nucleon respectively, for which we take the BBA-2003 parametrization~\cite{Budd:2003wb}. It uses recent electron scattering data from JLab to account for deviations from the dipole $Q^2$-dependence.
By assuming pion pole dominance, we can use the partially conserved axial current hypothesis (PCAC) to relate $F_A$ and $F_P$,
\be
F_P(Q^2)=\frac{2 M_N^2}{Q^2+m_{\pi}^2}F_A(Q^2). \label{eq:pseudoff}
\ee
The axial form factor is given by the standard dipole form
\be
F_A(Q^2)=g_A \left(1+\frac{Q^2}{M_A^2}\right)^{-2};  \label{eq:axialff}
\ee
the axial vector constant $g_A=-1.267$ is obtained from $\beta$ decay. The axial mass $M_A$ has to be measured in neutrino experiments. We use $M_A=1.00 \myunit{GeV}$ which was extracted from data by the analysis of Budd et al.~\cite{Budd:2003wb}.

\begin{figure}
\begin{center}
\includegraphics[scale=.75]{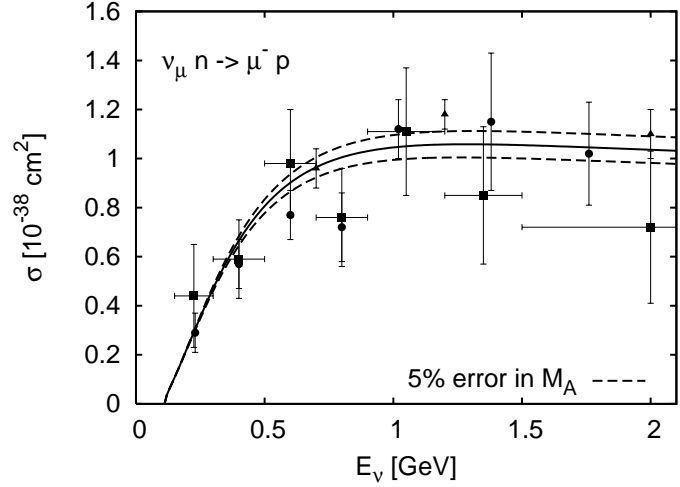}
\end{center}
\caption{Total QE cross section as a function of the neutrino energy (solid line). The datapoints are taken from Refs.~\cite{Barish:1977qk} ($\bullet$),  \cite{Mann:1973pr} ($\blacksquare$), \cite{Baker:1981su} ($\blacktriangle$).  The dashed lines are the results of a 5\% change in $M_A$ in \refeq{eq:axialff}.  \label{fig:QE_sigmatot}}
\end{figure}

With these ingredients we obtain the QE cross section on the neutron, which is presented in \reffig{fig:QE_sigmatot} in comparison with the available data on H$_2$ and D$_2$ targets.
The main source of uncertainty comes from the axial form factor whose $Q^2$ dependence has to be extracted from $\nu$ scattering data. Experimental analyses (see table 3 of \refcite{Budd:2003wb} and references therein) give $M_A$ with an accuracy of about 5\%. The impact of this error on the cross section is shown by the dashed lines. It is clear that more precise measurements are required to constrain more this parameter.

\subsubsection{$\Delta$ production \label{subsec:delta}} 

For a theoretical treatment of the neutrino-induced charged current $\Delta$ production we use a fully relativistic formalism (cf.~\refscite{Schreiner:1973mj,Fogli:1979cz,Alvarez-Ruso:1997jr,Singh:1998ha,Lalakulich:2005cs}). 
The hadronic current for the reaction in \refeq{eq:deltapl} is given by
\bea
J_{\alpha}^{\Delta}&=& \langle \Delta^{+} |J_{\alpha}^{\Delta}(0) | n \rangle  \nonumber \\
          &=&  \bar{\psi}^{\beta}(p') B_{\beta \alpha } u(p)
\eea
with
\bea
B_{\beta \alpha } &=& \left[ \frac{C_3^V}{M_N} (g_{\alpha \beta} q\hspace{-2mm}/ \, - q_{\beta} \gamma_{\alpha})+
  \frac{C_4^V}{M_N^2} (g_{\alpha \beta} q\cdot p' - q_{\beta} p'_{\alpha}) 
  + \frac{C_5^V}{M_N^2} (g_{\alpha \beta} q\cdot p - q_{\beta} p_{\alpha}) + g_{\alpha \beta} C_6^V\right] \gamma_{5} \nonumber  \\  
  &&\; \;\;\;\;+ \frac{C_3^A}{M_N} (g_{\alpha \beta} q\hspace{-2mm}/ \, - q_{\beta} \gamma_{\alpha})+
  \frac{C_4^A}{M_N^2} (g_{\alpha \beta} q\cdot p' - q_{\beta} p'_{\alpha})+
 {C_5^A} g_{\alpha \beta}
  + \frac{C_6^A}{M_N^2} q_{\beta} q_{\alpha},
\eea
where $\bar{\psi}^{\beta}(p')$ is the Rarita-Schwinger spinor for the $\Delta$, and $u(p)$ is the Dirac spinor for the nucleon; $q=k-k'$ is the momentum transfer. 
This yields the following hadronic tensor
\be
H^{\alpha \beta}_{\Delta}=\frac12 {\rm Tr}\left[ (p \hspace{-2mm}/ \, + M) \tilde{B}^{\alpha \rho} \Lambda_{\rho \sigma} B^{\sigma \beta}  \right]  \label{eq:hadrtensorDELTA}
\ee
with 
\be
\tilde{B}_{\alpha\beta}=\gamma_0 B_{\alpha\beta}^{\dagger} \gamma_0
\ee
and, for free nucleons, $M=M_N$.
The spin $3/2$ projection operator is given by 
\be
\Lambda_{\rho \sigma}=- \left(p'\hspace{-3mm}/ \, + \sqrt{p\,'^2} \right) \left( g_{\rho \sigma} - \frac{2}{3} \frac{p'_{\rho } p'_{\sigma }}{p\,'^2} 
     + \frac{1}{3} \frac{p'_{\rho } \gamma_{\sigma} - p'_{\sigma } \gamma_{\rho}}{\sqrt{p\,'^2}} - \frac{1}{3} \gamma_{\rho} \gamma_{\sigma} \right).
\ee
By isospin relations, we obtain for $\Delta^{++}$ production: 
\be
\langle \Delta^{++} |J_{\alpha}(0) | p \rangle = \sqrt{3} \langle \Delta^{+} |J_{\alpha}(0) | n \rangle. 
\ee

Basically, two approaches are considered in the literature for the $N-\Delta$ transition form factors $C_i^{V,A}$ with $i=3, \ldots, 6$. 
They can be phenomenological with parameters extracted from neutrino and electron scattering data or calculated using quark models. Early attempts for the latter are reviewed in \refcite{Schreiner:1973mj} (see also references therein); more recent ones are summarized in \refcite{Liu:1995bu}. In their model of resonance production, Rein and Sehgal~\cite{Rein:1980wg} adopted the quark model of Feynman, Kislinger and Ravndal~\cite{Feynman:1971wr}. 
A more recent calculation was done by Liu et al.~\cite{Liu:1995bu} who applied the Isgur-Karl quark model and by Sato et al.~\cite{Sato:2003rq} who developed a dynamical model including pion cloud effects.

As in QE scattering, we choose the first approach and use phenomenological form factors. 
The vector form factors $C_i^V$ can be related  to the ones obtained in electron scattering by applying CVC. 
This, together with the assumption of $M_{1+}$ dominance of the electroproduction amplitude, yields~\cite{Fogli:1979cz}
\be
C_6^{V}(Q^2)=0, \quad C_5^{V}(Q^2)=0 \quad \text{and} \quad C_4^{V}(Q^2)=-\frac{M_N}{\sqrt{p\,'^2}}C_3^{V}(Q^2).
\ee
This leaves only one independent vector form factor, $C_3^{V}$, which can be parametrized in various ways to describe pion electroproduction data. The data show that the $Q^2$-dependence is steeper than that of a dipole. Thus we use a modified dipole form and adopt a parametrization used in \refcite{Paschos:2003qr}
\be
C_3^{V}(Q^2)=C_3^V(0) \left( 1+\frac{Q^2}{M_V^2} \right)^{-2}  \left(1+\frac{Q^2}{4 M_V^2}\right)^{-1},
\ee
with $C_3^V(0)=1.95$ and $M_V=0.84 \myunit{GeV}$.

In the axial sector, we apply similar techniques as in the QE case. Pion pole dominance yields for $C_6^{A}$~\cite{Schreiner:1973mj}
\bea
C_6^{A}(Q^2)= \frac{g_{\Delta N \pi} f_{\pi}}{\sqrt{6}M_N} \frac{M_N^2}{Q^2+m_{\pi}^2} F_{\pi}(Q^2)
\eea
with $m_{\pi}$ being the pion mass . $g_{\Delta N \pi}$ is the $\Delta^{++} \to p \pi^+$ coupling constant and $f_{\pi}$, the pion decay constant. $F_{\pi}(Q^2)$ is the vertex form factor with $F_{\pi}(Q^2=m_{\pi}^2)=1$.
This relation together with the assumption of PCAC connects $C_6^{A}$ and $C_5^{A}$ in the axial current
\bea
C_6^{A}(Q^2)=C_5^{A} (Q^2)\frac{M_N^2}{Q^2+m_{\pi}^2}.
\eea
In the limit $Q^2=0$ and with the assumption that $F_{\pi}(Q^2)$ is a slowly variating function with $F_{\pi}(Q^2=m_{\pi}^2) \approx F_{\pi}(Q^2=0) =1$, we obtain the off-diagonal Goldberger-Treiman relation
\bea
C_5^{A} (0)= \frac{g_{\Delta N \pi} f_{\pi}}{\sqrt{6}M_N} \simeq 1.2.
\eea
This coupling was extracted from the BNL data in \refcite{Alvarez-Ruso:1998hi} and found to be consistent with the PCAC prediction. 

Since there are no other theoretical constraints for $C_3^{A}(Q^2), C_4^{A}(Q^2)$ and $C_5^{A}(Q^2)/C_5^{A}(0)$, they have to be fitted to neutrino scattering data. The available information comes mainly from two bubble chamber experiments, the 12-foot bubble chamber at Argonne (ANL)~\cite{Barish:1978pj,Radecky:1981fn} and the 7-foot bubble chamber at Brookhaven (BNL)~\cite{Kitagaki:1990vs}. The fits adopted the Adler model~\cite{Adler:1968tw} where 
\be
C_4^{A}(Q^2)=-\frac{C_5^{A}(Q^2)}{4} \quad \text{and} \quad  C_3^{A}(Q^2)= 0.
\ee
For $C_5^{A}$ we have taken, as in \refcite{Paschos:2003qr}, again a modified dipole
\be
C_5^{A}(Q^2)=C_5^A(0) \left( 1+\frac{Q^2}{M_A^2} \right)^{-2} \left(1+\frac{Q^2}{3 M_A^2}\right)^{-1},
\ee
with $M_A=1.05 \myunit{GeV}$. We conclude that the neutrino-induced $\Delta$ excitation is fully described by two independent form factors, $C_3^V$ and $C_5^{A}$. 

The finite width of the $\Delta$ resonance is accounted for in the cross section of \refeq{eq:crosssection} by replacing  
\be
\delta\left(p'^2-M_{\Delta}^2\right) \to  - \frac{1}{\pi} \mathcal{I}m \left(\frac{1}{p'^2-M_{\Delta}^2+i \sqrt{p'^2} \Gamma}\right),\label{eq:spectraldelta}
\ee
where we used $M'=M_{\Delta}$ with $M_{\Delta}=1.232 \myunit{GeV}$.

For the free decay width $\Gamma$ we use an energy dependent $P$-wave width as required by angular momentum conservation. It can be parametrized by using Blatt-Weisskopf functions~\cite{manley}
\be
\Gamma=\Gamma_0 \frac{\beta(W)}{\beta(M_{\Delta})}
\ee
with $\Gamma_0=0.118$ GeV, $W=\sqrt{p'^2}$ and 
\be
\beta(W)=\frac{q_{CM}(W)}{W} \frac{(q_{CM}(W) R)^2}{1+( q_{CM}(W)R)^2}.
\ee
For the interaction length $R$ we take $R=1 \myunit{fm}$.
$q_{CM}$ is the pion momentum in the rest frame of the resonance, for which we obtain
\be
q_{CM}(W)=\frac{\sqrt{(W^2-m_{\pi}^2-M_N^2)^2 - 4 m_{\pi}^2 M_N^2}}{2 W}.
\ee

\begin{figure}[tb]
  \begin{minipage}[t]{.78\textwidth}
      \begin{center}
        \includegraphics[height=5.5cm]{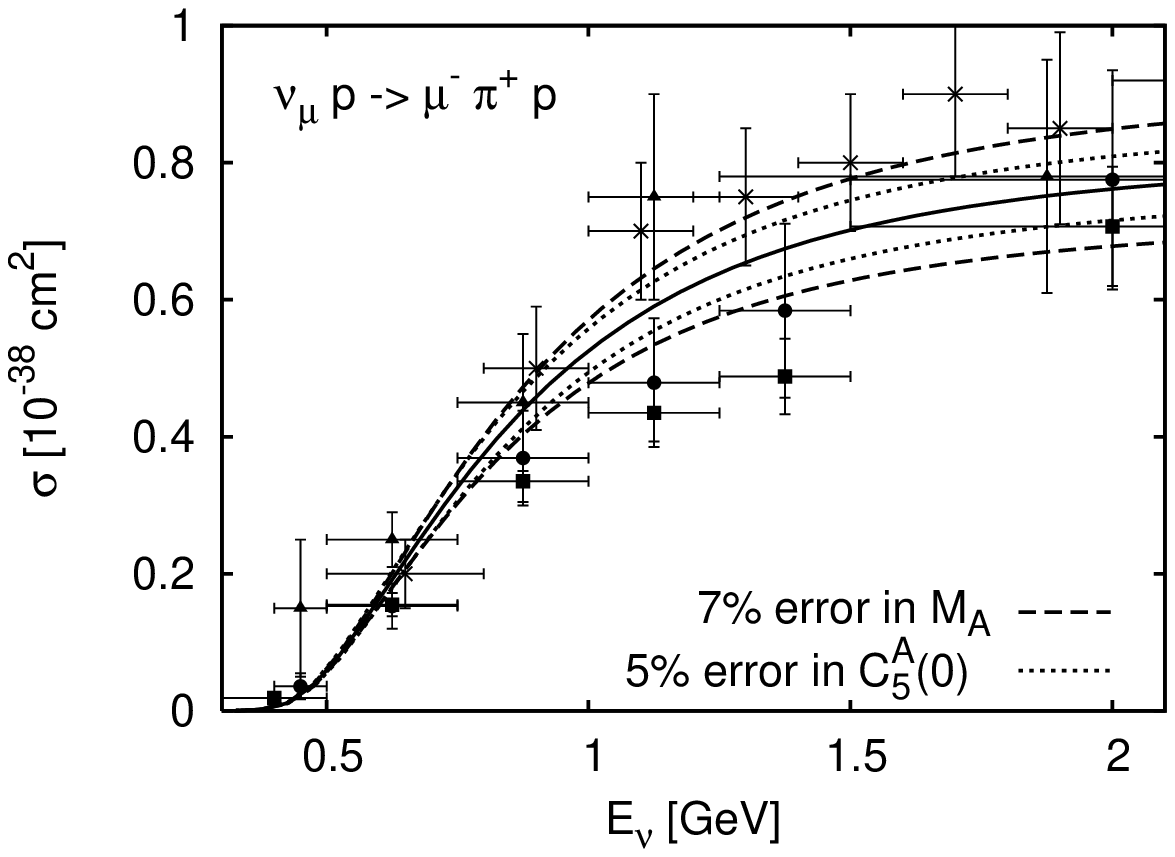}
       \end{center}
  \end{minipage}
 \\ \hfill \\
  \begin{minipage}[t]{.48\textwidth}
      \begin{center}
        \includegraphics[height=5.5cm]{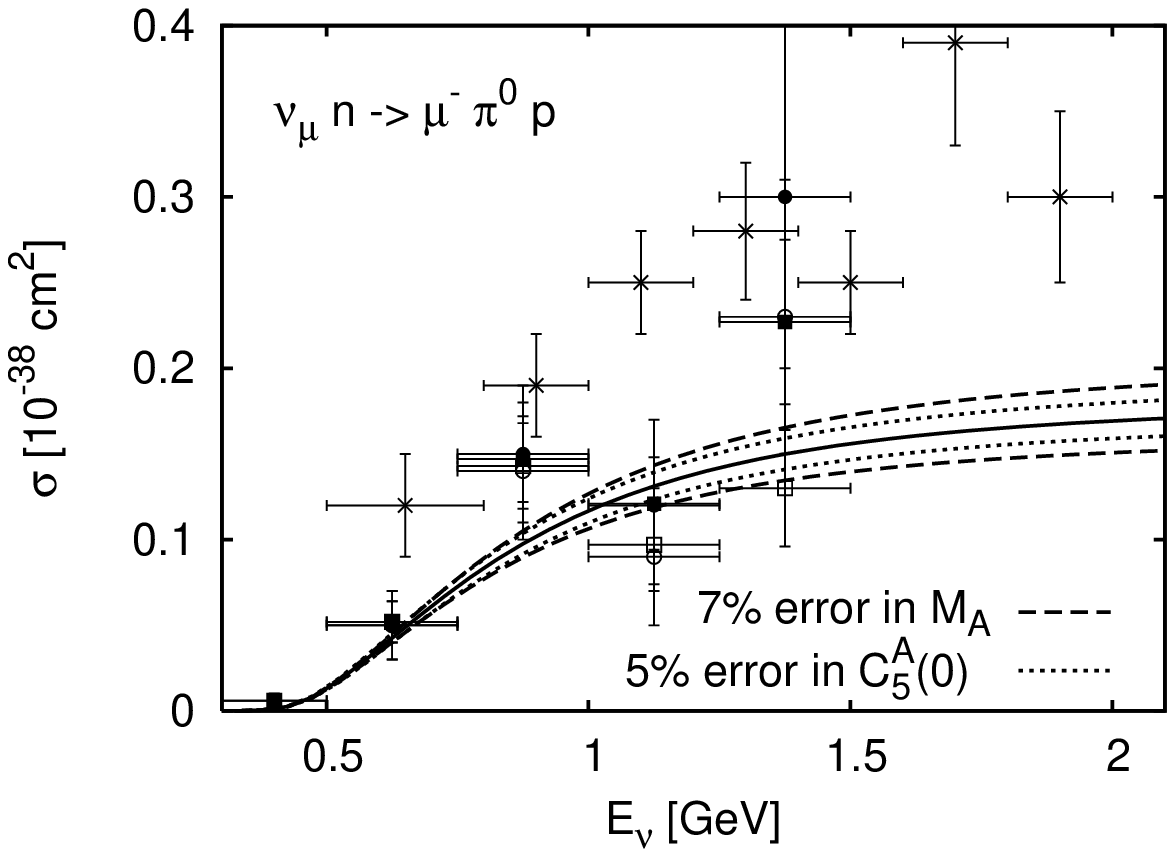}
      \end{center}
  \end{minipage}
  \begin{minipage}[t]{.48\textwidth}
      \begin{center}
        \includegraphics[height=5.5cm]{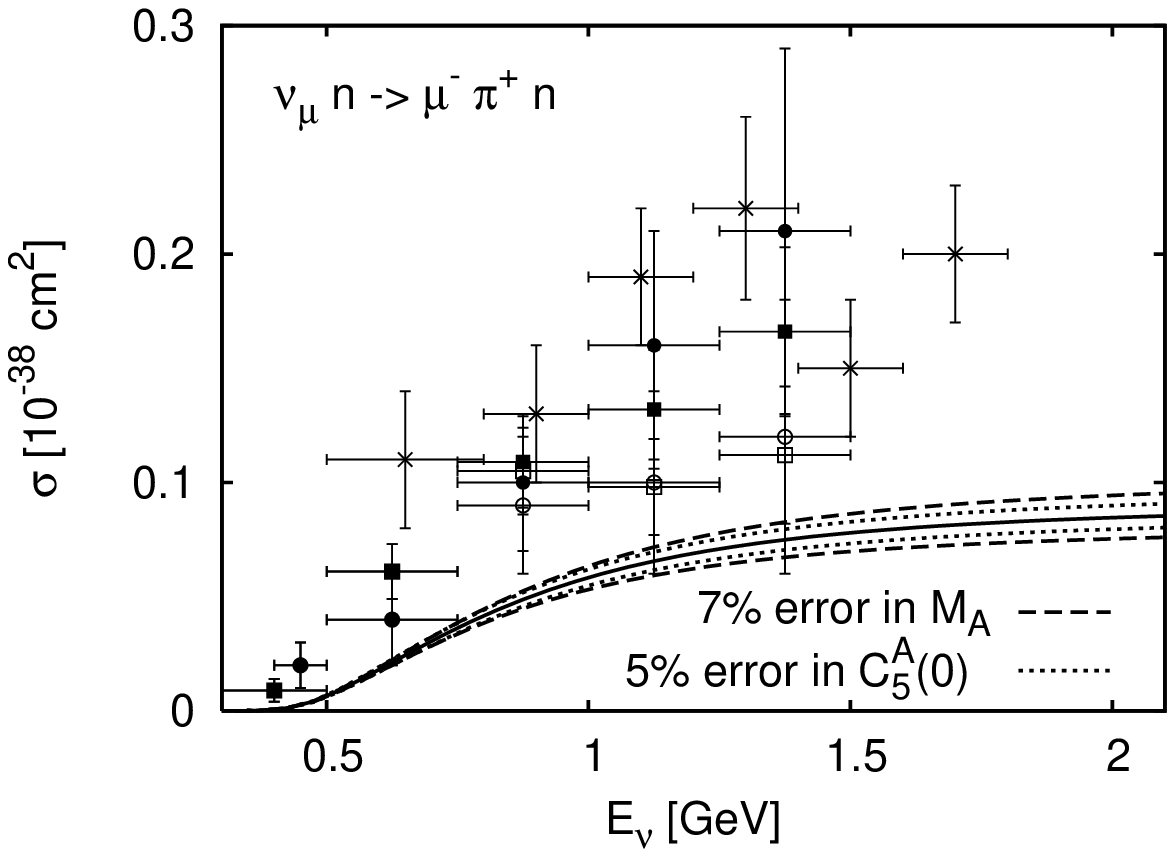}
      \end{center}
  \end{minipage}
\caption{Total pion production cross sections through $\Delta$ excitation as a function of the neutrino energy (solid lines) compared to the pion production data of Refs.~\cite{Barish:1978pj} ($\bullet$ without and $\circ$ with mass cut at $W < 1.4$~GeV), \cite{Radecky:1981fn} ($\blacksquare$  without and $\square$ with mass cut at $W < 1.4$~GeV), \cite{Kitagaki:1983px} ($\times$), \cite{Campbell:1973wg} ($\blacktriangle$).  Dashed and dotted lines reflect the uncertainties of $C_5^A(0)$ and $M_A$, respectively. See text for the discussion of the discrepancies between the calculation and the data in the lower panels. Note also the different scales.    \label{fig:Delta_sigmatot}}  
\end{figure}

With these ingredients, we obtain the cross section for CC $\Delta^{++}$ and $\Delta^+$ production on proton and neutron, respectively. Once produced, the $\Delta$ decays into $\pi N$ pairs. For the amplitudes $\mathcal{A}$ of pion production the following isospin decomposition applies
\begin{eqnarray}
\mathcal{A}(\nu_l \, p \to l^- p \,\pi^+)&=& \mathcal{A}_3, \\
\mathcal{A}(\nu_l \, n \to l^- n \,\pi^+)&=& \frac{1}{3}\mathcal{A}_3 + \frac{2 \sqrt{2}}{3} \mathcal{A}_1, \label{eq:nppl} \\
\mathcal{A}(\nu_l \, n \to l^-  p \, \pi^0)&=& -\frac{\sqrt{2}}{3}\mathcal{A}_3 + \frac{2}{3} \mathcal{A}_1, \label{eq:ppnull}
\end{eqnarray}
with $\mathcal{A}_3$ being the amplitude for the isospin $3/2$ state of the $\pi N$ system, predominantly $\Delta$, and $\mathcal{A}_1$, the amplitude for the isospin $1/2$ state which is not considered here. Those three channels are plotted for $\nu_{\mu}$ in \reffig{fig:Delta_sigmatot} together with pion production data from ANL and BNL on H$_2$ and D$_2$ targets.
Since $\pi^+ p$ production is dominated by $\Delta$ excitation, this channel serves as a quality check of our model, and a good agreement with the data is reached. In the case of $\pi^+ n$ and $\pi^0 p$ our results lie systematically below the data. This is mainly due to the non-negligible contribution from the heavier isospin $1/2$ resonances like $N(1440)$, $N(1520)$ and $N(1535)$, as can be seen in \refcite{Fogli:1979cz} where those states are considered. We achieve a good agreement with the ANL data when higher invariant masses are excluded ($W < 1.4$~GeV --- open squares and open circles) \cite{Barish:1978pj, Radecky:1981fn}.
For the BNL experiment, data with cuts are not available, however, their analysis indicates, that at energies of 0.5~-~6~GeV approximately half of the $\pi^+ n$ and $\pi^0 p$ events correspond to invariant masses $W > 1.4$~GeV (cf.~Fig.~6 of \refcite{Kitagaki:1983px}).  The underestimate in the (small) $\pi^+ n$ channel translates into an error of about 10\% in the isospin-averaged total $\pi^+$ production cross section at $E_{\nu}=1$ GeV. This is well within any experimental uncertainties.

As in the QE case, the main source of error for $\Delta$ production is contained in the axial form factors and, in particular, in $C_5^A$ which gives the dominant contribution to the cross section at low $Q^2$.
We have investigated the sensitivity of the cross section to the value of $C_5^A$ at $Q^2=0$ assuming an error of 5\% as given in \refcite{Alvarez-Ruso:1998hi} (dotted lines). Furthermore, we show the impact of the uncertainties in the $Q^2$ dependencies  of the form factors by assuming a 7\% error in the determination of $M_A$ as reported in the ANL and BNL  analyses  \cite{Kitagaki:1990vs, Radecky:1981fn} (dashed lines).

\subsection{In-medium modifications of the $\nu N$ reaction  \label{subsec:inmed}}

If the nucleon is not free but bound in the nucleus we must account for nuclear effects such as Fermi motion of the initial nucleons and Pauli blocking of the final ones. In local density approximation (LDA), the local Fermi momenta of the nucleons are given by 
\be
p_F(\myvec{r})=\left( \frac{3}{2} \pi^2 \rho(\myvec{r}) \right)^{1/3}.
\ee
For the density distribution we take a Woods-Saxon parametrization
\be
\rho(r)=\rho_0 \left(1+\exp\frac{r-r_0}{\alpha}\right)^{-1}, \label{eq:woodssaxon}
\ee
with parameters extracted from Hartree-Fock calculations~\cite{lenske}. The corresponding values for $^{56}$Fe are 
$r_0=4.22 \myunit{fm}$, $\alpha=0.477 \myunit{fm}$ and $\rho_0=0.158 \myunit{fm}^{-3}$.
In neutrino-induced reactions, the phase-space density $f$ is given by
\be
f(\myvec{r},\myvec{p}) \sim \Theta \left( p_F(\myvec{r})- |\myvec{p}|\right).
\ee
We can then approximate the probability that a final state of a nucleon is not Pauli blocked by
\be
P_{Pauli}=1-\Theta \left( p_F(\myvec{r}) - \left| \myvec{p} \right|\right).
\ee

Furthermore, we take into account that the nucleons are bound in a density and momentum dependent scalar potential 
\be
U=U(\myvec{r},\myvec{p}),
\ee
which can be related to an effective mean-field potential as outlined in the following.
The general expression for a relativistic one-particle Hamiltonian for a particle of mass $m$ is given as
\be
H= \sqrt{\left( m + S \right)^2 + \left(\myvec{p} - \vec{U}_V\right)^2} + U_V^0,
\ee
where $S$ is a scalar potential and $(U_V^0,\vec{U}_V)$ a vector potential.
The spatial components of $\vec{U}_V$ vanish in the rest frame of the nucleus (NRF). $S$ is set to zero, and we identify the effective potential $V$ with the zeroth component $U_V^0$, which leads to 
\be
H_{NRF}=\sqrt{m^2 + \myvec{p}_{NRF}^{\;2}} + V.
\ee
The scalar potential $U$ in any frame is now defined as
\be
U=\sqrt{H_{NRF}^2 - \myvec{p}_{NRF}^{\;2}} -m.
\ee
The effective potential $V$, which  describes many-body interactions of the nucleons, can be parametrized as a sum of a Skyrme part, only depending on the density $\rho$, and a momentum dependent part. We take a parametrization of \refcite{welke},
\be
V(\myvec{r},\myvec{p})=A \frac{\rho(\myvec{r})}{\rho_0} + B \left( \frac{\rho(\myvec{r})}{\rho_0}\right)^{\tau} 
+ \frac{2 C}{\rho_0}g \int \frac{{\rm d}^3p'}{(2 \pi)^3}\frac{f(\myvec{r},\myvec{p})}{1+\left( \frac{\myvec{p}-\myvec{p'}}{\Lambda}\right)^2}. \label{eq:meanfield}
\ee
Throughout this work we use $A=-29.3 \myunit{MeV}$, $B=57.2 \myunit{MeV}$, $C=-63.5 \myunit{MeV}$, $\tau=1.76$ and $\Lambda=2.13 \myunit{fm}^{-1}$.
These parameters are fitted to the saturation density of nuclear matter, and also to the momentum dependence of the nucleon optical potential as measured in $pA$ collisions~\cite{Effenberger:1999ay}.

Note that the introduced quantities describe the structure of the nucleus and do not depend on the particular nuclear reaction. Therefore, there are no free nuclear parameters for $\nu A$ scattering.

If a particle is bound in the nucleus it acquires an effective mass, defined as
\be
M_{eff}=M + U(\myvec{r},\myvec{p}).  \label{eq:intromeff}
\ee
As consequence of the density and momentum dependence of the scalar potential 
the effective masses of initial and final particles are different even if their masses at rest are equal. 
This fact is taken into account for QE scattering by replacing $M$ and $M'$ in the cross section formula (\refeq{eq:crosssection}) and in the QE hadronic tensor (\refeq{eq:hadrtensorQE}) with
\bea
M &\to& M_N + U(\myvec{r},\myvec{p}), \\
M' &\to& M_N + U(\myvec{r},\myvec{p}').
\eea

Analogously, for $\Delta$ excitation, the effect of the binding potential is considered by substituting  
$M$ and $M'$ in the cross section formula (\refeq{eq:crosssection}, cf.~\refeq{eq:spectraldelta}) and in the $\Delta$ hadronic tensor (\refeq{eq:hadrtensorDELTA}) by
\bea
M &\to& M_N + U(\myvec{r},\myvec{p}), \\
M' &\to& M_{\Delta} + U_{\Delta}(\myvec{r},\myvec{p}').
\eea
We consider that the $\Delta$ is less bound in the nucleus than the nucleons by setting the $\Delta$ potential $U_{\Delta}$ to $2/3$ of the nucleon potential $U$. This choice is motivated by the phenomenological value of $U_{\Delta}= -30 \myunit{MeV}$ at normal nuclear density $\rho_0$~\cite{Ericson:1988gk}.

Furthermore, we take into account that the width of the $\Delta$ is modified in the nuclear medium.
While the nucleons inside the nucleus are constrained to have momenta below the Fermi momentum, there is no such constraint for the production of the resonances. Their decay, however, is influenced by Pauli blocking, e.~g.~a resonance decaying into a pion nucleon pair is Pauli blocked if the nucleon's momentum is below the Fermi momentum. Therefore, the width of the resonance inside the nuclear medium is lowered due to Pauli blocking. On the other hand, it is increased by collisions inside the medium. The collisional width $\Gamma_{coll}$ accounts for additional decay channels of the $\Delta$ inside the nucleus. Through two-body and three-body absorption processes like $\Delta N \to N N$ or $\Delta N N \to N N N$, the $\Delta$ can disappear without producing a pion, while via $\Delta N \to \pi N N$ additional pions can be produced. Also elastic scattering $\Delta N \to \Delta N$ contributes to the collisional broadening. For $\Gamma_{coll}$ we use the results of Oset and Salcedo~\cite{Oset:1987re} that have been conveniently parametrized.
Therefore, we have for the total in-medium width
\be
\Gamma^{med}_{tot} = \tilde{\Gamma} + \Gamma_{coll}, \label{eq:delta_mod_width}
\ee
which then replaces $\Gamma$ in \refeq{eq:spectraldelta}. 
Note that even though the Pauli blocked decay width $\tilde{\Gamma}$ reduces the total in-medium width, the overall effect is a broadening of the $\Delta$ in the medium (cf.~e.~g.~\refcite{Buss:2006vh}).
Numerical results, which illustrate the influence on the cross section of the different in-medium modifications in comparison to the elementary case, are presented in \refsubch{subsec:inclusivecs}.

Notice that we do not yet include any modifications of the form factors due to the nuclear medium.  Moreover, it has been shown for QE scattering by Singh et al.~\cite{Athar:2005hu} and Nieves et al.~\cite{Nieves:2004wx, Nieves:2005rq} that the renormalization of the weak transition strength in the nuclear medium has a non-negligible impact on the cross section. We shall consider this additional modification in the future.

\subsection{Final-state interactions}

Final-state interactions (FSI) of the produced particles --- in our case nucleons and $\Delta$ resonances --- are implemented by means of the semi-classical Boltzmann-Uehling-Uhlenbeck (BUU) transport model. This approach allows a full coupled-channel treatment of the FSI.
Originally developed to describe heavy-ion collisions~\cite{Teis:1996kx, Hombach:1998wr, Wagner:2004ee}, the Giessen BUU model (GiBUU) has been extended to describe interactions of pions, photons and electrons with nuclei~\cite{Weidmann:1997vj, Effenberger:1999ay, Lehr:1999zr, Falter:2004uc, Buss:2006vh}. In particular, it has been successfully tested against the existing data on the interaction of photons and electrons with nuclei~\cite{Falter:2003uy, Alvarez-Ruso:2004ji} --- an important prerequisite for any model aiming at the description of the interaction of neutrinos with nuclei. 
The applicability to and test in many different nuclear reactions is 
a strength of our model, which describes many different processes using the same physics input. In this respect, the study of $\nu A$ scattering is a natural extension of our previous work and does not introduce any new free parameter. Here we review only the main ingredients; details of the Giessen BUU model can be found in the references given above.

The BUU equation describes the space-time evolution of a many-particle system under the influence of a mean-field potential and a collision term. A separate transport equation for each particle species is required. For a particle of type $i$
it is given by
\be 
\left({\partial_t}+\vec\nabla_p H\cdot\vec\nabla_r
  -\vec\nabla_r H\cdot\vec\nabla_p\right)F_i(\myvec{r},\myvec{p}, m; t)
= I_{coll}[F_i,F_N,F_\pi,F_{\Delta},...], \nonumber
\ee
where $F_i(\myvec{r},\myvec{p}, m; t)$ is the phase-space density at time $t$; $\myvec{r},\myvec{p}$ are the coordinates and the momentum of this particle. $H$ is the relativistic Hamilton function of a particle of mass $m$ in a scalar potential $U$, which is given by
\be
H=\sqrt{\left[ m + U(\myvec{r},\myvec{p})\right]^2 + \myvec{p}^{\,2} }.  \label{eq:hamilton}
\ee
For nucleons, the scalar potential $U$ was explicitly discussed in \refsubch{subsec:inmed}. The same potential is used for all baryons except the $\Delta$ resonance, for which we take $U_{\Delta} = \frac23 U$. We include neither potentials for the mesons nor Coulomb corrections. 

$I_{coll}$ is the collision integral which accounts for changes in the phase-space density due to elastic and inelastic collisions between the particles. If the respective particle decays into other hadrons, this also accounts for changes in $F_i$. 
Thus, $I_{coll}$ consists of a gain and a loss term~\cite{Effenberger:1999ay,Falter:2004uc} which include also Pauli blocking.  
All BUU equations are coupled through $I_{coll}$ and, with less strength, also through the potential in $H$. In this coupled-channel treatment of the FSI, our model differs from standard Glauber approaches since the collision term allows not only for absorption but also for side-feeding and rescattering. 

In between the collisions, all particles are propagated in their mean-field potential according to the BUU equation. Our model allows for off-shell transport of resonances as described in \refscite{Effenberger:1999ay,Falter:2004uc}.

In the non-strange mesonic sector we include $\pi$, $\eta$, $\rho$, $\omega$, $\sigma$, $\phi$, $\eta'$. A full list of all mesons (also those containing strangeness and charm) with their properties is given in \refcite{Falter:2004uc}.
Besides the nucleon and the $\Delta$ resonance, we include in the baryonic sector 29 other nucleon resonances~\cite{Effenberger:1999ay}. Their properties are taken from an analysis done in \refcite{manley}. They can couple to the channels $N \pi$, $N \eta$, $\Lambda K$, $N \omega$, $\Delta \pi$, $N \rho$, $N \sigma$, $N^*(1440) \pi$ and $\Delta \rho$. 
Also in the baryonic sector, strange and charmed hadrons are included and we refer to \refscite{Effenberger:1999ay,Falter:2004uc} for a full list. Finally, we take into account the corresponding antiparticle of each baryon.
More details on the cross sections in the resonance region and their in-medium modification are given in \refscite{Effenberger:1999ay,Lehr:1999zr, Buss:2006vh}. Above invariant energies of $\sqrt{s}>2.6 \myunit{GeV}$ for baryon-baryon collisions or $\sqrt{s}>2.2 \myunit{GeV}$ for meson-baryon collisions, the FRITIOF string fragmentation model is used~\cite{Andersson:1992iq}. Thus, our transport model is also able to describe the final-state interactions of high-energy particles and can therefore be extended in the future to the study of neutrino scattering in the DIS regime, where it has already been successfully applied to the electroproduction of hadrons in nuclei~\cite{Falter:2004uc}.

However, the most important particles for this work are the nucleon, the $\Delta$ resonance and the pion. 
For the $NN$ cross section and its angular dependence we use a parametrization of \refcite{Cugnon:1996kh} which has been fitted to available data. The $\Delta$ resonance is propagated off-shell in our model --- its decay is isotropic. Absorption processes of the $\Delta$ are implemented by an absorption probability depending on the in-medium width given in \refeq{eq:delta_mod_width}.  The treatment of pions, and the whole $\pi N \Delta$ dynamics, in our BUU model has undergone numerous previous tests in $A \, A \to \pi \, X$~\cite{Teis:1996kx}, $\pi \, A \to X$~\cite{Engel:1993jh, Buss:2006vh} and $\gamma \, A \to \pi \, X$~\cite{Lehr:1999zr} reactions (see \refcite{Buss:2006vh} for the cross sections used in the present article). In particular, in \refcite{Krusche:2004uw} quantitative comparisons of calculated $\gamma \, A \to \pi^0 \, X$ cross sections to experiment have been given. These results, which contain the side-feeding of the $\pi^0$ channel from originally produced charged pion channels, show that the measured pion momentum distributions are described very well by our theory. The method has, because of its semiclassical nature, a lower limit of validity of about 20 - 30 MeV pion kinetic energy~\cite{Buss:2006vh}; this limitation is of no concern for the results to be discussed later. The comparison in \refcite{Krusche:2004uw} to data obtained with a few hundred MeV photon beam, roughly corresponding to the energy regime treated in this article, also shows that the deviations from experiment are typically of the order of 20\%. We thus expect a similar systematic uncertainty in the pion results reported in this article.

As we shall see in \refsubch{subsec:exclusivech}, FSI lead to absorption, charge exchange, a redistribution of energy and momentum as well as to the production of new particles.

\section{Results\label{sec:results}}

We now present different results for CC neutrino-induced reactions on nuclei at neutrino energies up to 2~GeV where 
we have studied, as a representative example, the reaction $\nu_{\mu}\, ^{56}\text{Fe} \to \mu^- X$.
Iron will be one of the targets of the upcoming experiment MINER$\nu$A~\cite{minervaprop} which aims at a precision measurement of the neutrino cross sections. 
In order to compare with other calculations, some results are also presented for $\nu_{\mu}$ scattering off $^{16}\text{O}$ and $^{40}\text{Ar}$. However, we want to emphasize that our model is applicable for CC and NC scattering of $\nu$ and $\bar{\nu}$ of all flavors off any nucleus (from $^{12}$C on). 

In this section we first discuss the inclusive cross section for QE scattering and $\Delta$ production and its sensitivity to the in-medium modifications introduced in \refsubch{subsec:inmed}. By accounting for FSI within the Giessen BUU model we can investigate in detail pion production and nucleon knockout --- results are presented in \refsubch{subsec:exclusivech}.

\subsection{Inclusive cross sections \label{subsec:inclusivecs}}

By inclusive we understand the $\nu A$ reaction in which a muon is produced, regardless of the rest of the outcome. 
The cross section for the $\nu A$ reaction is given by an integration over the contributions of all nucleons.

\begin{figure}
	\begin{center}
	\includegraphics{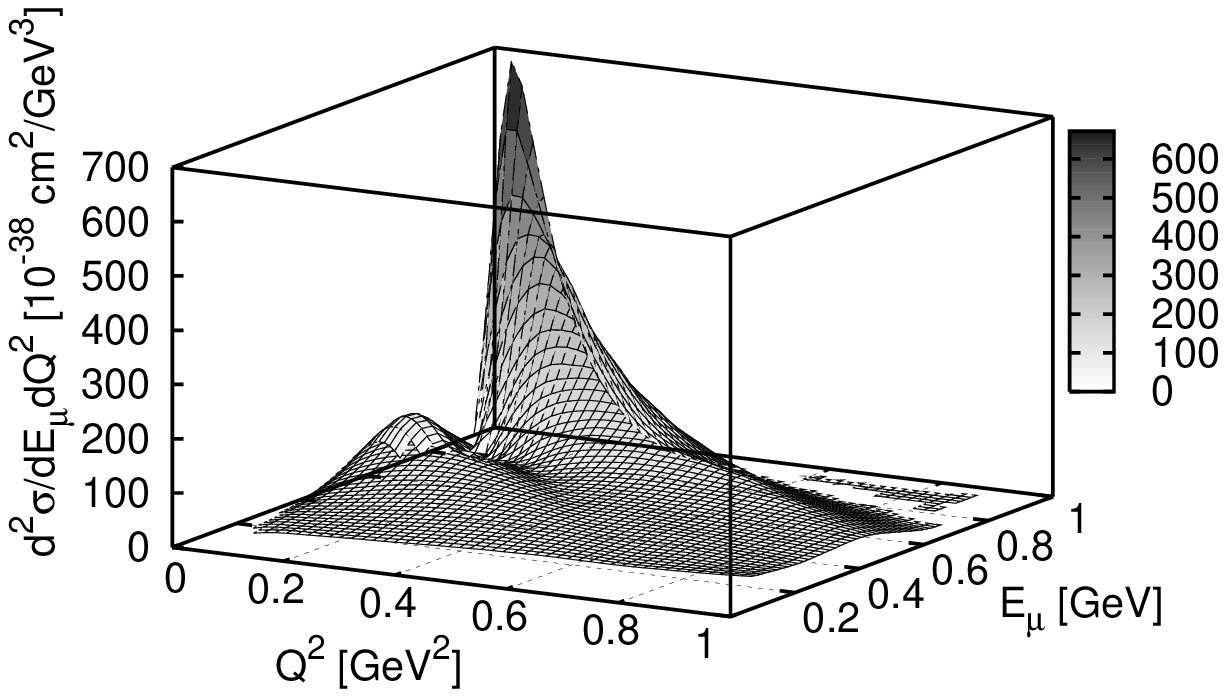}
	\end{center}
	\vspace{1cm}
	\begin{center}
	\includegraphics[height=8cm]{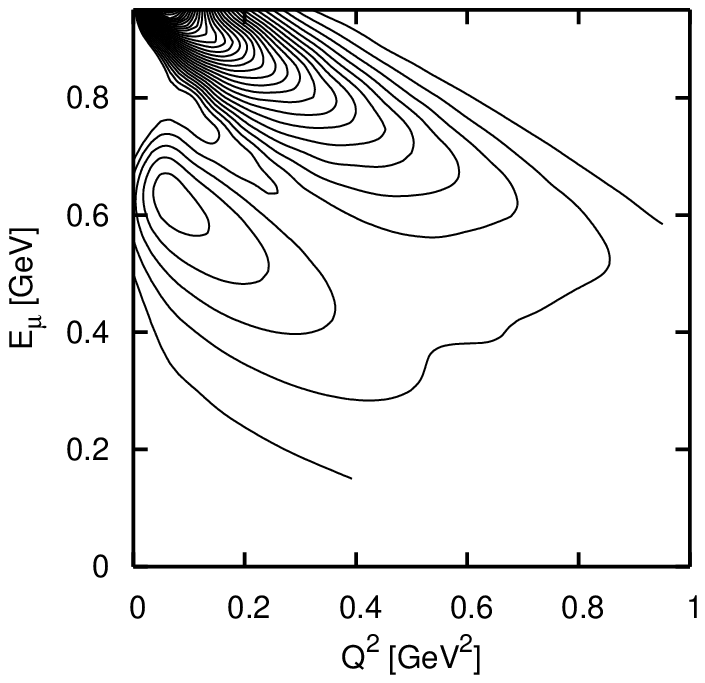}
	\end{center}
\caption{Inclusive double differential cross section ${\rm d}^2\sigma/({\rm d}Q^2 {\rm d} E_{\mu})$ on $^{56}\text{Fe}$ at $E_{\nu}=1 \myunit{GeV}$.  The higher peak is due to QE scattering, the lower one is due to $\Delta$ excitation. Included are all in-medium modifications: Fermi motion, Pauli blocking, binding energies and the modification of the $\Delta$ width in the medium. The 3-dimensional plot of the upper panel is projected to the $Q^2-E_{\mu}$ plane in the lower panel where the contour lines increase from $0$ to $680 \times 10^{-38} \text{cm}^2/\text{GeV}^3$ equidistantly by $20 \times 10^{-38} \text{cm}^2/\text{GeV}^3$ per contour. \label{fig:incl3d}}
\end{figure}

In \reffig{fig:incl3d}, we show  the inclusive double differential cross section on $^{56}\text{Fe}$ at $E_{\nu}=1 \myunit{GeV}$ taking into account Fermi motion, Pauli blocking, binding potentials and in-medium widths. 
At $Q^2<0.4 \myunit{GeV}$ two peaks can be clearly distinguished. The one at higher $E_{\mu}$ corresponds to quasielastic events, whereas the one at lower $E_{\mu}$ results from $\Delta$ production. 
At higher $Q^2$, the two peaks overlap, the distinct peak structure vanishes and the inclusive cross section tends to zero.

\begin{figure}
	\begin{center}
	\includegraphics[scale=.75]{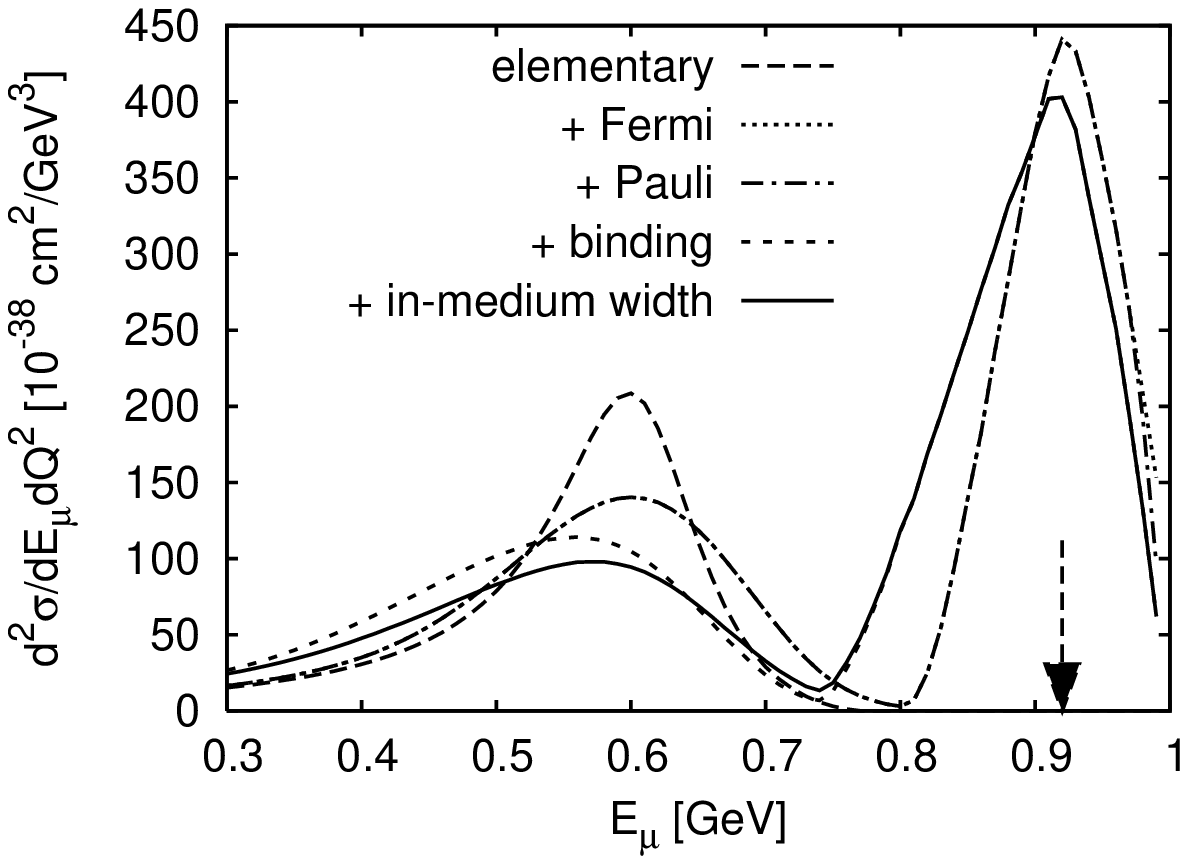}
	\caption{Inclusive double differential cross section ${\rm d}^2\sigma/({\rm d}Q^2 {\rm d} E_{\mu})$ on $^{56}\text{Fe}$ at $E_{\nu}=1 \myunit{GeV}$ and $Q^2=0.15\myunit{GeV}^2$. The left peak is due to $\Delta$ excitation, the right one to QE scattering. 
	The dashed line shows the elementary cross section for $\Delta$ production. The position of the $\delta$-function of the QE cross section is indicated by the arrow. We subsequently include Fermi motion (dotted line, this line overlaps with the dash-dotted line except at $E_{\mu} \to 1$ GeV), Pauli blocking (dash-dotted line), binding energies (short-dashed line) and the in-medium modification of the $\Delta$ width (solid line).\label{fig:incldoublediff}}
	\end{center}
\end{figure}

To understand the influence of the different in-medium effects, we cut the double differential cross section at a fixed value of $Q^2=0.15\myunit{GeV}^2$ (see \reffig{fig:incldoublediff}). The solid line corresponds to the full model calculation as in \reffig{fig:incl3d}.

First, we want to draw attention to the QE peak (right peak in \reffig{fig:incldoublediff}). 
The line denoted with "elementary" is the vacuum cross section which can be obtained from \refeq{eq:crosssection}.
In the case of QE scattering, the differential cross section is a $\delta$-function in $E_{\mu}$ whose position is marked by an arrow in \reffig{fig:incldoublediff}. 
This position, at which QE scattering occurs, is given by
\be
E_{\mu}=E_{\nu}-\frac{Q^2}{2 M_N}.
\ee 
If we "switch on" Fermi motion, denoted as "+ Fermi" in \reffig{fig:incldoublediff} (it overlaps with the dash-dotted line), we find that QE scattering is now possible for a range of values of $E_{\mu}$ with $Q^2$ and $E_{\nu}$ fixed. This can be understood from the on-shell condition
\be
s=(k+p)^2=M_N^2-Q^2+2E_q E - 2 \myvec{p} \cdot \myvec{q} \equiv M_N^{2},
\ee
which has multiple solutions for $E_{\mu}=E_{\nu}-E_q$ due to the Fermi momentum $\myvec{p}$ of the nucleons. 
Pauli blocking (dash-dotted line in \reffig{fig:incldoublediff}) causes a reduction of the cross section when $E_{\mu}$ gets close to $E_{\nu}$. The effect is actually more apparent when the neutrino energy is smaller.
Furthermore, we take into account the binding of the nucleons in a mean-field potential (short-dashed line, at the QE peak this line overlaps with the solid line) by introducing effective masses as outlined in \refsubch{subsec:inmed}.
The inclusion of the potential lowers and shifts the peak to smaller energies. 

Turning to the $\Delta$ excitation (left peak in \reffig{fig:incldoublediff}), we obtain the vacuum cross section $A \times {\rm d}^2\sigma_{\nu N}/({\rm d}Q^2 {\rm d} E_{\mu})$ from \refeq{eq:crosssection} (denoted by the long-dashed line) by replacing the $\delta$-function according to \refeq{eq:spectraldelta} which then leads to the peak structure at lower $E_{\mu}$. 
For the same reason as for QE scattering, we obtain a broadening of the peak when Fermi motion is included (dotted line, which coincides with the dash-dotted line). The inclusion of Pauli blocking has of course no impact on the $\Delta$ production cross section. Taking into account the nuclear binding of the initial nucleon and also the potential of the final $\Delta$, we observe a broadening and a shift of the peak.  
Finally, we include the in-medium modification of the width of the $\Delta$ resonance, labeled in \reffig{fig:incldoublediff} as "+ in-medium width". The vacuum width is replaced by a sum of the vacuum width, modified due to Pauli blocking, and a collisional width accounting for additional channels in the medium, as discussed before. 
Since the cross section scales with the inverse of the width this also lowers the peak.

\begin{figure}[tb]
    \begin{minipage}[t]{.48\textwidth}
       \begin{center}
         \includegraphics[height=5.1cm]{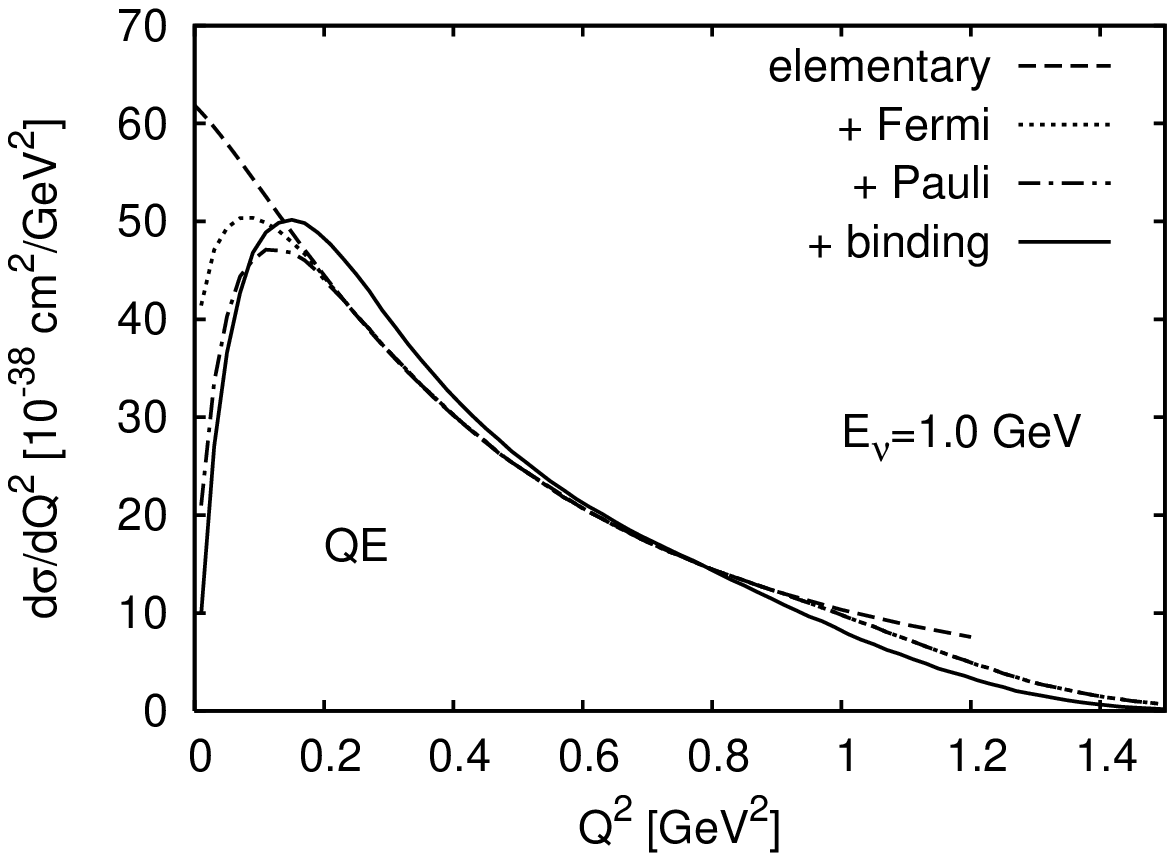}
          \end{center}
   \end{minipage}
     \begin{minipage}[t]{.48\textwidth}
       \begin{center}
         \includegraphics[height=5.1cm]{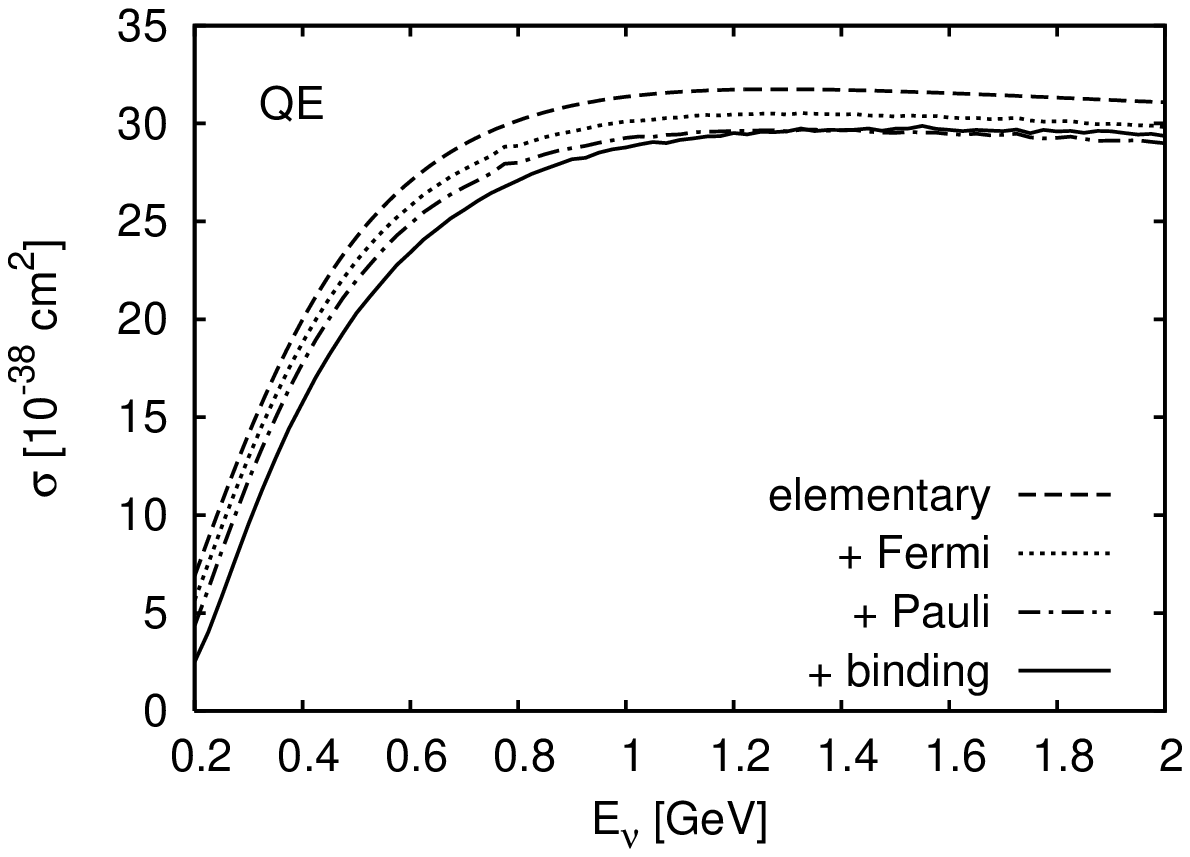}
       \end{center}
   \end{minipage}
   \caption{Inclusive QE cross section on $^{56}\text{Fe}$. The left panel shows the differential cross section ${\rm d}\sigma/{\rm d}Q^2$ at $E_{\nu}=1 \myunit{GeV}$, the right panel the total cross section $\sigma$ as a function of $E_{\nu}$. The in-medium modifications as labeled in the plot are added subsequently to the calculation. The solid line shows the full model. "Elementary" denotes the vacuum cross section multiplied by the mass number $A$. \label{fig:inclQE}}
    \end{figure}

\begin{figure}[tb]
     \begin{minipage}[t]{.48\textwidth}
       \begin{center}
         \includegraphics[height=5.1cm]{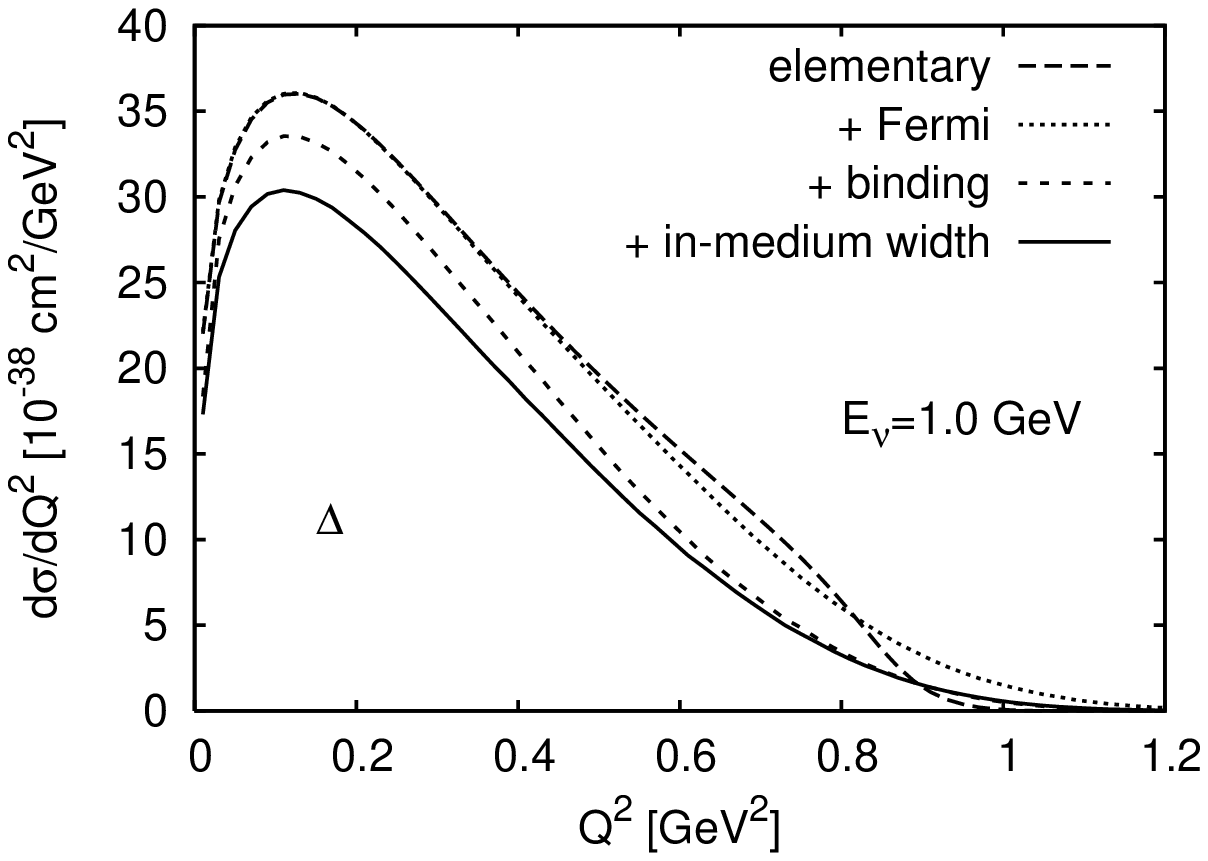}
          \end{center}
   \end{minipage}
     \begin{minipage}[t]{.48\textwidth}
       \begin{center}
         \includegraphics[height=5.1cm]{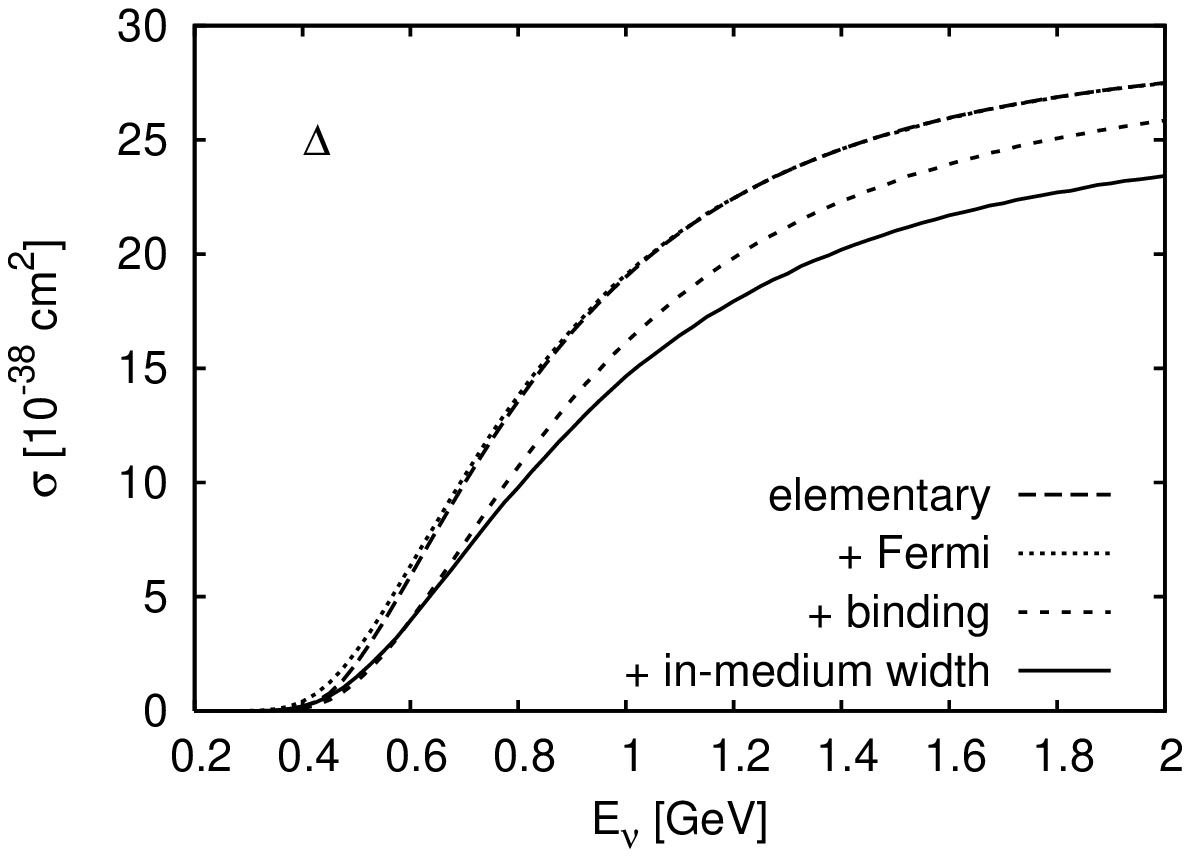}
       \end{center}
   \end{minipage}
   \caption{Same as shown in \reffig{fig:inclQE} for $\Delta$ excitation. \label{fig:inclDEL}}
   \end{figure}

The integration over leptonic degrees of freedom yields the differential cross section ${\rm d}\sigma / {\rm d}Q^2$ and the total cross section $\sigma$. They are shown in \reffig{fig:inclQE} for QE scattering and in \reffig{fig:inclDEL} for $\Delta$ excitation. 
In both figures, the in-medium modifications are included subsequently as indicated in the plots --- their explanation was given above. In both cases the total cross section $\sigma$ is reduced in the medium (solid line) compared to the one for scattering on a free nucleon ($\sigma_N$; long-dashed line): $\sigma < A \times \sigma_N$.  

Our result for the QE total cross section can be compared with previous calculations on $^{56}\text{Fe}$. Paschos et al.~\cite{Paschos:2001np} only include a "Pauli" factor and neglect other nuclear effects such as Fermi motion and binding. In our calculation, the latter lead to a significant reduction of the free cross section that can not be obtained with Pauli blocking alone. Athar et al.~\cite{Athar:2005hu} consider, as we do, all these effects and further account for renormalization of the strength in the medium. Thus, they obtain a cross section even lower than ours (approx.~30\% versus 10\%). 
Juszczak et al.~\cite{Juszczak:2005wk} apply a Fermi gas in the local density approximation with Pauli blocking. Furthermore, they use a momentum dependent binding potential. They find --- as we do --- a reduction of the free cross section due to the Fermi momenta and Pauli blocking of the nucleons and a further decrease due to the binding.

The total cross section for $\Delta$ excitation can be compared to Athar et al.~\cite{Athar:2005hu} who find a reduction of 5 - 10\% with respect to the elementary ones. As in our model the authors have included in-medium modifications of the width of the $\Delta$ but not any nuclear binding. However, through the consistent inclusion of binding energies, we obtain an even bigger reduction of about 20\%. 

\begin{figure}[tb]
   \begin{minipage}[t]{.48\textwidth}
       \begin{center}
         \includegraphics[height=5.1cm]{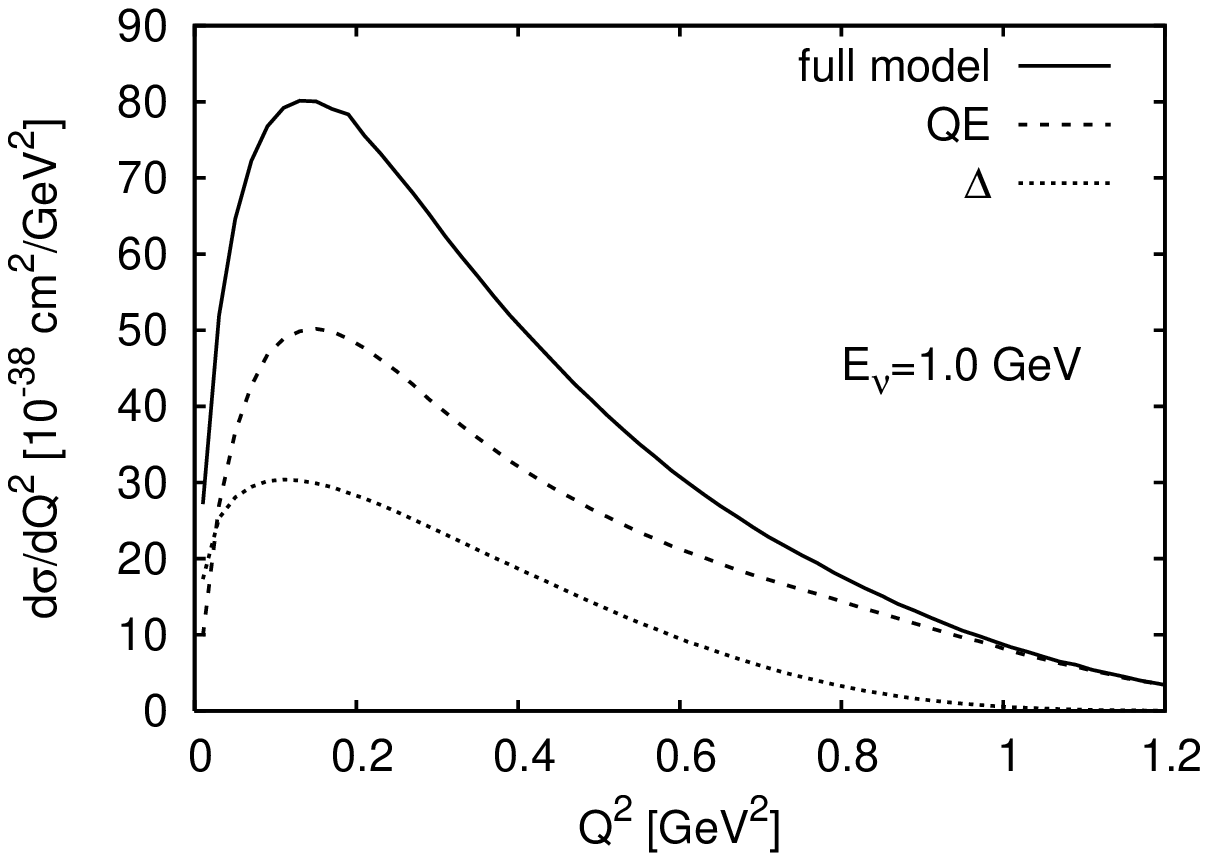}
          \end{center}
   \end{minipage}
   \begin{minipage}[t]{.48\textwidth}
       \begin{center}
         \includegraphics[height=5.1cm]{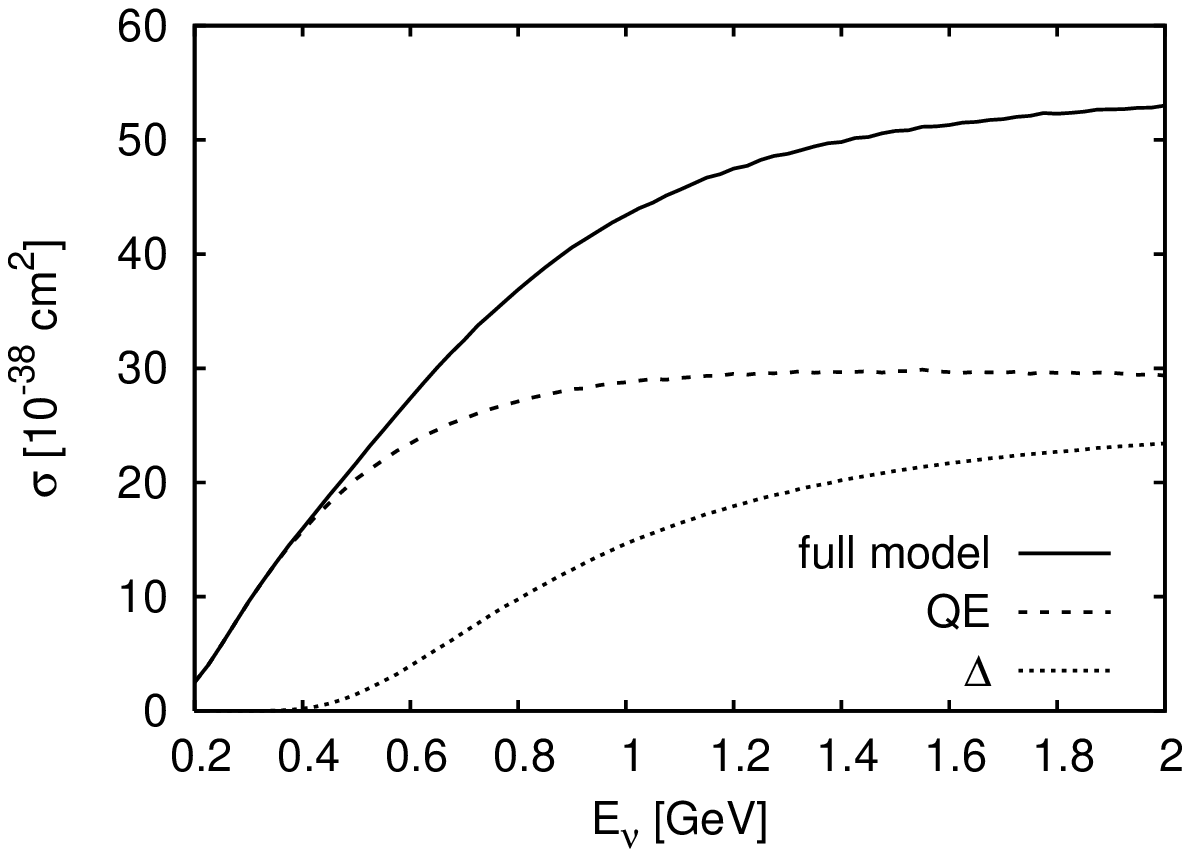}
       \end{center}
   \end{minipage}
   \caption{Inclusive cross sections on $^{56}\text{Fe}$.\label{fig:incl}}
 \end{figure}

Finally, we show in \reffig{fig:incl} the differential cross section ${\rm d}\sigma/{\rm d}Q^2$ at a neutrino energy of 1 GeV (left panel) and the total inclusive cross section $\sigma$ as a function of $E_{\nu}$ (right panel).
The solid line represents the full model with all in-medium effects as described above. We plot also the contributions from QE scattering and $\Delta$ excitation which add up to the full result. 

Apart from H$_2$ and D$_2$ scattering data from ANL and BNL discussed in the previous section, data were also taken for quasielastic and inelastic scattering on heavy (Neon, Propane and Freon) targets e.~g.~at Gargamelle at CERN~\cite{Pohl:1979zm,Pohl:1979fw} and at the Serpukhov bubble chamber SKAT~\cite{Belikov:1983kg,Brunner:1989kw}. These experiments were limited in statistics with large neutrino flux uncertainties. Unfortunately, the large error bars do not allow to draw any conclusion about the role of nuclear effects on those measurements. For this reason we do not show them here. In any case, our results for the total QE cross section agree with these data within their error bars. In the case of pion production on heavy nuclei, there are scarce data points only at energies above the region considered here. 
The data will become more sensitive to in-medium processes with the cross section experiments MINER$\nu$A~\cite{minervaprop} and FINeSSE~\cite{finesseprop} with such a precision measurement of both the quasielastic and the inelastic neutrino-nucleus cross sections, that a meaningfull comparison with theoretical models will be possible.

\subsection{Exclusive channels \label{subsec:exclusivech}}
In the case of exclusive channels, even if the cross section for the $\nu A$ reaction is again given by an integration over all nucleons, for each distinct channel, e.~g.~$\pi^+$~production, the contributions have to be weighted with the multiplicity of the given final state. This multiplicity is calculated with the coupled-channel BUU transport model.

\subsubsection{Pion Production} 

\begin{figure}[tb]
  \begin{minipage}[t]{.48\textwidth}
      \begin{center}
        \includegraphics[height=7cm]{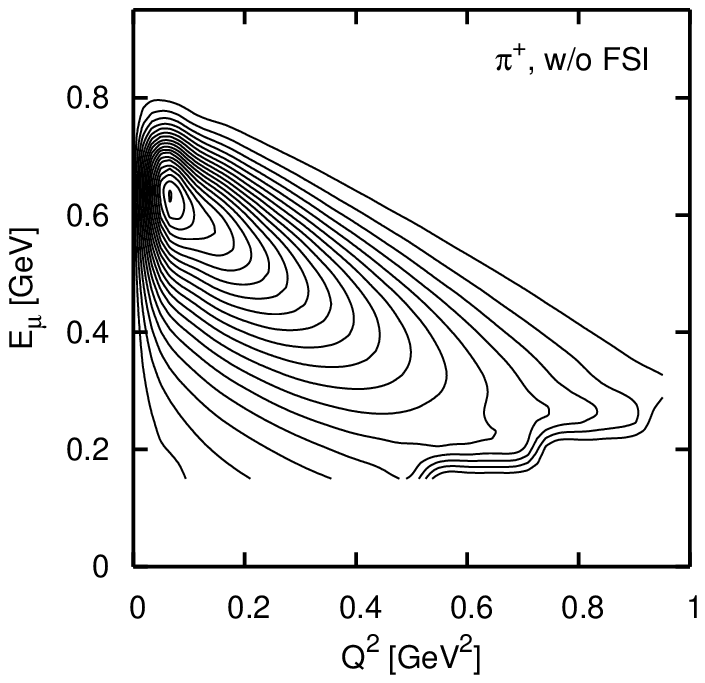}
      \end{center}
  \end{minipage}
  \begin{minipage}[t]{.48\textwidth}
      \begin{center}
        \includegraphics[height=7cm]{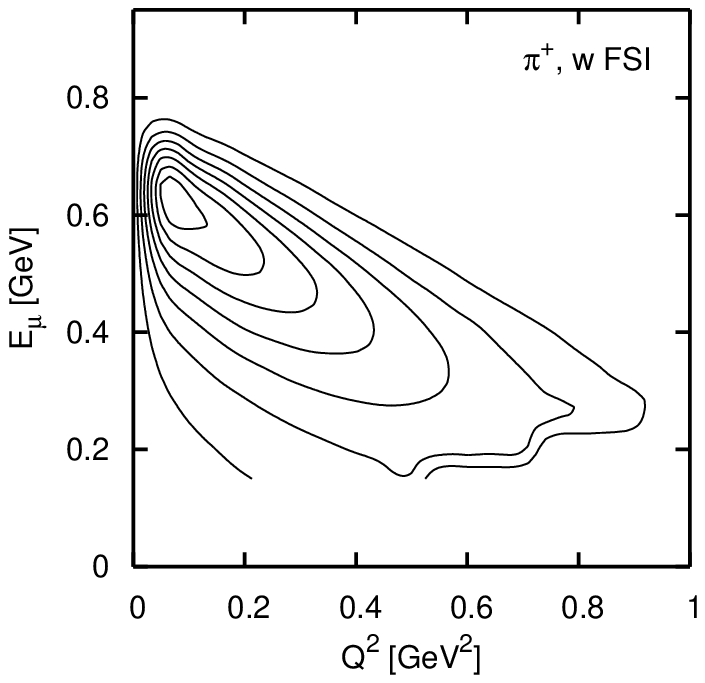}
      \end{center}
  \end{minipage}
 \caption{Double differential cross section ${\rm d}\sigma/({\rm d}Q^2 {\rm d} E_{\mu})$ for $\pi^+$ production on $^{56}\text{Fe}$ at $E_{\nu}=1 \myunit{GeV}$.  The cross section is mapped to the $Q^2-E_{\mu}$ plane. In this result all in-medium modifications of the elementary cross section are included. The right panel additionally includes FSI. The contour lines are equidistant every $5 \times 10^{-38} \text{cm}^2/\text{GeV}^3$ from $0$ to $95$ (left panel) and $35 \times 10^{-38} \text{cm}^2/\text{GeV}^3$ (right panel), respectively. \label{fig:excl3d_pipl}}  
\end{figure}
\begin{figure}[tb]
  \begin{minipage}[t]{.48\textwidth}
      \begin{center}
        \includegraphics[height=7cm]{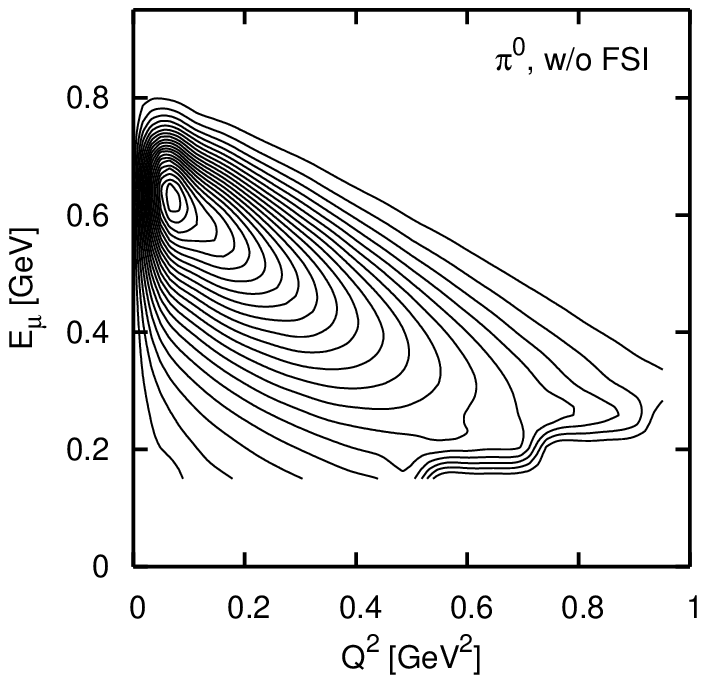}
      \end{center}
  \end{minipage}
  \begin{minipage}[t]{.48\textwidth}
      \begin{center}
        \includegraphics[height=7cm]{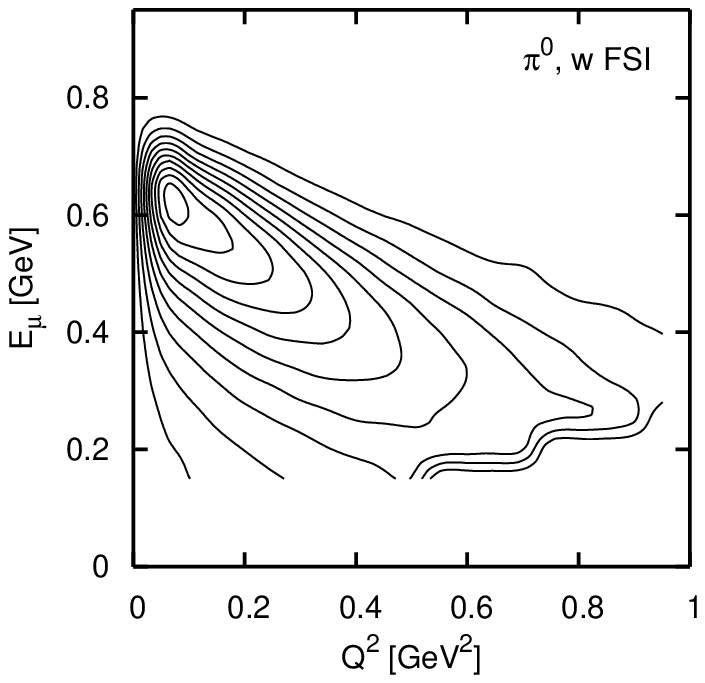}
      \end{center}
  \end{minipage}
 \caption{Same as shown in \reffig{fig:excl3d_pipl} for $\pi^0$ production.  The contour lines are equidistant every $1 \times 10^{-38} \text{cm}^2/\text{GeV}^3$ from $0$ to $21$ (left panel) and $10 \times 10^{-38} \text{cm}^2/\text{GeV}^3$ (right panel), respectively.\label{fig:excl3d_pinull}}  
\end{figure}
We start our discussion of pion production with the double differential cross section shown in \reffig{fig:excl3d_pipl} and \reffig{fig:excl3d_pinull}. 
There we plot the exclusive cross section for neutrino-induced $\pi^+$ and $\pi^0$ production on $^{56}\text{Fe}$ at $E_{\nu}=1 \myunit{GeV}$ as a function of $E_{\mu}$ and $Q^2$. 
In this calculation we included all medium modifications of the elementary cross section as explained in \refsubch{subsec:inmed}. 
The left panels shows the results without FSI while in the right panels they are included.
Once produced the $\Delta$ can decay or interact via $\Delta N \to N N$, $\Delta N N \to N N N$, $\Delta N \to \pi N N$ or $\Delta N \to \Delta N$. The produced pions interact through $\pi N \to \pi N$, $NN\pi \to NN$ and $\pi N \to \Delta$, i.~e.~they can scatter elastically, undergo charge exchange or be absorbed.
This results in the creation of additional pions or their absorption. 

\begin{figure}[tb]
   \begin{minipage}[t]{.48\textwidth}
       \begin{center}
         \includegraphics[height=5.1cm]{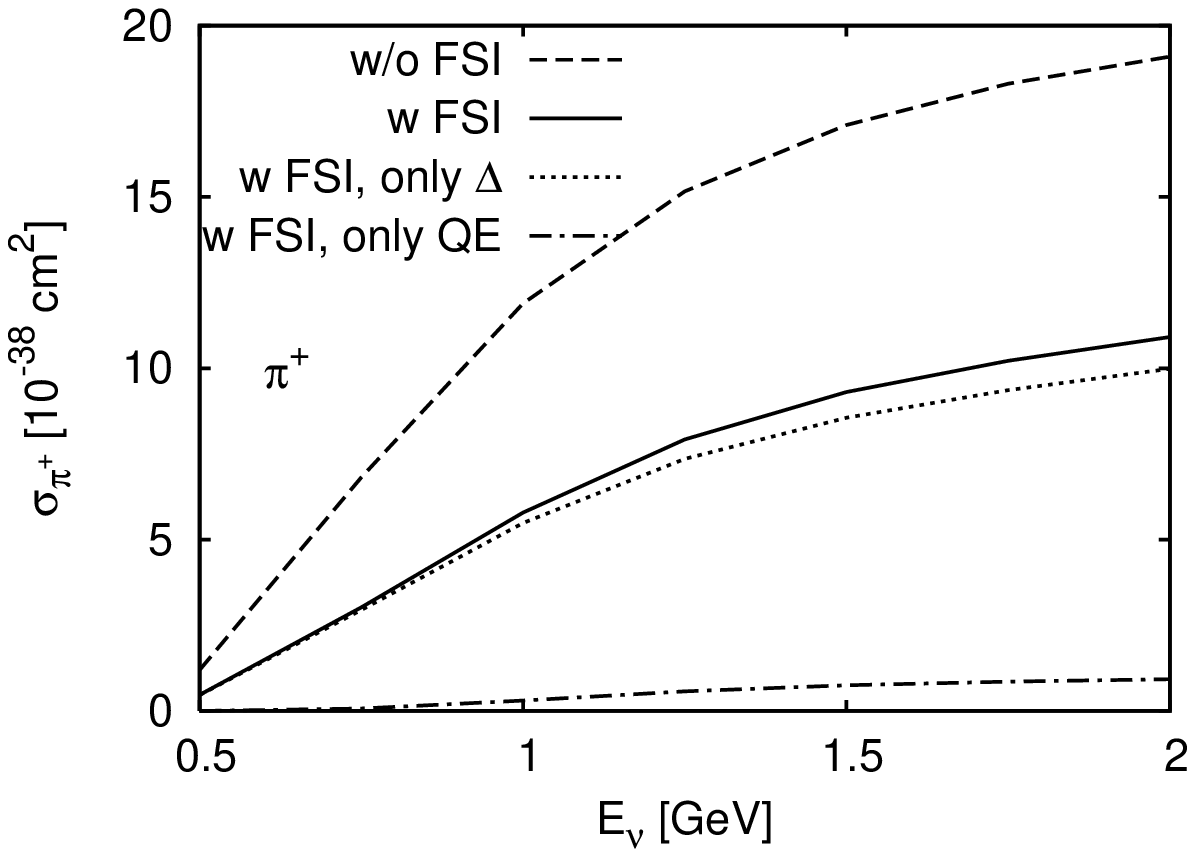}
          \end{center}
   \end{minipage}
   \begin{minipage}[t]{.48\textwidth}
       \begin{center}
         \includegraphics[height=5.1cm]{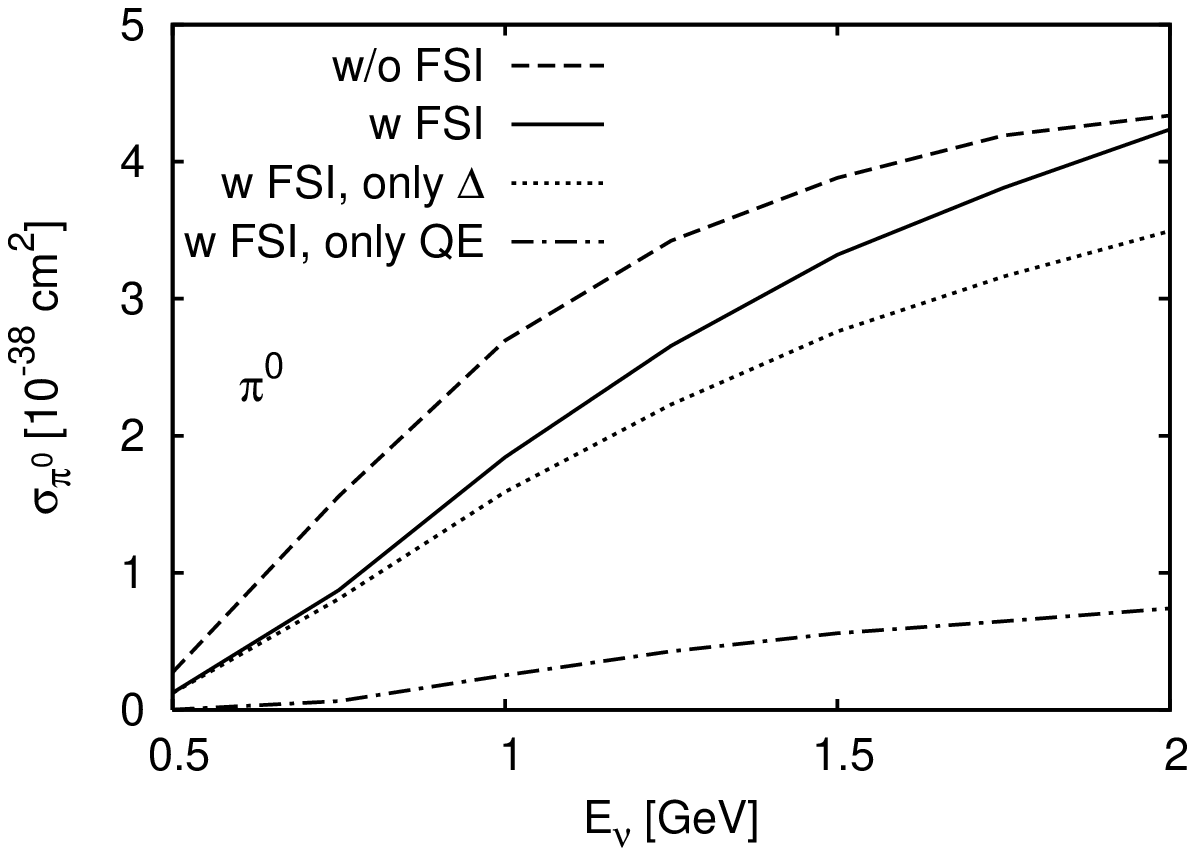}
       \end{center}
   \end{minipage}
   \caption{Total cross section for $\pi^+$ (left) and $\pi^0$ (right panel) production on $^{56}\text{Fe}$. The dashed line shows the results without FSI (only the decay of the $\Delta$ is possible); the results denoted by the solid line include FSI. Furthermore, the origin of the pions is indicated (QE or $\Delta$ excitation). \label{fig:pplpnull_tot}}
   \end{figure}

The cross sections in \reffig{fig:excl3d_pipl} and \reffig{fig:excl3d_pinull} peak clearly at the position of the $\Delta$ (compare to the lower panel of \reffig{fig:incl3d}) which indicates that most of the produced pions come from the initially produced $\Delta$.
Pions can also originate in multistep processes initiated by QE scattering. There, the produced nucleon can scatter in the nucleus and create pions through $NN \to N \Delta$ or $NN \to NN \pi$. 
However, this effect, which can occur only at high momentum transfer $Q^2$, is less important as can be seen in \reffig{fig:pplpnull_tot}. 
There, we show the total cross section for $\pi^+$ (left) and $\pi^0$ production (right panel). 
The dashed lines show the result without final-state interactions, while the solid lines denote the result of the full calculation. Furthermore, we show the contribution from initial $\Delta$~excitation (dotted line) and from initial QE events (dash-dotted).
The cross sections without FSI for $\pi^0$ production is significantly lower than the one for $\pi^+$. This difference is a consequence of the primary interaction mechanism:  
\bea
\nu p &\to& l^- \Delta^{++}, \\
\nu n &\to& l^- \Delta^{+}. 
\eea
The first process is enhanced by an isospin factor of three.
These $\Delta$s decay into pions by
\bea
\Delta^{++} &\to& p \pi^+,\\
\Delta^{+} &\to& p \pi^0, \; \; n \pi^+.
\eea
With the isospin amplitudes of these processes, we obtain a ratio of $\pi^+:\pi^0 =\left[ Z + \left(1/3\right)^2 N \right]\left[\left(-\sqrt{2}/3\right)^2 N\right]^{-1}= 4.4 : 1$ ($N$ and $Z$ are the proton and neutron numbers)
for the cross sections without final-state interactions. FSI, however, change this ratio. Indeed, the comparison of the $\pi^+$ channel to the $\pi^0$ channel (left and right panel of \reffig{fig:pplpnull_tot}) reveals big differences.
For $\pi^+$ we find a strong reduction of the cross section due to FSI, while in the $\pi^0$ channel this reduction is much smaller (compare dotted and dashed lines). 
The additional strength in the $\pi^0$ channel is a consequence of the "disappearance" in the dominant $\pi^+$ channel: $\pi^+$ undergo charge exchange reactions like $\pi^+ n \to \pi^0 p$ contributing in this way to the $\pi^0$ channel. This leads to the observed side-feeding. 
Side-feeding in the opposite direction is strongly suppressed by the ratio of $\pi^+$ to $\pi^0$ production on the nucleon. 

As said before, $\pi^0$ and $\pi^+$ production through FSI of QE scattering is not very sizable and happens only if the neutrino energy is high enough.
However, the effect is relatively more important in the $\pi^0$ channel than in the $\pi^+$ one. This follows from the fact that, while the production of both $\pi^0$ and $\pi^+$ from initial quasielastic scattering is basically the same, the $\pi^0 : \pi^+$ ratio from initially produced $\Delta$ resonances is roughly a factor of $4$ smaller as just outlined. Thus, this effect also enhances the $\pi^0$ channel due to FSI.

\begin{figure}
\begin{center}
\includegraphics[scale=.75]{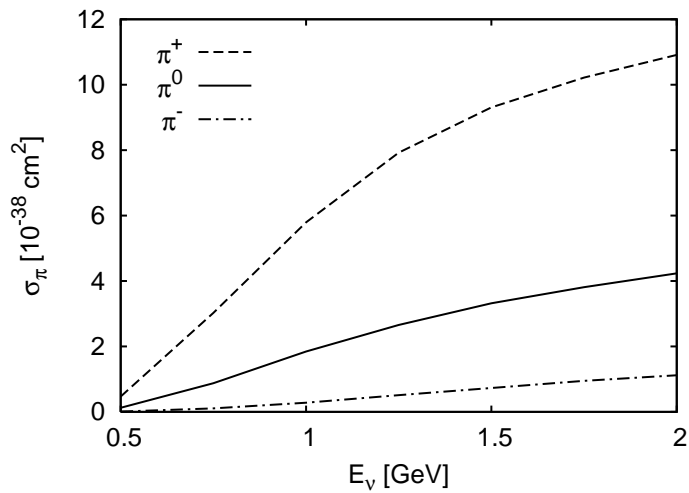}
\end{center}
\caption{Total cross section for $\pi^+$ (dashed), $\pi^0$ (solid) and $\pi^-$ (dash-dotted) production on $^{56}\text{Fe}$ versus the neutrino energy. These results were obtained with a full-model calculation. \label{fig:pion_tot}}
\end{figure}

$\pi^-$ cannot be produced directly in the neutrino-nucleon reactions, but only via final-state interactions. Thus, they play only a minor role as can be seen in \reffig{fig:pion_tot} where we plot the total cross section for $\pi^+$, $\pi^0$ and $\pi^-$ production including FSI. Note that this situation is reversed in antineutrino reactions where only $\Delta^-$ and $\Delta^0$ can be produced in the initial interaction.

\begin{figure}[tb]
  \begin{minipage}[t]{.48\textwidth}
      \begin{center}
        \includegraphics[height=5.1cm]{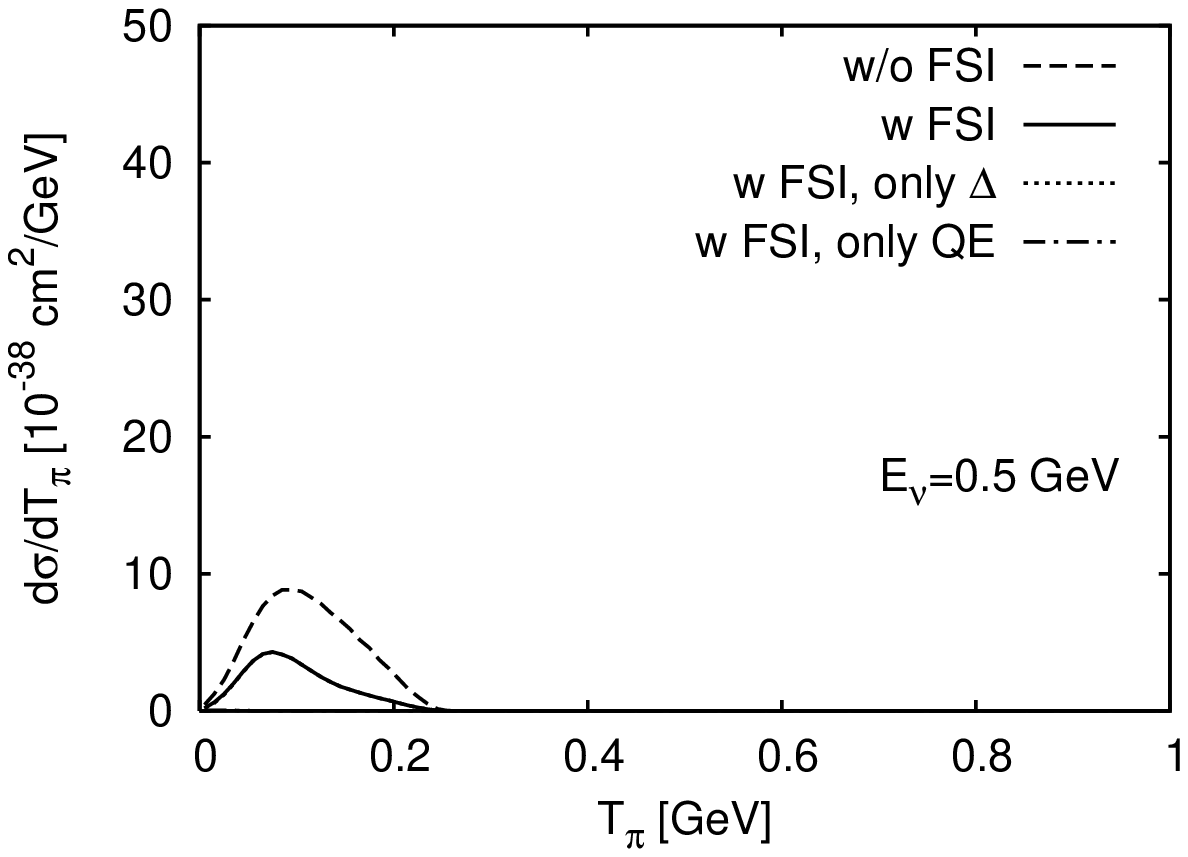}
       \end{center}
  \end{minipage}
  \begin{minipage}[t]{.48\textwidth}
      \begin{center}
        \includegraphics[height=5.1cm]{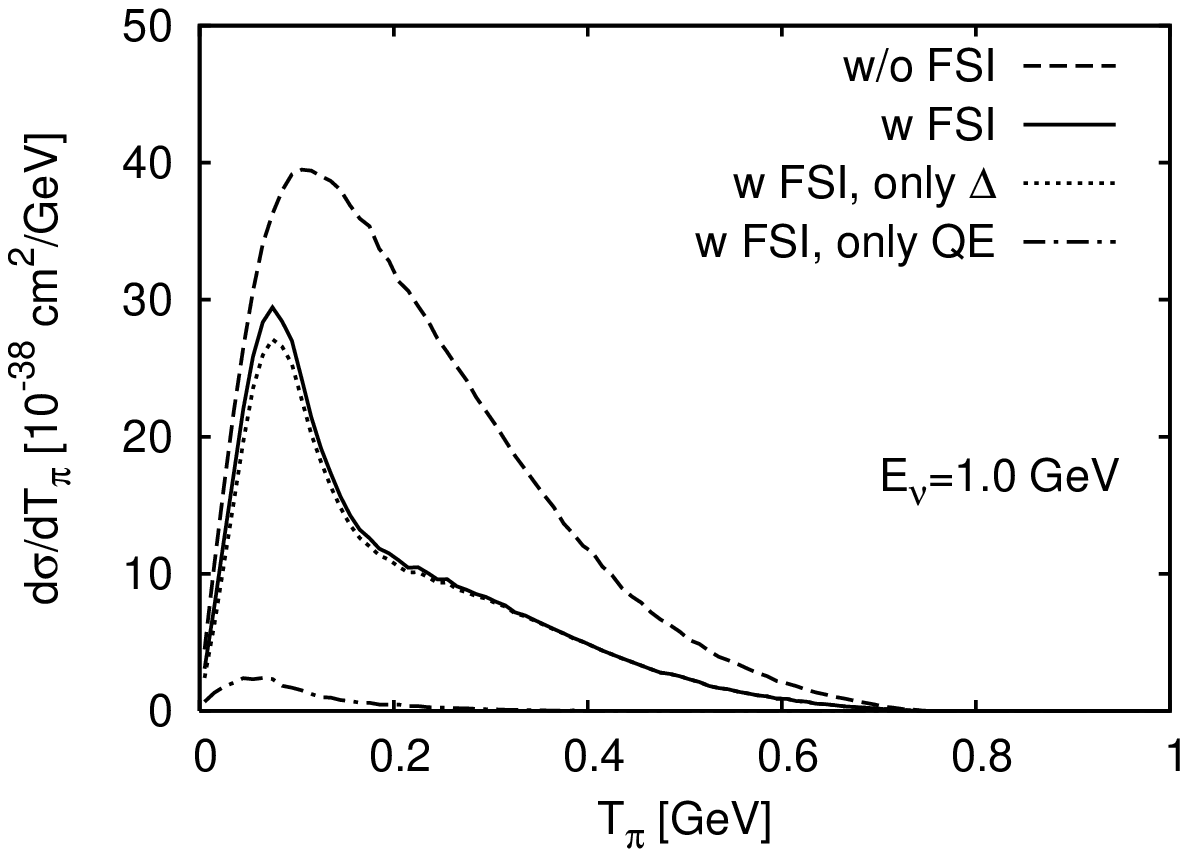}
      \end{center}
  \end{minipage}
 \\ \hfill \\
  \begin{minipage}[t]{.48\textwidth}
      \begin{center}
        \includegraphics[height=5.1cm]{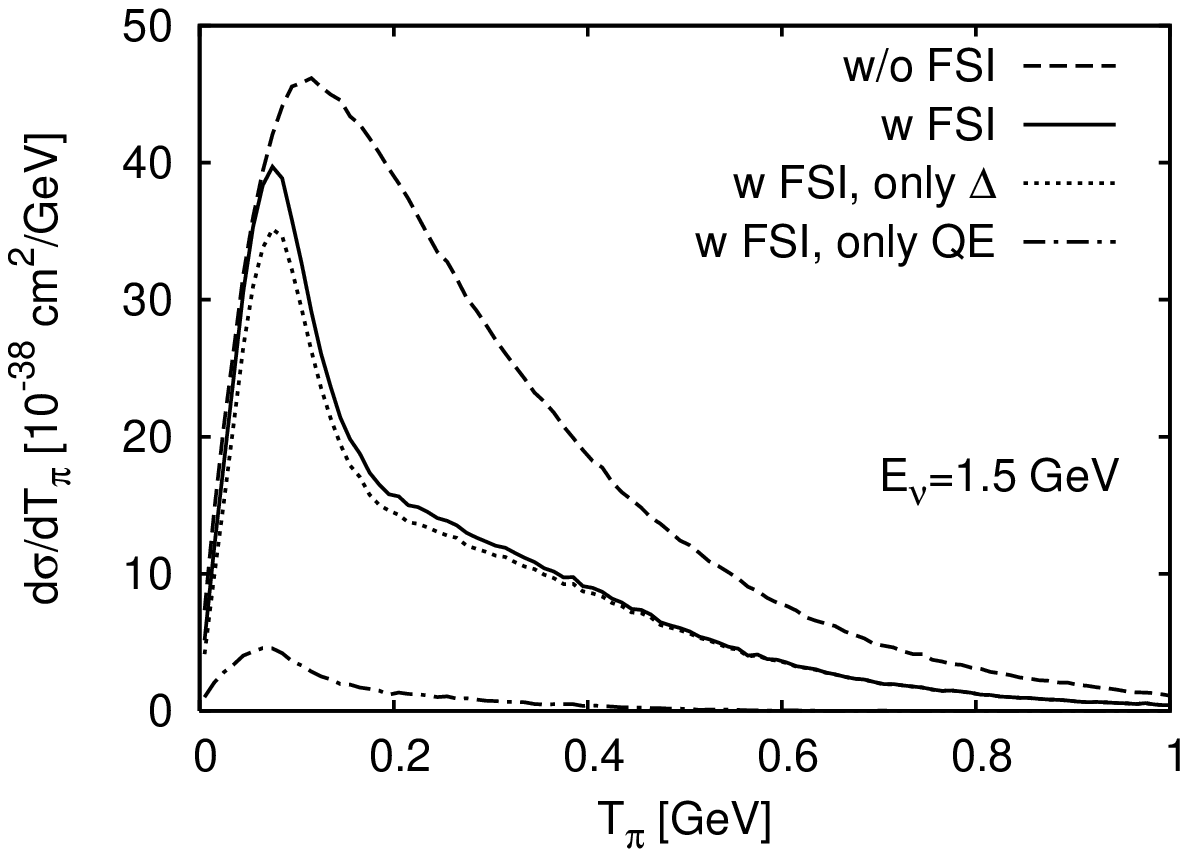}
      \end{center}
  \end{minipage}
  \begin{minipage}[t]{.48\textwidth}
      \begin{center}
        \includegraphics[height=5.1cm]{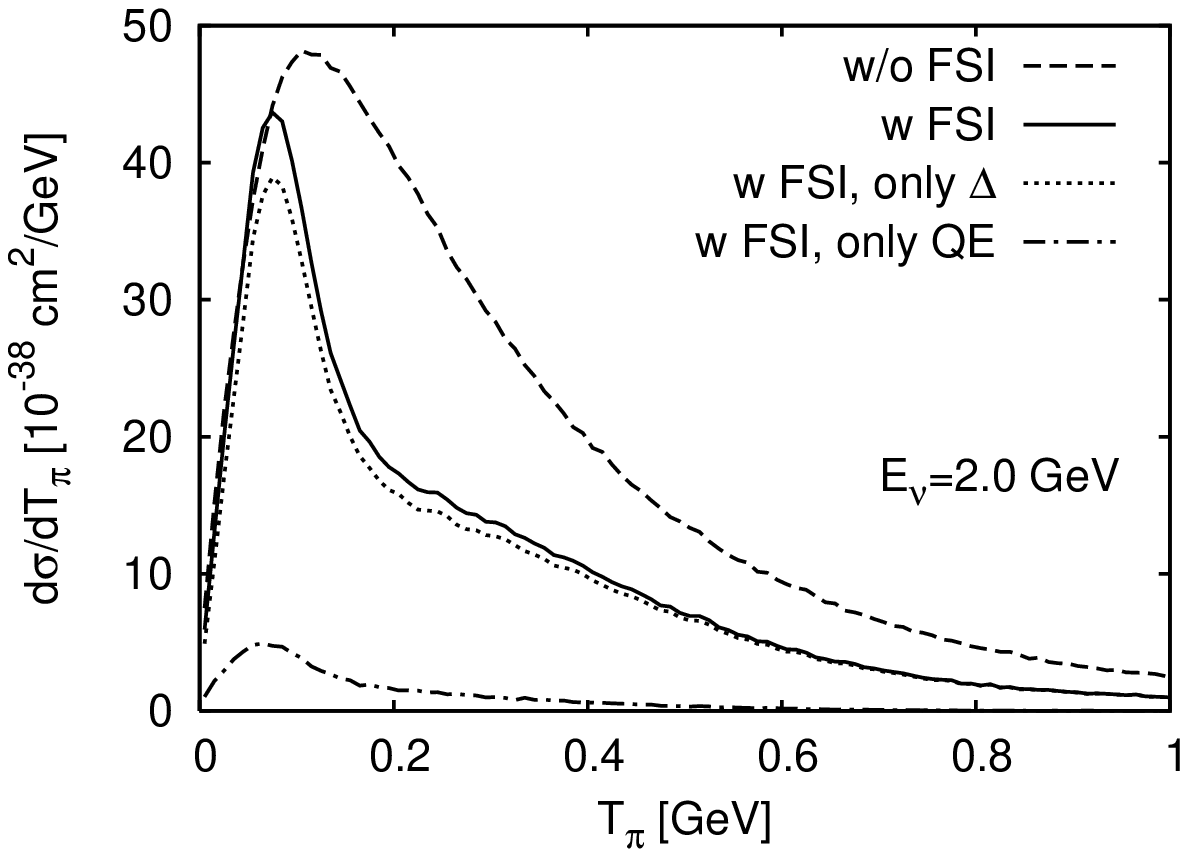}
      \end{center}
  \end{minipage}
  \caption{Kinetic energy differential cross section for $\pi^+$ production on $^{56}\text{Fe}$ versus the pion kinetic energy $T_{\pi}$ at different values of $E_{\nu}$ integrated over $Q^2$ and $E_{\mu}$. The dashed line shows the results without FSI interactions (only the decay of resonances is possible), the results denoted by the solid line include FSI. Furthermore, the origin of the $\pi^+$ is indicated (QE or $\Delta$ excitation).  \label{fig:pipl}}  
\end{figure}

\begin{figure}[tb]
  \begin{minipage}[t]{.48\textwidth}
      \begin{center}
        \includegraphics[height=5.1cm]{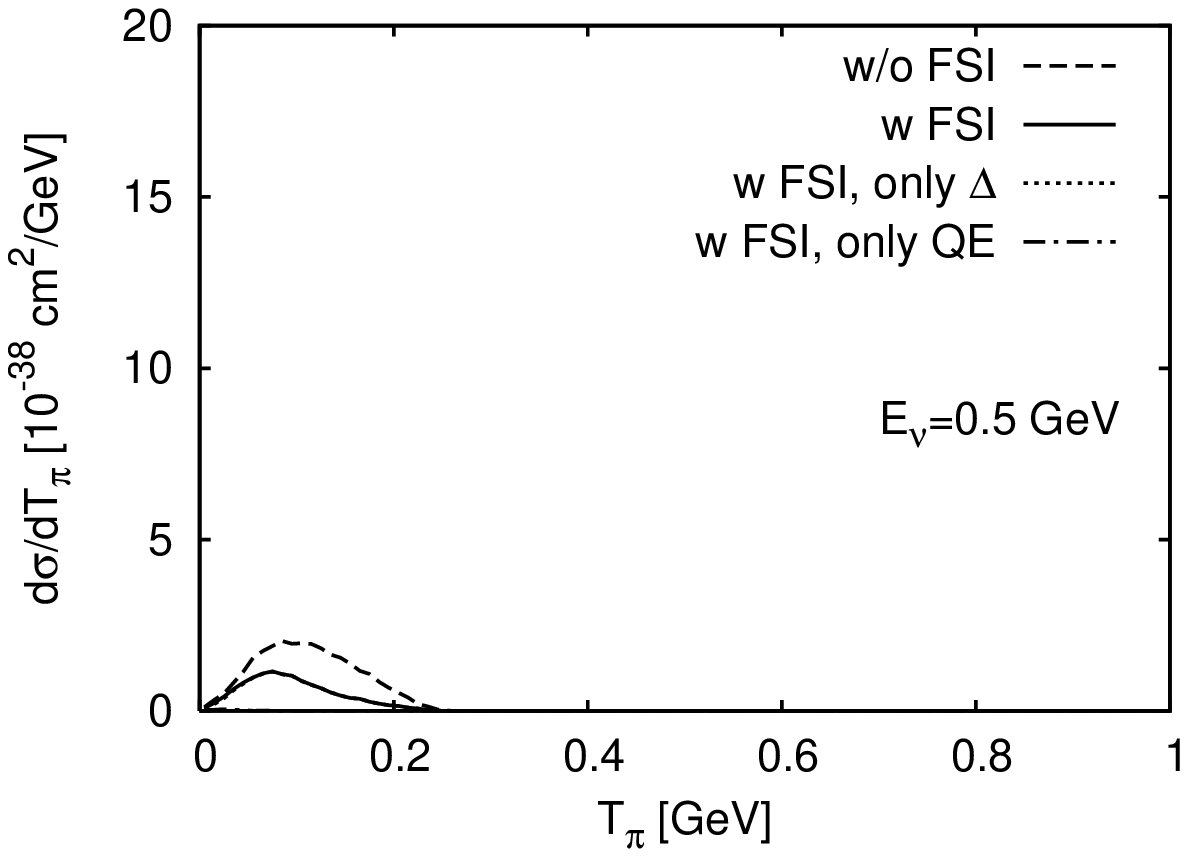}
       \end{center}
  \end{minipage}
  \begin{minipage}[t]{.48\textwidth}
      \begin{center}
        \includegraphics[height=5.1cm]{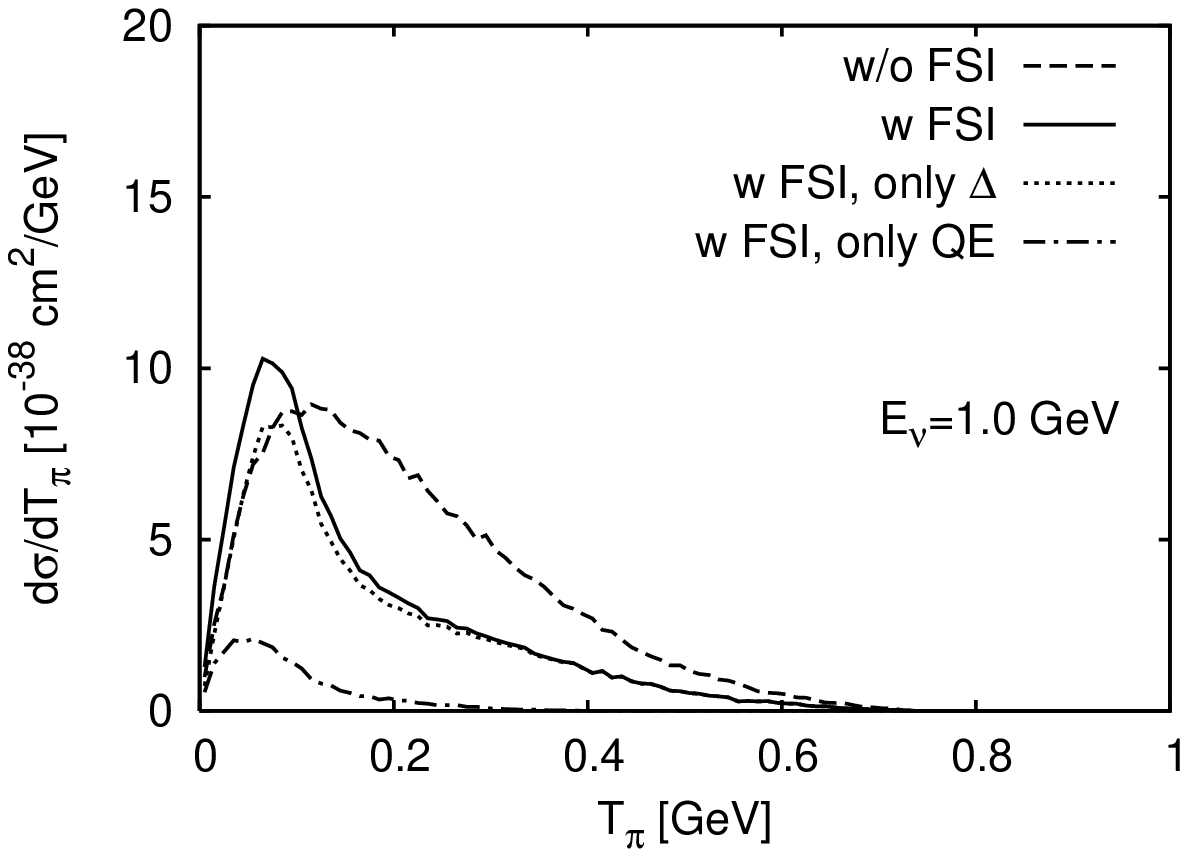}
      \end{center}
  \end{minipage}
 \\ \hfill \\
  \begin{minipage}[t]{.48\textwidth}
      \begin{center}
        \includegraphics[height=5.1cm]{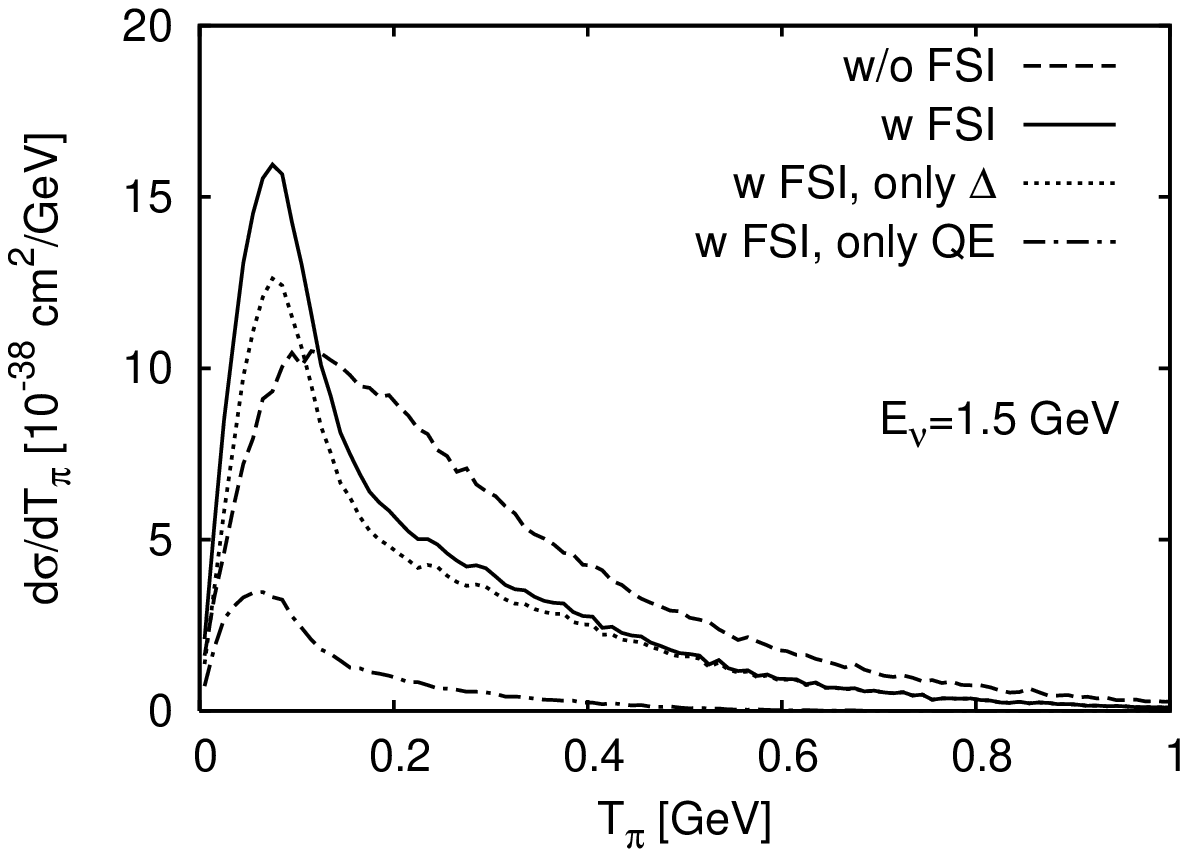}
      \end{center}
  \end{minipage}
  \begin{minipage}[t]{.48\textwidth}
      \begin{center}
        \includegraphics[height=5.1cm]{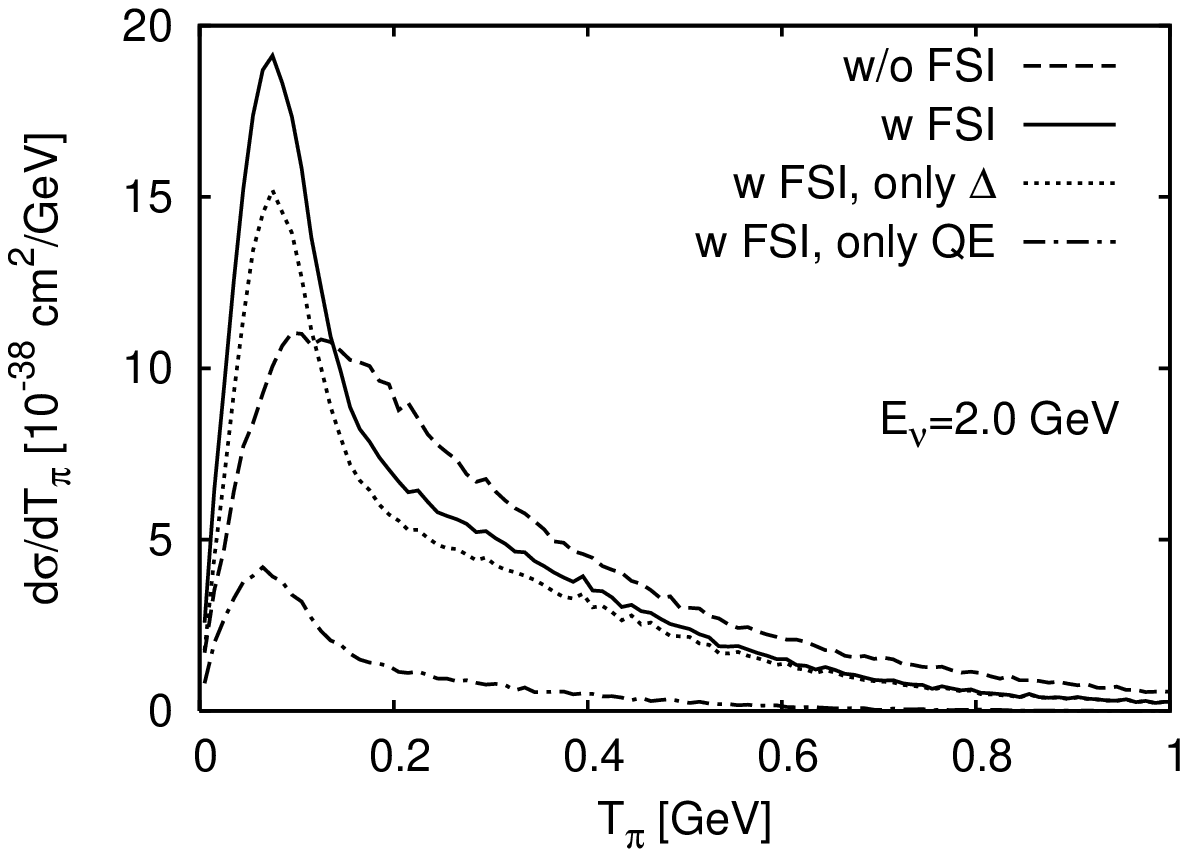}
      \end{center}
  \end{minipage}
  \caption{Same as shown in \reffig{fig:pipl} for $\pi^0$ production. \label{fig:pinull}}  
\end{figure}

Further details can be brought up by studying the pion kinetic energy distributions. They are shown in \reffig{fig:pipl} (\reffig{fig:pinull}) for $\pi^+$ ($\pi^0$) production at different values of $E_{\nu}$.
The dashed lines show again the result without final-state interactions and the solid lines the result of the full calculation.
The contributions from initial $\Delta$ excitation (dotted line) and from initial QE events (dash-dotted) are also plotted. $\pi^0$ and $\pi^+$ production through FSI of QE scattering contributes mostly to the low energy region of the pion spectra because of the energy redistribution in the collisions.

The maximum of the solid curve (i.~e.~the calculation with final-state interactions) peaks at 0.05~-~0.1~GeV in all cases shown in \reffig{fig:pipl} and in \reffig{fig:pinull}. This is due to the energy dependence of the pion absorption. The absorption is higher in the resonance region where 
the pions are mainly absorbed through the reaction $\pi N \to \Delta$, followed by $\Delta N \to N N$. This strong reduction for high energy pions and the corresponding shift of the maximum to lower energies can be seen by comparing the dashed and the solid lines. These absorption processes equally affect $\pi^+$ and $\pi^0$ yields. But pions do not only undergo absorption when propagating through the nucleus. Of particular importance for pions of all energies is elastic scattering $\pi N \to \pi N$ which redistributes the kinetic energies, again shifting the distribution to lower energies. 
It is important to stress that similar patterns are obtained within our BUU model for $\pi^0$ photoproduction in nuclei in a good agreement with data as can be seen in Fig.~14 of \refcite{Krusche:2004uw}.

The different scale of $\pi^+$ and $\pi^0$  (\reffig{fig:pipl} and \reffig{fig:pinull}) is a consequence of their different production rates in the neutrino-nucleon reaction. This leads to the already discussed side-feeding from the dominant $\pi^+$ channel to the $\pi^0$ channel. Also, pions produced from initial QE events, contribute relatively more to the $\pi^0$ channel.  For this reason, at $E_{\nu} \gtrsim 1 \myunit{GeV}$ we even get an enhancement of $\pi^0$ at low kinetic energies compared to the calculation without final-state interactions.

\begin{figure}[tb]
  \begin{minipage}[t]{.48\textwidth}
      \begin{center}
        \includegraphics[height=5.1cm]{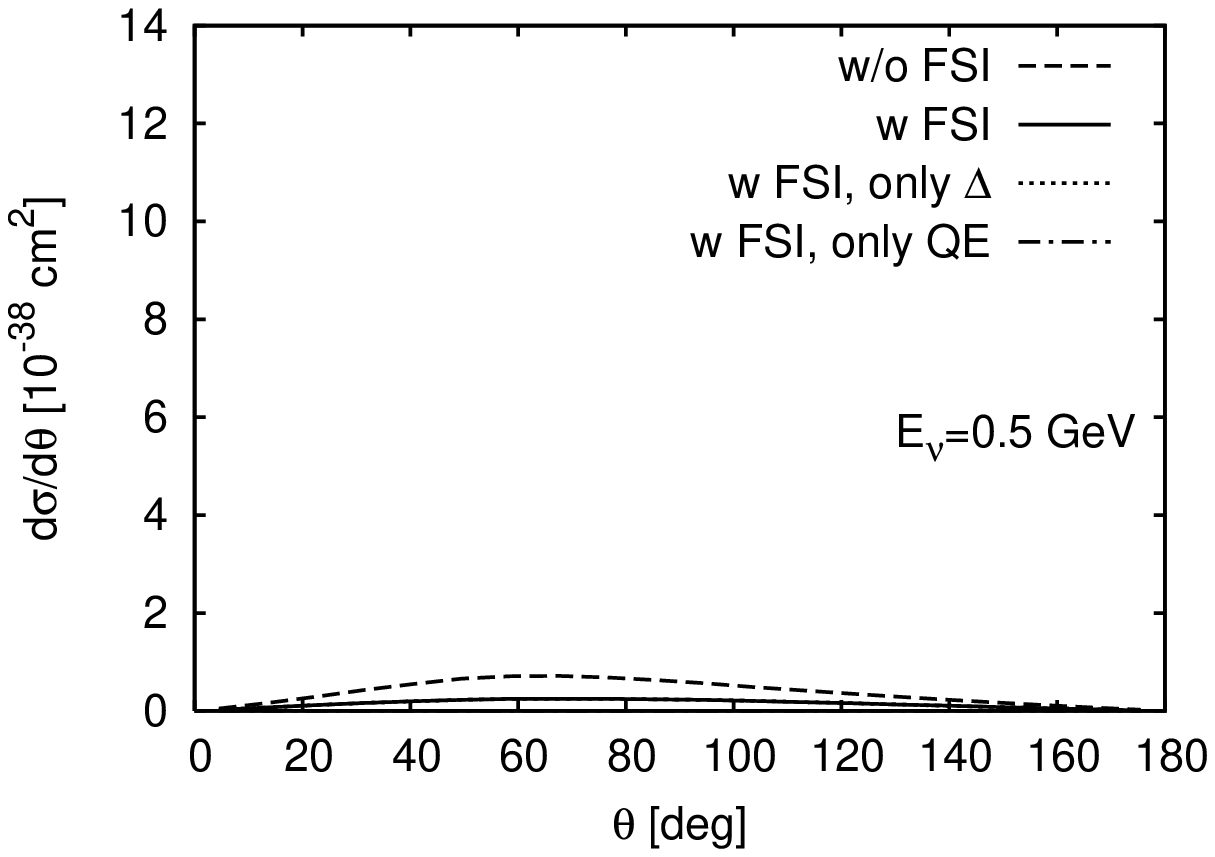}
       \end{center}
  \end{minipage}
  \begin{minipage}[t]{.48\textwidth}
      \begin{center}
        \includegraphics[height=5.1cm]{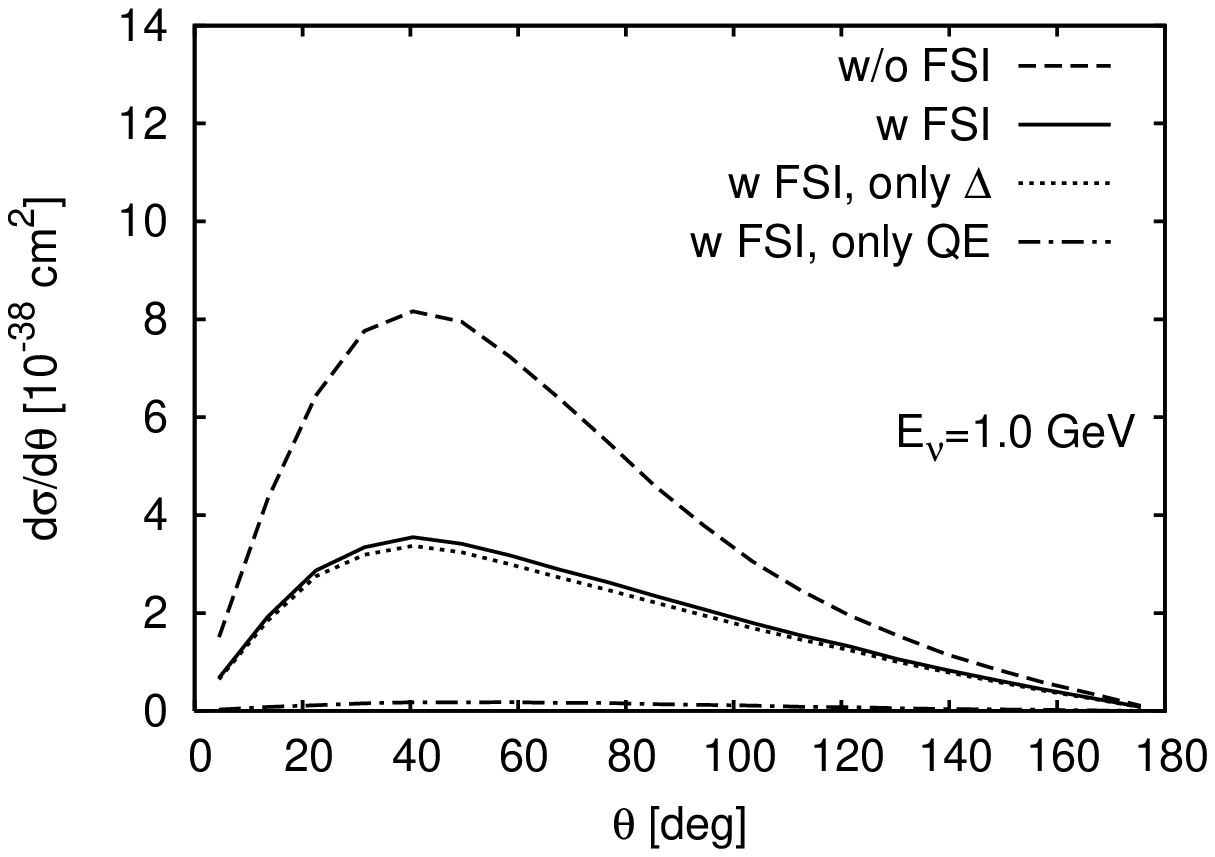}
      \end{center}
  \end{minipage}
 \\ \hfill \\
  \begin{minipage}[t]{.48\textwidth}
      \begin{center}
        \includegraphics[height=5.1cm]{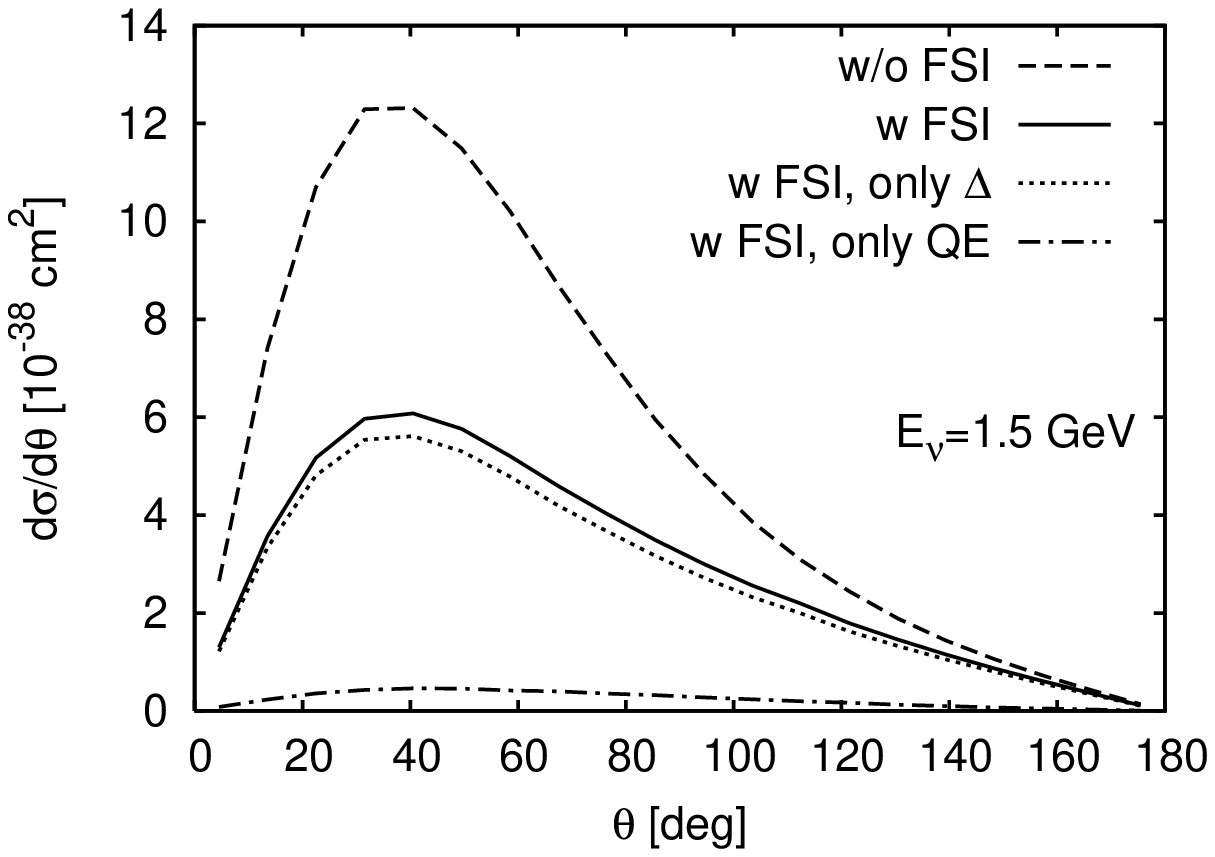}
      \end{center}
  \end{minipage}
  \begin{minipage}[t]{.48\textwidth}
      \begin{center}
        \includegraphics[height=5.1cm]{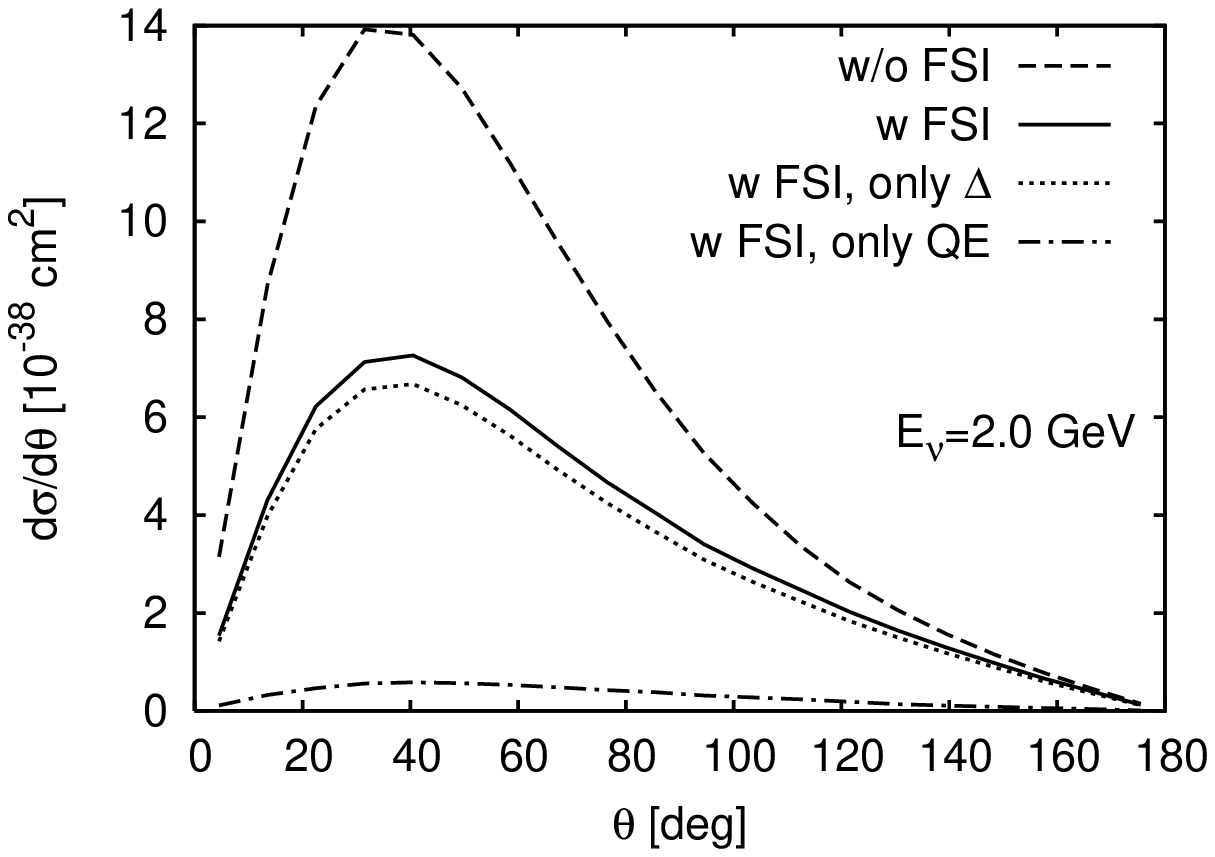}
      \end{center}
  \end{minipage}
  \caption{Angular distribution for $\pi^+$ production on $^{56}\text{Fe}$ at various neutrino energies. The angle $\theta$ is measured with respect to the direction of the incoming neutrino. Labels are as in \reffig{fig:pipl} and \reffig{fig:pinull}. \label{fig:angularpipl}}  
\end{figure}

\begin{figure}[tb]
  \begin{minipage}[t]{.48\textwidth}
      \begin{center}
        \includegraphics[height=5.1cm]{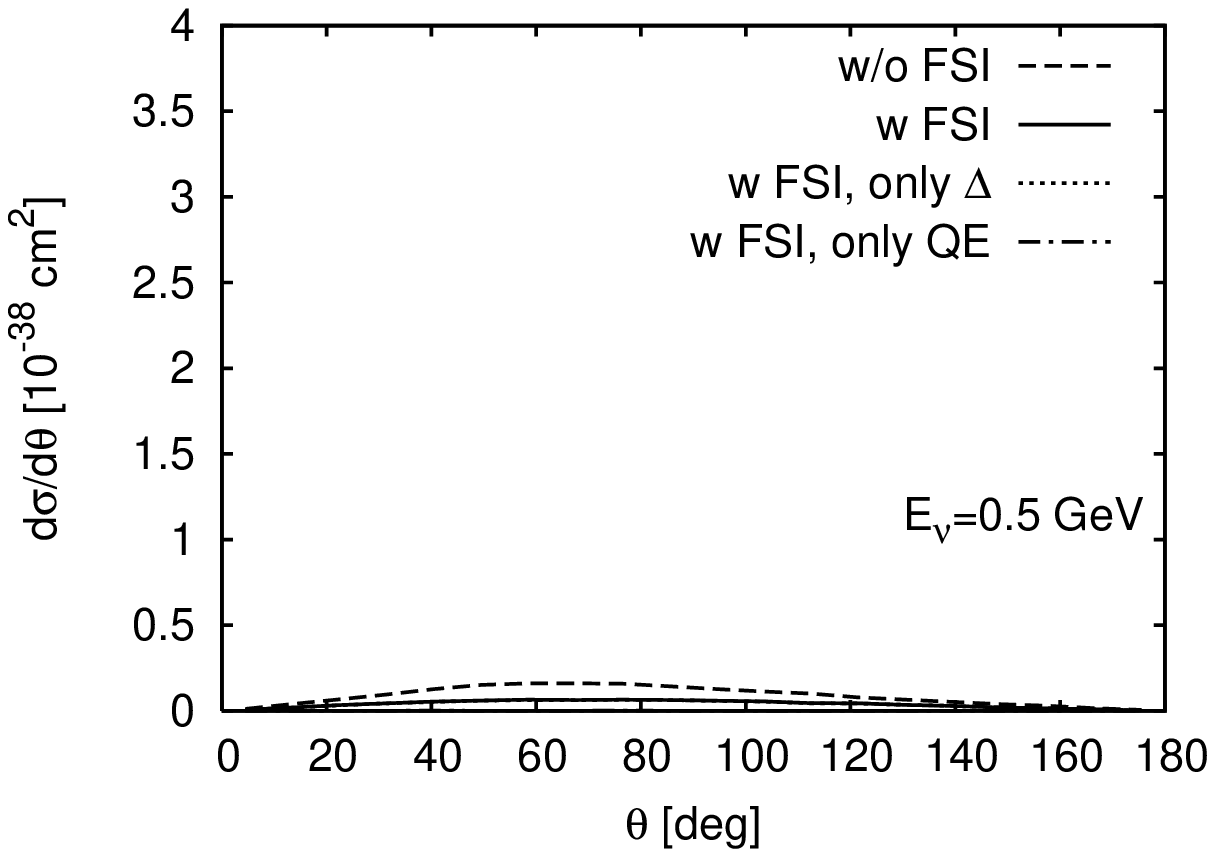}
       \end{center}
  \end{minipage}
  \begin{minipage}[t]{.48\textwidth}
      \begin{center}
        \includegraphics[height=5.1cm]{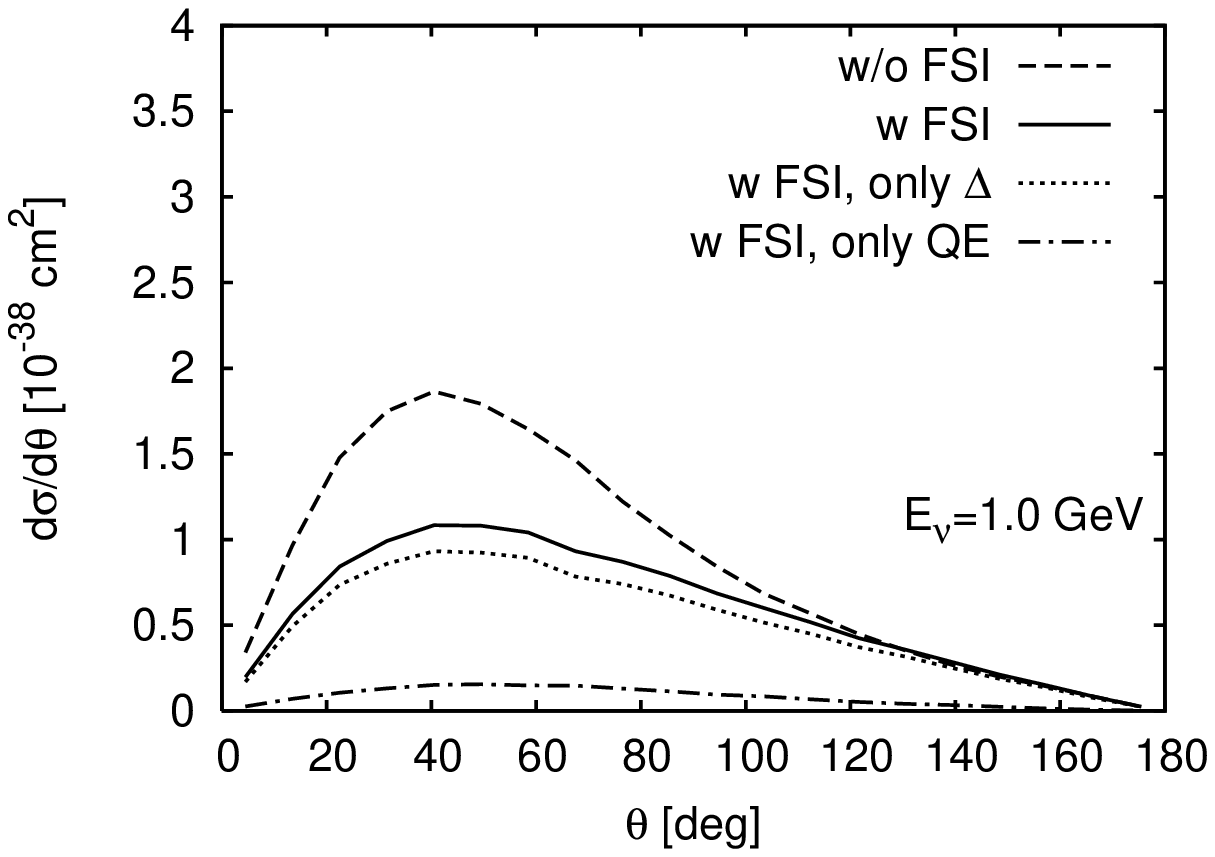}
      \end{center}
  \end{minipage}
 \\ \hfill \\
  \begin{minipage}[t]{.48\textwidth}
      \begin{center}
        \includegraphics[height=5.1cm]{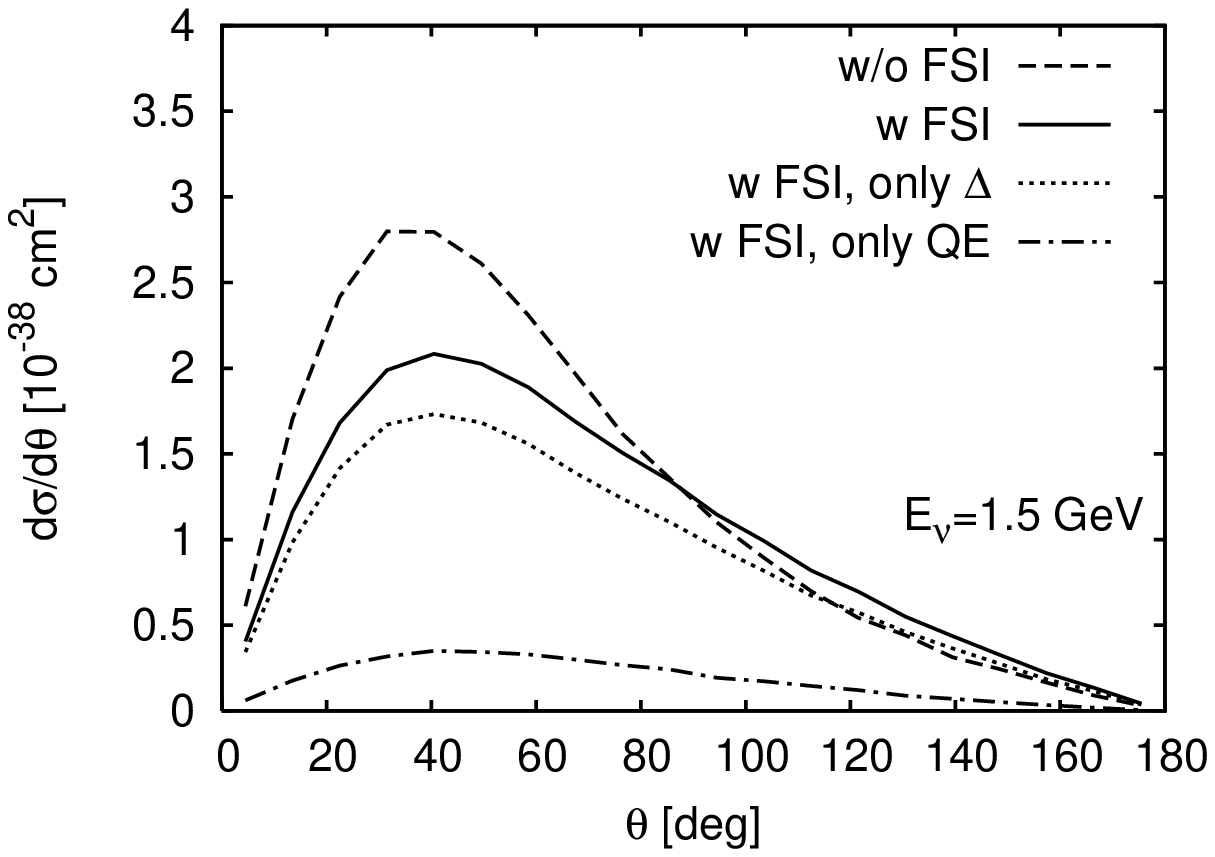}
      \end{center}
  \end{minipage}
  \begin{minipage}[t]{.48\textwidth}
      \begin{center}
        \includegraphics[height=5.1cm]{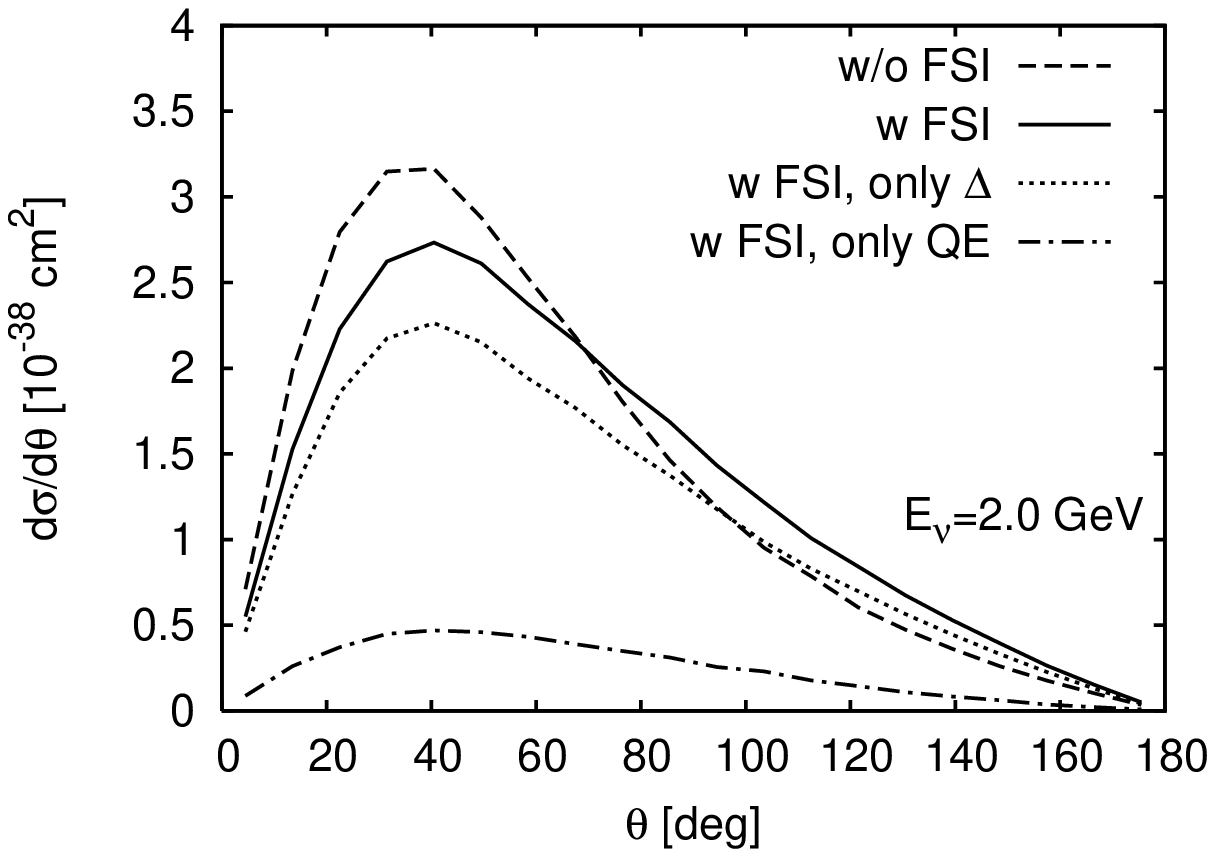}
      \end{center}
  \end{minipage}
  \caption{Same as shown in \reffig{fig:angularpipl} for $\pi^0$ production. \label{fig:angularpipnull}}  
\end{figure}

In \reffig{fig:angularpipl} and \reffig{fig:angularpipnull}, we show the angular distribution for $\pi^+$ and $\pi^0$ production.
Notice that the forward peaked $\pi^+$ are absorbed and thus a flatter distribution is obtained. 
Pions produced in FSI from initial QE events do not have any preferred direction, which leads to a rather flat contribution (dash-dotted line).
The $\pi^0$ channel gains strength from the explained side-feeding, therefore, the absorption is less and the cross section is even enhanced for large scattering angles.

Finally, we compare with calculations available in the literature. In \refcite{Fluka}, the FLUKA cascade model predicts a large pion absorption in the $\Delta$ region. Namely, for 1~GeV $\nu_{\mu}$ energy they find that only 55\% of the charged pions leave a Fe nucleus. We obtain a very similar result of 51\% as can be seen in the left panel of \reffig{fig:pplpnull_tot}.
Large in-medium effects are also reported by Singh et al.~\cite{Singh:1998ha}. They obtain an overall reduction of 40\% in the CC $\Delta$ production of pions at $E_{\nu}=0.75$~GeV on $^{16}$O. By comparing our free cross section shown in the right panel of \reffig{fig:inclDEL} with the sum of all pion cross sections of \reffig{fig:pion_tot} at the same energy we find a reduction of 67\% on a considerably heavier nucleus ($^{56}$Fe). For $^{16}$O we obtain approximately 50\%.
Using the model of pion rescattering of \refscite{Adler:1974qu, Paschos:2000be} Yu \cite{yuphd} finds for $^{16}$O a total reduction of 40~-~60\% in the $\pi^+$ channel and 0~-~25\% in the $\pi^0$ channel depending on the absorption model. The smaller reduction in the $\pi^0$ channel is also due to side-feeding from the dominant $\pi^+$ 	channel followed by charge exchange.
At 2~GeV, we obtain an overall reduction on $^{56}$Fe of 43\% and 2\% for $\pi^+$ and $\pi^0$, respectively (compare dashed and solid line of \reffig{fig:pplpnull_tot}). In the case of $^{16}$O we find also at 2~GeV a reduction of 28\% for $\pi^+$ and an enhancement of 11\% in the $\pi^0$ cross section.

Here we shall recall that all the results obtained for pion production carry an uncertainty related to the poor knowledge of the $\Delta$ form factors as discussed in \refsubch{subsec:delta}.
It is also clear from \reffig{fig:Delta_sigmatot} that the contribution from the higher resonances will affect the pion yields. In the $\pi^+$ channel we expect a minor impact of the order of 10\% since most of them come from the $\Delta$ dominated $\pi^+ p$ channel, while there could be a change of up to 30\% in the $\pi^0$ channel at higher energies due to the smaller cross section. We shall account for this missing strength in the future.

\subsubsection{Nucleon knockout}

\begin{figure}[tb]
  \begin{minipage}[t]{.48\textwidth}
      \begin{center}
        \includegraphics[height=7cm]{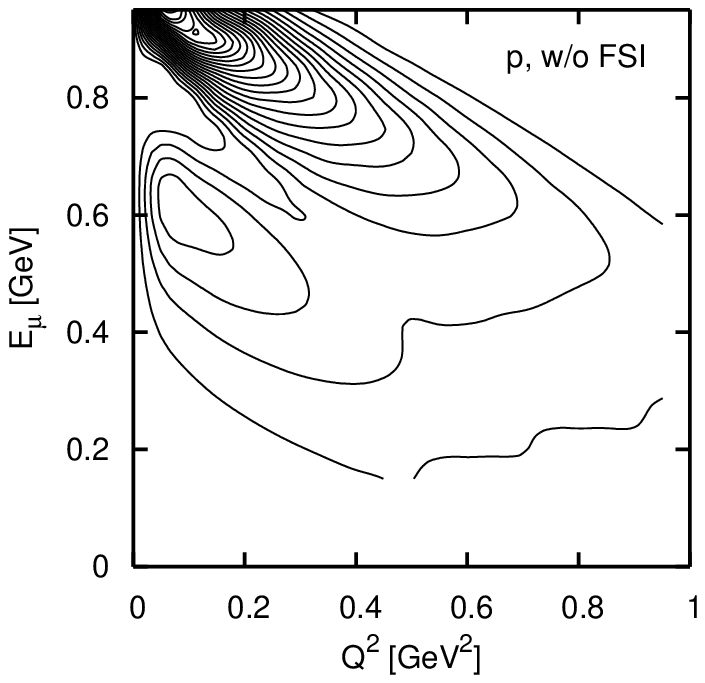}
      \end{center}
  \end{minipage}
  \hfill
  \begin{minipage}[t]{.48\textwidth}
      \begin{center}
        \includegraphics[height=7cm]{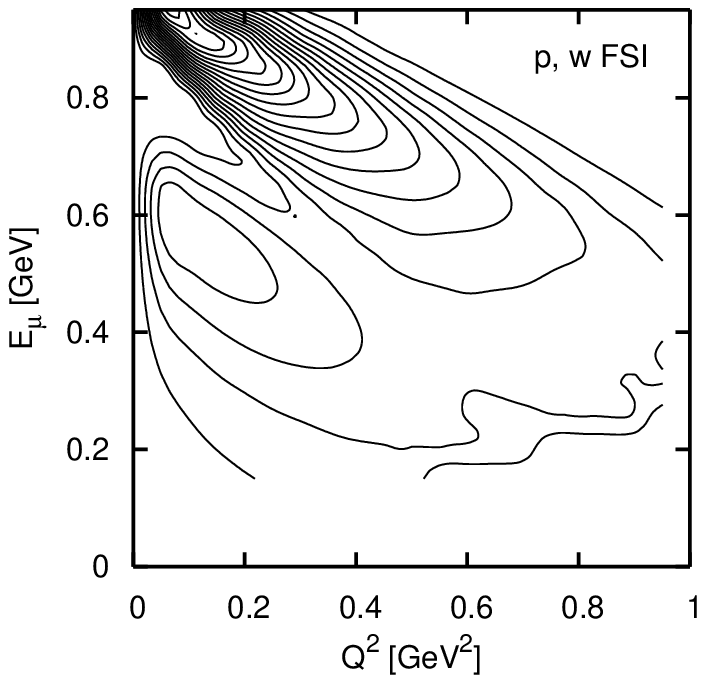}
      \end{center}
  \end{minipage}
 \hfill
\caption{Double differential cross section ${\rm d}\sigma/({\rm d}Q^2 {\rm d} E_{\mu})$ for proton knockout on $^{56}\text{Fe}$ at $E_{\nu}=1 \myunit{GeV}$. The cross section is mapped to the $Q^2-E_{\mu}$ plane. In this result all in-medium modifications are included. The right panel additionally includes FSI. The contour lines are equidistant every $20 \times 10^{-38} \text{cm}^2/\text{GeV}^3$ from $0$ to $480$ (left panel) and $360 \times 10^{-38} \text{cm}^2/\text{GeV}^3$ (right panel), respectively.\label{fig:excl3d_p}}  
\end{figure}

\begin{figure}[tb]
  \begin{minipage}[t]{.48\textwidth}
      \begin{center}
        \includegraphics[height=7cm]{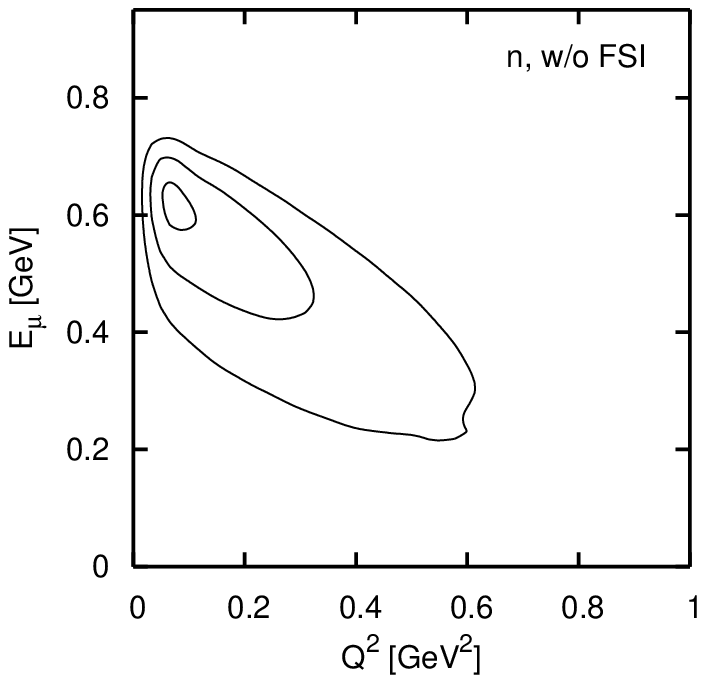}
      \end{center}
  \end{minipage}
  \begin{minipage}[t]{.48\textwidth}
      \begin{center}
        \includegraphics[height=7cm]{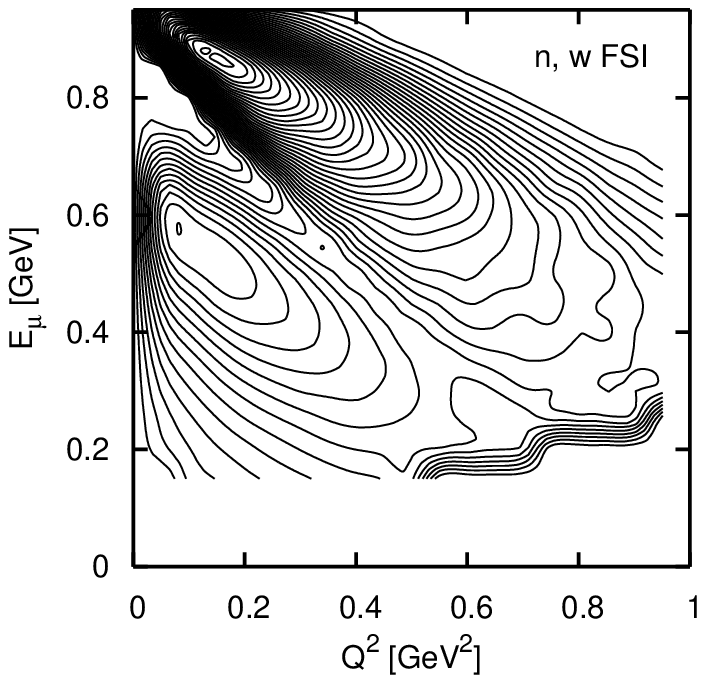}
      \end{center}
  \end{minipage}
 \caption{Same as shown in \reffig{fig:excl3d_p} for neutron knockout. The contour lines are equidistant every $3 \times 10^{-38} \text{cm}^2/\text{GeV}^3$ from $0$ to $9$ (left panel) and $108 \times 10^{-38} \text{cm}^2/\text{GeV}^3$ (right panel), respectively.\label{fig:excl3d_n}}  
\end{figure}

The channel under investigation now is nucleon knockout. We take into account all nucleons which leave the nucleus due to the $ \nu A$ reaction. 
In \reffig{fig:excl3d_p} (\reffig{fig:excl3d_n}) we plot the exclusive cross section for proton (neutron) knockout on $^{56}\text{Fe}$ as a function of $Q^2$ and $E_{\mu}$ for $E_{\nu}=1\myunit{GeV}$. 
All medium modifications as explained in \refsubch{subsec:inmed} are included. In the right panels we additionally take into account all FSI.
Two clearly separated peaks can be seen at low $Q^2$ in both figures. The one at higher $E_{\mu}$ is due to the nucleons produced in initial quasielastic events, whereas the one at lower $E_{\mu}$ results from $\Delta$ production.
This clear separation is lost at higher momentum transfer. We have seen before that the inclusive cross section is smeared out with increasing $Q^2$ due to Fermi motion; the shape of the inclusive cross section (\reffig{fig:incl3d}) is reflected in the exclusive proton and neutron production cross section plotted here. 

The significant difference in scale for proton and neutron knockout is generated by the initial neutrino-nucleon production process: QE scattering produces only protons via
\bea
\nu n \to l^- p;
\eea
no neutrons are produced in the initial quasielastic interactions (cf.~left panel of \reffig{fig:excl3d_n}).
Also the $\Delta$ production mechanism favors protons
\bea
&&\nu p \to l^- \Delta^{++} \to l^- p \,\pi^+, \\
&&\nu n \to l^- \Delta^{+} \to l^- p \,\pi^0, \; \; l^- n \,\pi^+,
\eea
since the first process is enhanced by a factor of three.
Using isospin amplitudes we obtain a ratio of $p:n= \left[ Z + \left(-\sqrt{2}/3\right)^2 N \right]\left[\left(1/3\right)^2 N\right]^{-1}= 9.8:1$ ($N$ and $Z$ are the proton and neutron numbers) in the $\Delta$ region for $^{56}\text{Fe}$.
Therefore, in a calculation without final-state interactions, proton and neutron knockout differ by about a factor of ten in the $\Delta$ region. 

With final-state interactions, this scenario changes. On the right panel in \reffig{fig:excl3d_n} we can see that neutrons are produced in the quasielastic peak region through multistep processes. The initial neutrino-nucleon QE reaction produces only protons, which undergo elastic and inelastic nucleon-nucleon collisions in the medium via $NN \to NN$, $NN \to NN\pi$ or $NN \to N\Delta$. This leads to charge exchange, and thus, to neutron production. The situation is similar in the region of the $\Delta$.

\begin{figure}[tb]
   \begin{minipage}[t]{.48\textwidth}
       \begin{center}
         \includegraphics[height=5.1cm]{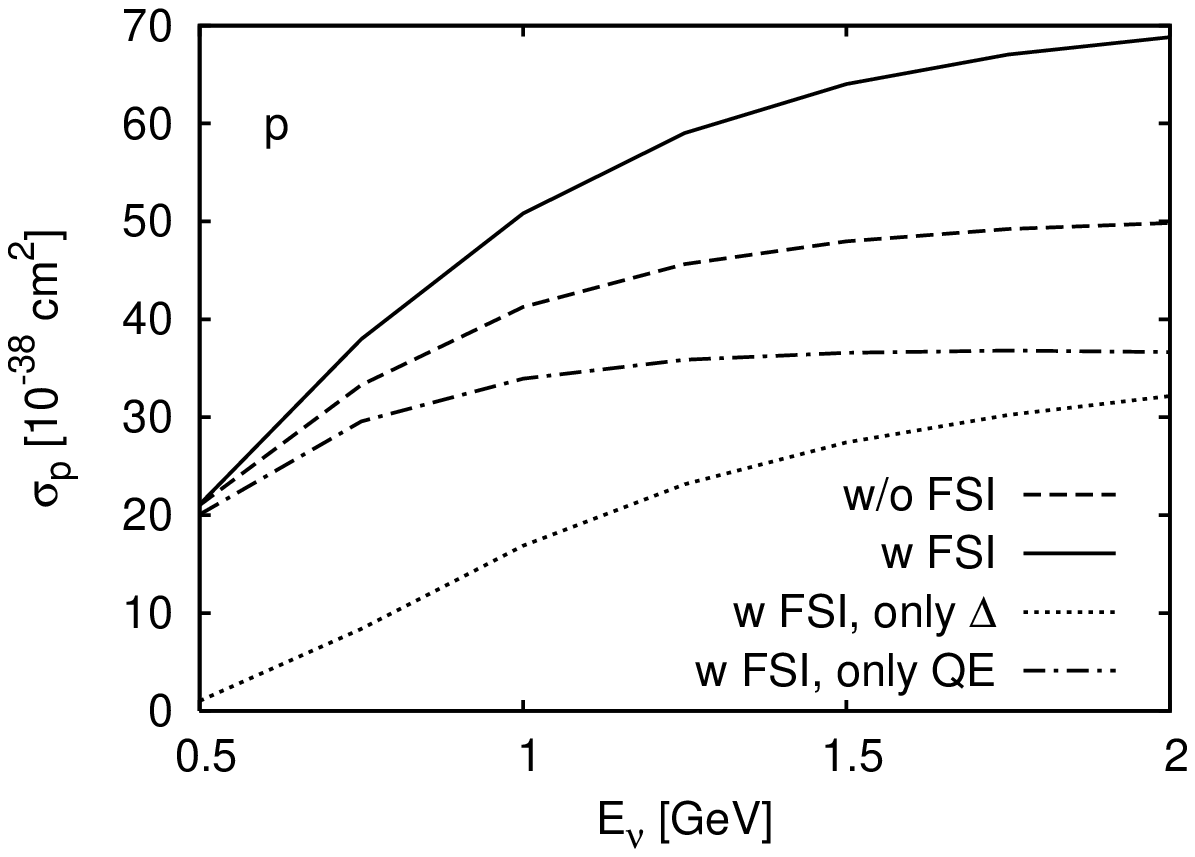}
          \end{center}
   \end{minipage}
   \begin{minipage}[t]{.48\textwidth}
       \begin{center}
         \includegraphics[height=5.1cm]{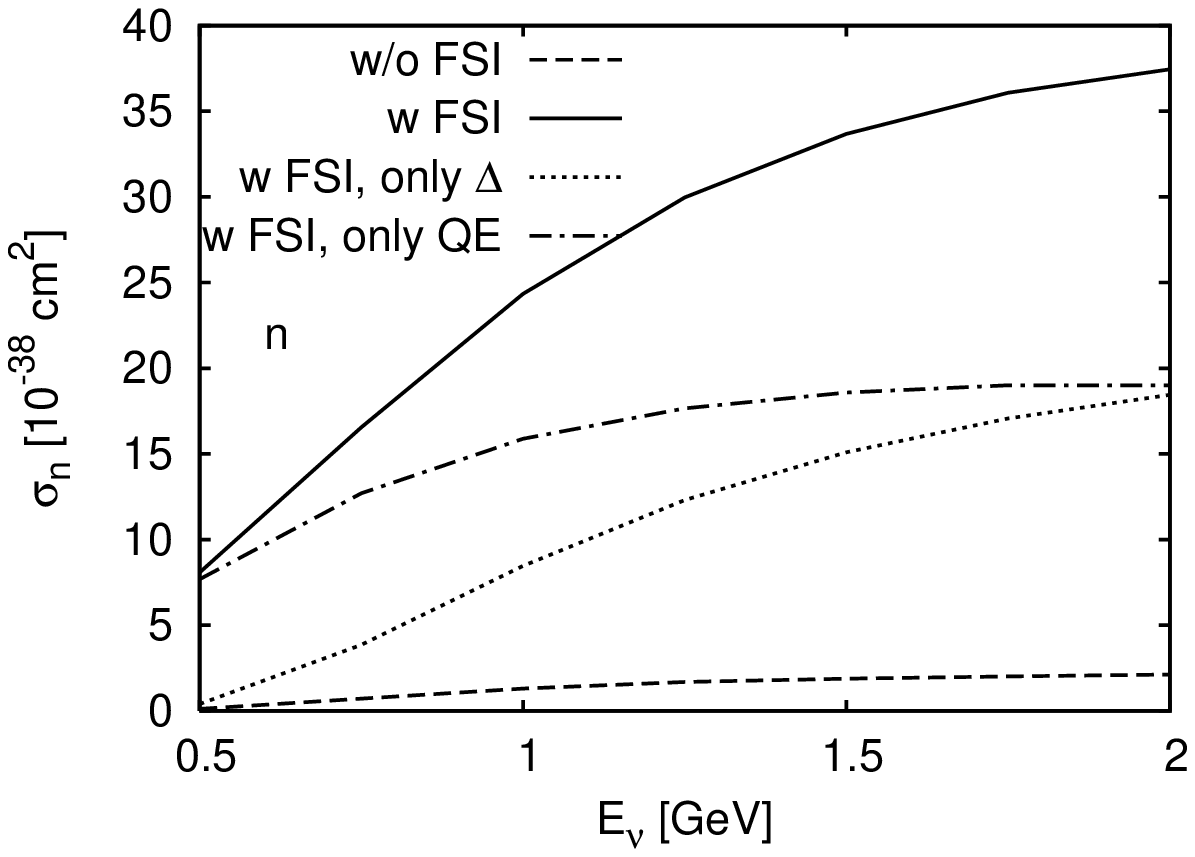}
       \end{center}
   \end{minipage}
   \caption{Total cross section for proton (left) and neutron (right panel) knockout on $^{56}\text{Fe}$ versus the neutrino energy. The dashed line shows the results without FSI interactions, the results denoted by the solid line include FSI. Furthermore, the origin of the nucleon is indicated (QE or $\Delta$ excitation).\label{fig:nucl_tot}}
 \end{figure}

The total cross sections for proton and neutron knockout are shown in \reffig{fig:nucl_tot}. 
The solid line, showing the result with all final-state interactions included, lies well above the one without FSI (dashed line) already for the protons, but even more so for the neutrons. This enhancement is entirely due to secondary interactions and cannot be obtained in a Glauber treatment.
Furthermore, it is indicated in \reffig{fig:nucl_tot} whether the knocked out nucleon stems from initial QE scattering or $\Delta$ excitation.
In contrast to the pion case, both contribute to the total cross section, even though with different weight depending on the neutrino energy. This weight follows simply from the total inclusive cross section (cf.~\reffig{fig:incl}) and thus explains the small contribution of the $\Delta$ at $E_{\nu}=0.5 \myunit{GeV}$, which gets larger at higher energies.

\begin{figure}[tb]
  \begin{minipage}[t]{.48\textwidth}
      \begin{center}
        \includegraphics[height=5.1cm]{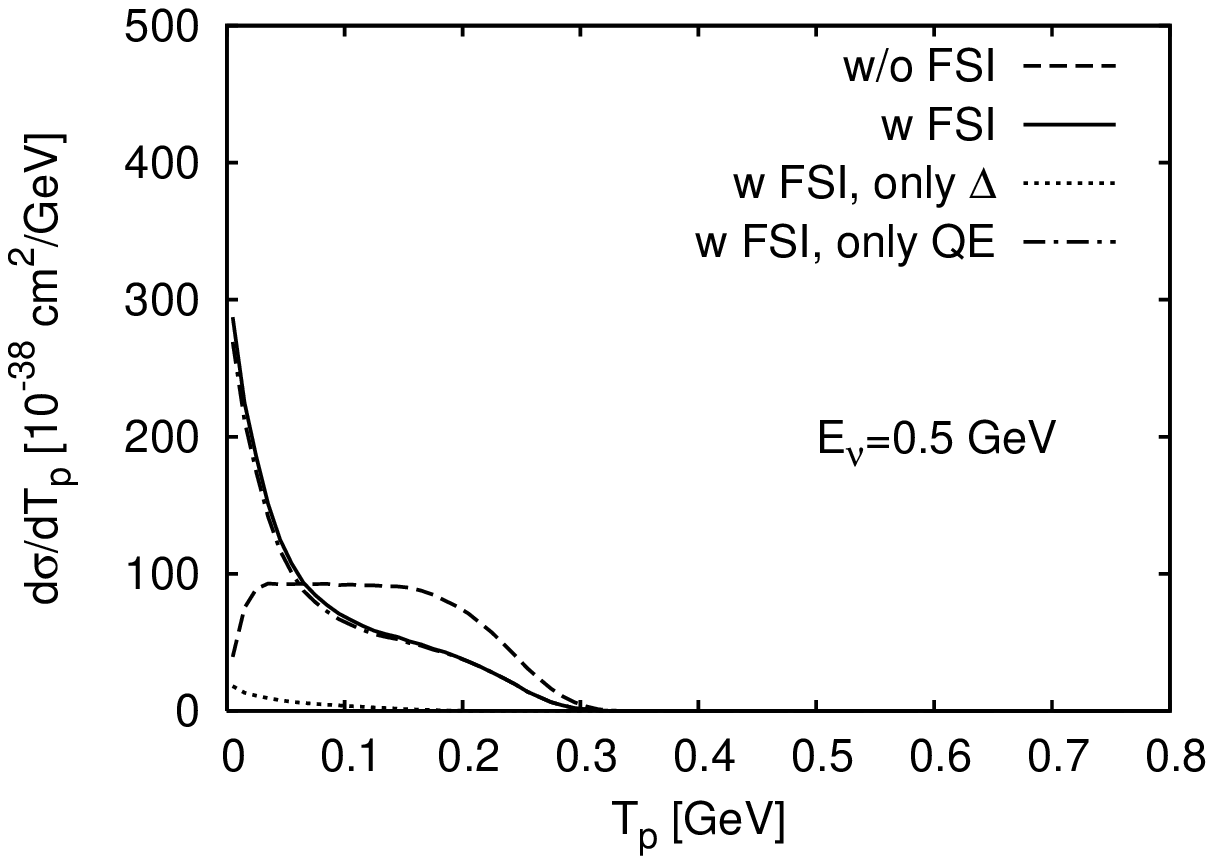}
       \end{center}
  \end{minipage}
  \begin{minipage}[t]{.48\textwidth}
      \begin{center}
        \includegraphics[height=5.1cm]{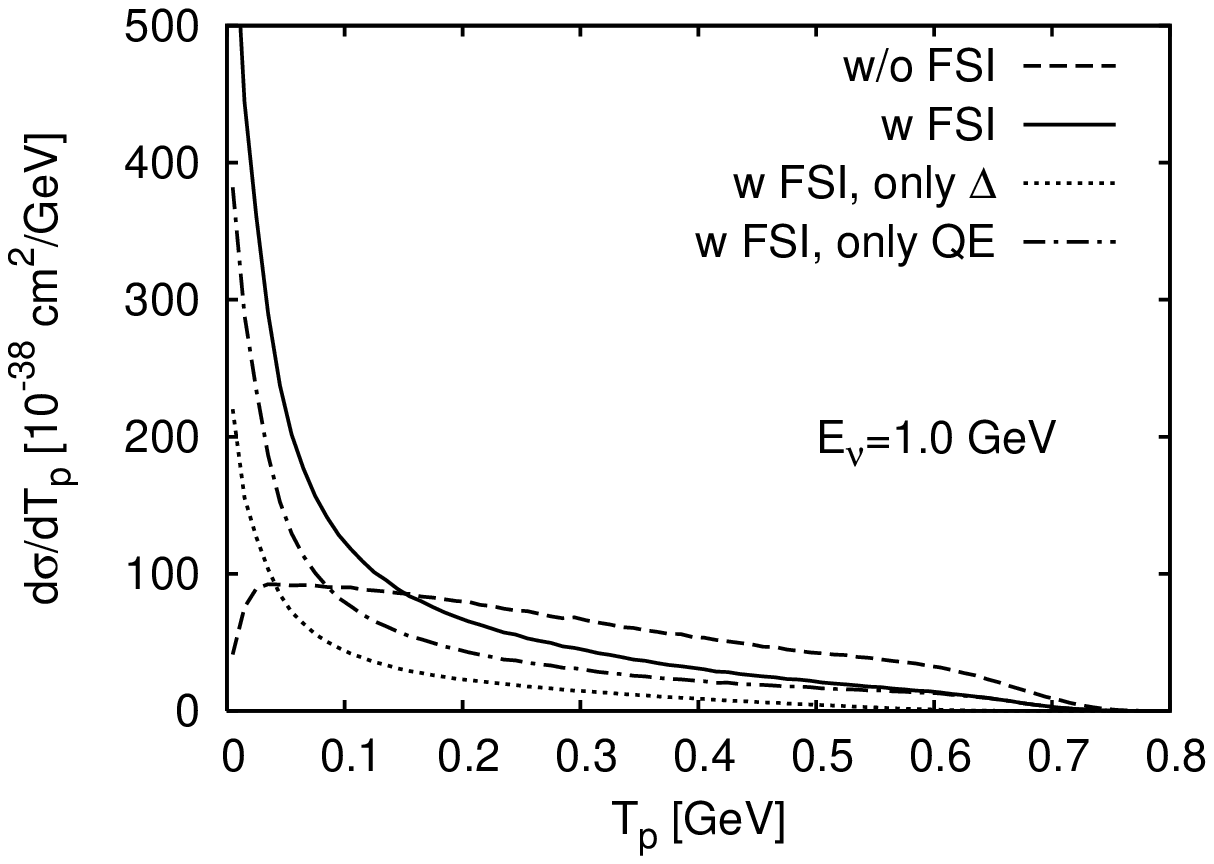}
      \end{center}
  \end{minipage}
 \\ \hfill \\
  \begin{minipage}[t]{.48\textwidth}
      \begin{center}
        \includegraphics[height=5.1cm]{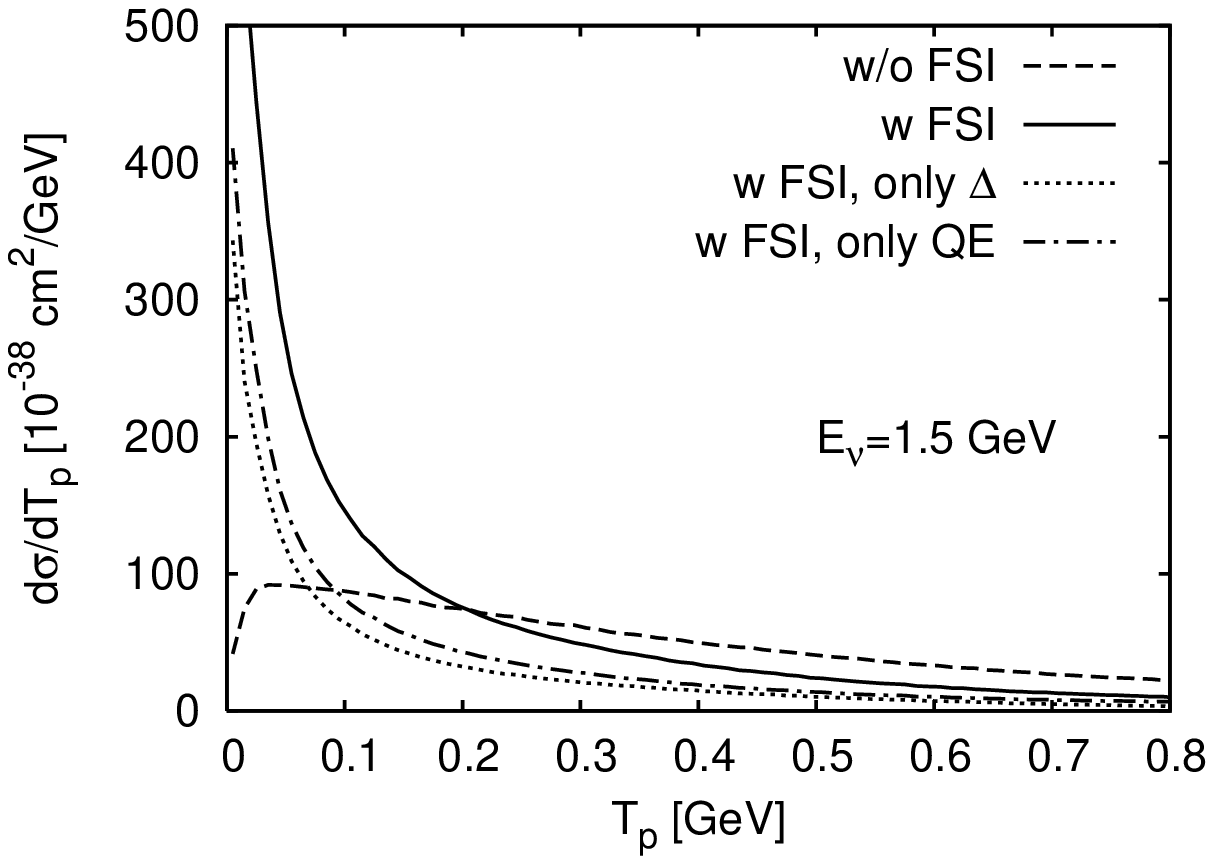}
      \end{center}
  \end{minipage}
  \begin{minipage}[t]{.48\textwidth}
      \begin{center}
        \includegraphics[height=5.1cm]{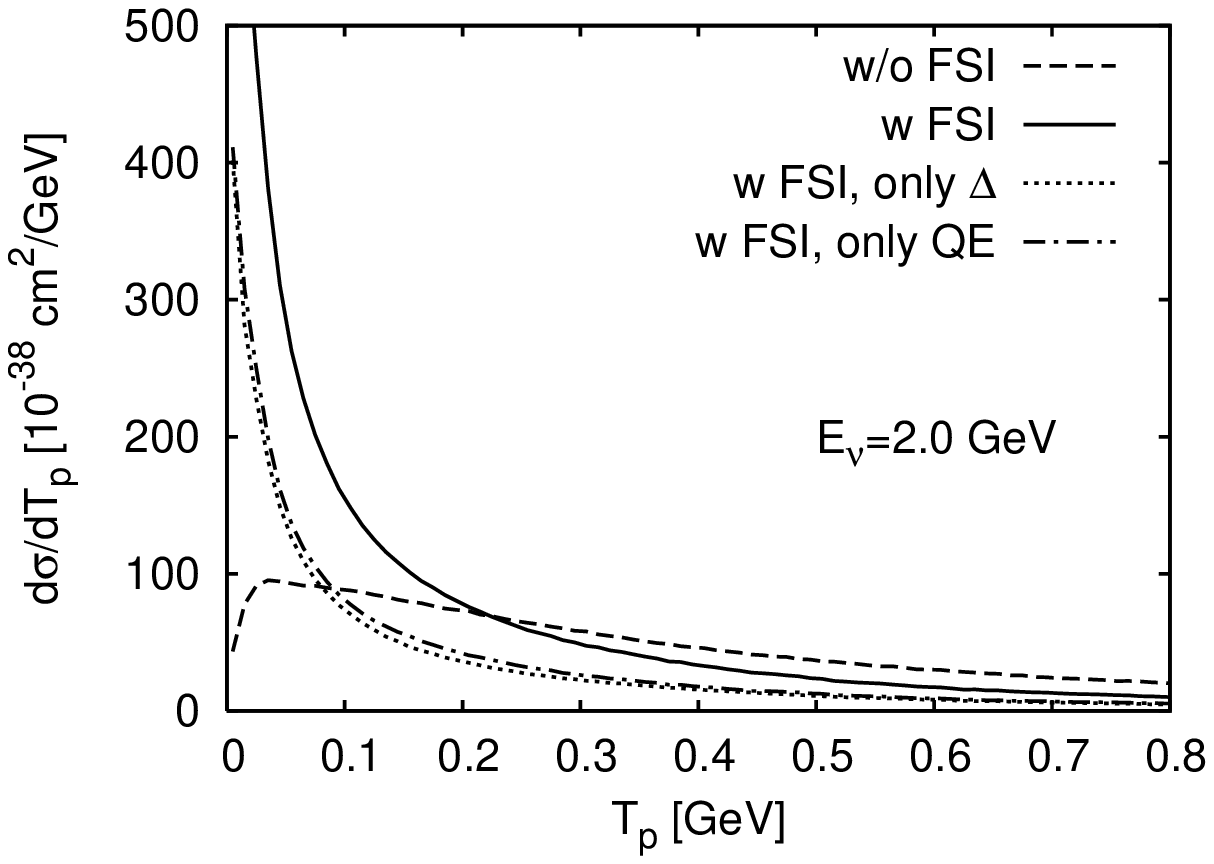}
      \end{center}
  \end{minipage}
  \caption{Kinetic energy differential cross section for proton knockout on $^{56}\text{Fe}$ versus the proton kinetic energy $T_{p}$ at different values of $E_{\nu}$. The dashed line shows the results without FSI interactions, the results denoted by the solid line include FSI. Furthermore, the origin of the proton is indicated (QE or $\Delta$ excitation). \label{fig:prot}}  
\end{figure}

\begin{figure}[tb]
  \begin{minipage}[t]{.48\textwidth}
      \begin{center}
        \includegraphics[height=5.1cm]{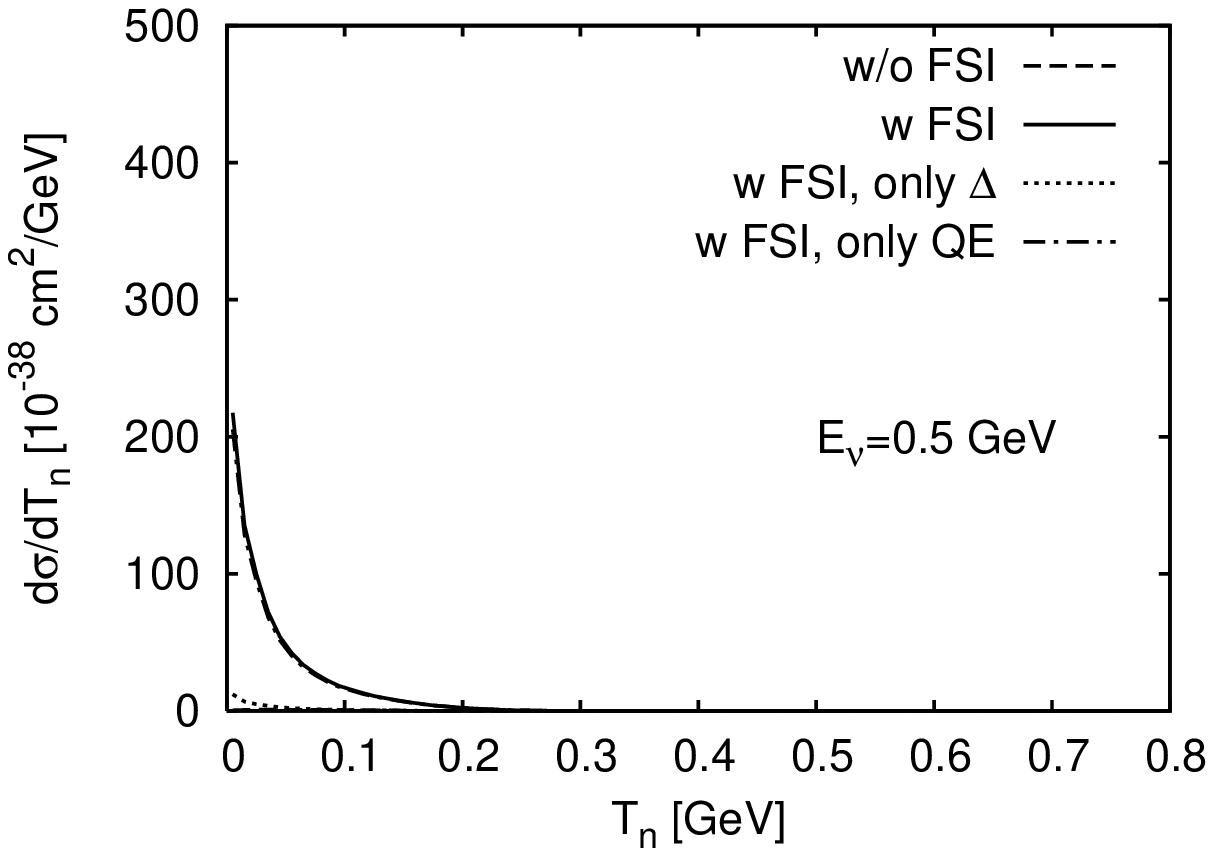}
       \end{center}
  \end{minipage}
  \begin{minipage}[t]{.48\textwidth}
      \begin{center}
        \includegraphics[height=5.1cm]{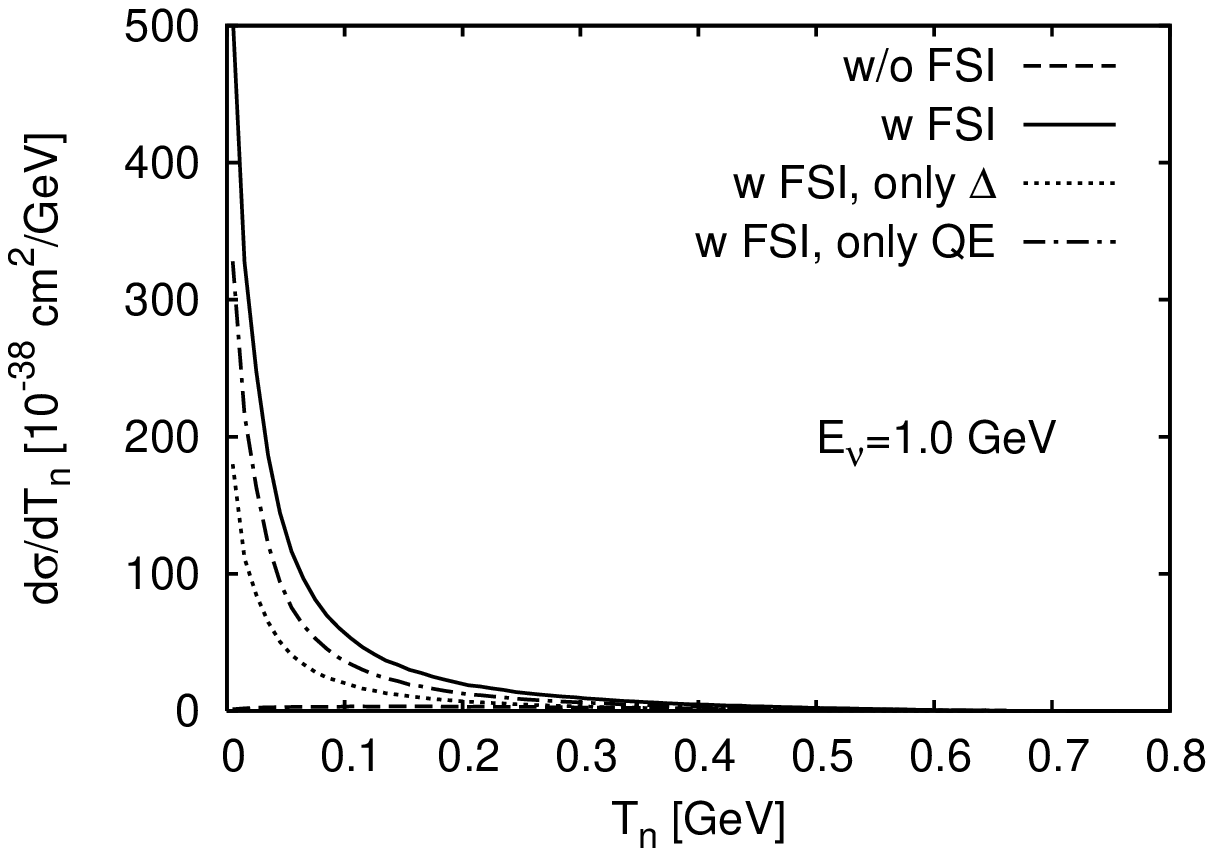}
      \end{center}
  \end{minipage}
 \\ \hfill \\
  \begin{minipage}[t]{.48\textwidth}
      \begin{center}
        \includegraphics[height=5.1cm]{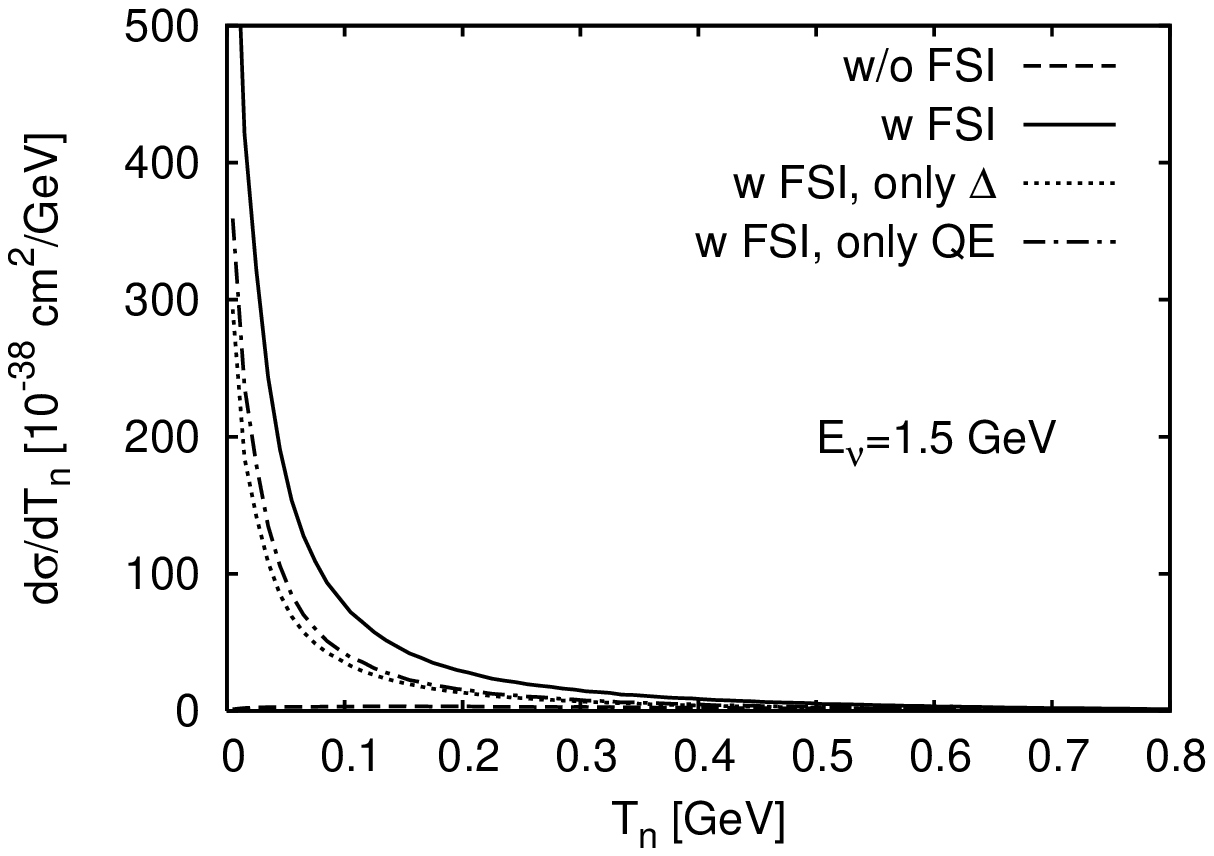}
      \end{center}
  \end{minipage}
  \begin{minipage}[t]{.48\textwidth}
      \begin{center}
        \includegraphics[height=5.1cm]{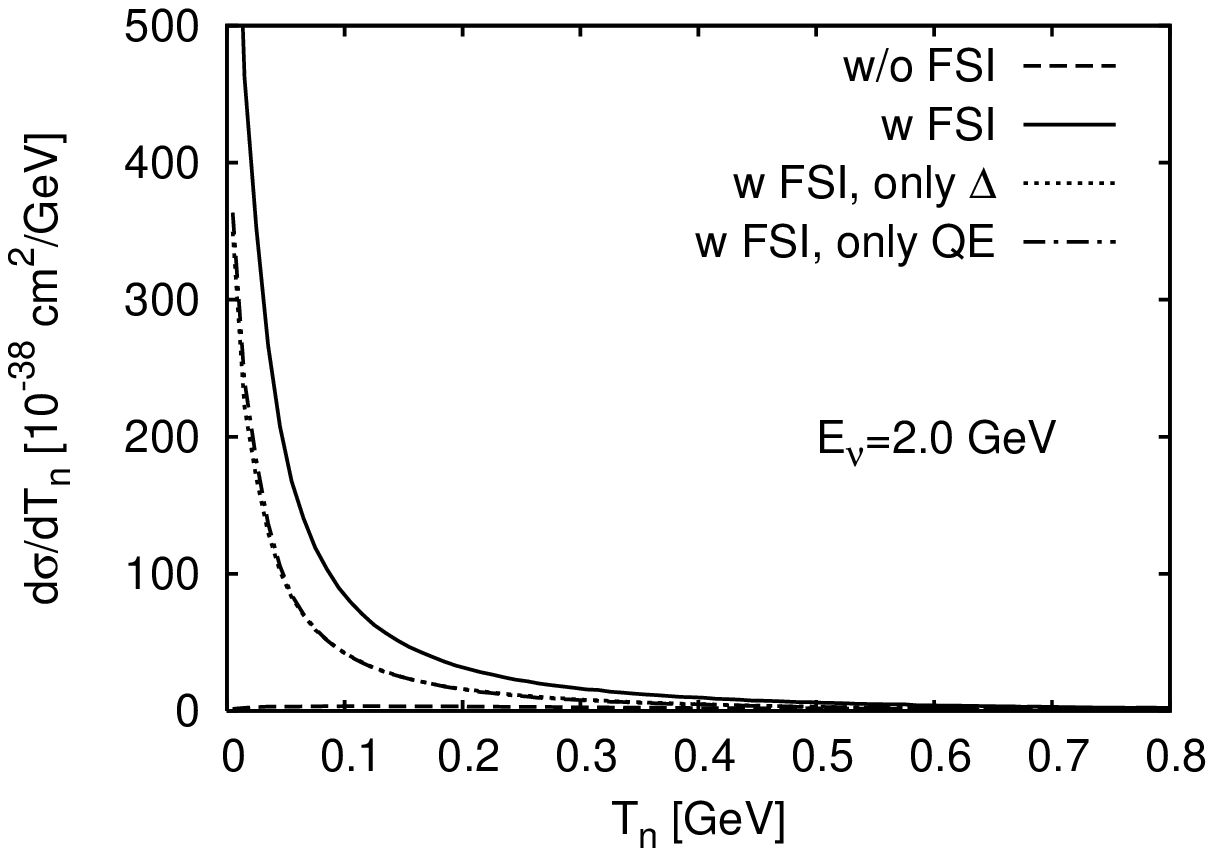}
      \end{center}
  \end{minipage}
  \caption{Same as shown in \reffig{fig:prot} for neutron knockout.\label{fig:neut}}  
\end{figure}

In \reffig{fig:prot} and \reffig{fig:neut} we present the kinetic energy differential cross section for proton and neutron knockout versus the kinetic energy for different values of $E_{\nu}$. 
The line styles are as in the previous figures.
FSI strongly modifies the shape of the distribution. High energy protons rescatter in the medium. As a consequence the flux at higher energies is reduced and a large number of secondary protons at lower energies appear (cf.~\reffig{fig:prot}).  
Also low energy neutrons are produced through charge changing FSI as can be seen in \reffig{fig:neut} where, in the case without FSI, the cross section almost vanishes.

\begin{figure}[tb]
  \begin{minipage}[t]{.48\textwidth}
      \begin{center}
        \includegraphics[height=5.1cm]{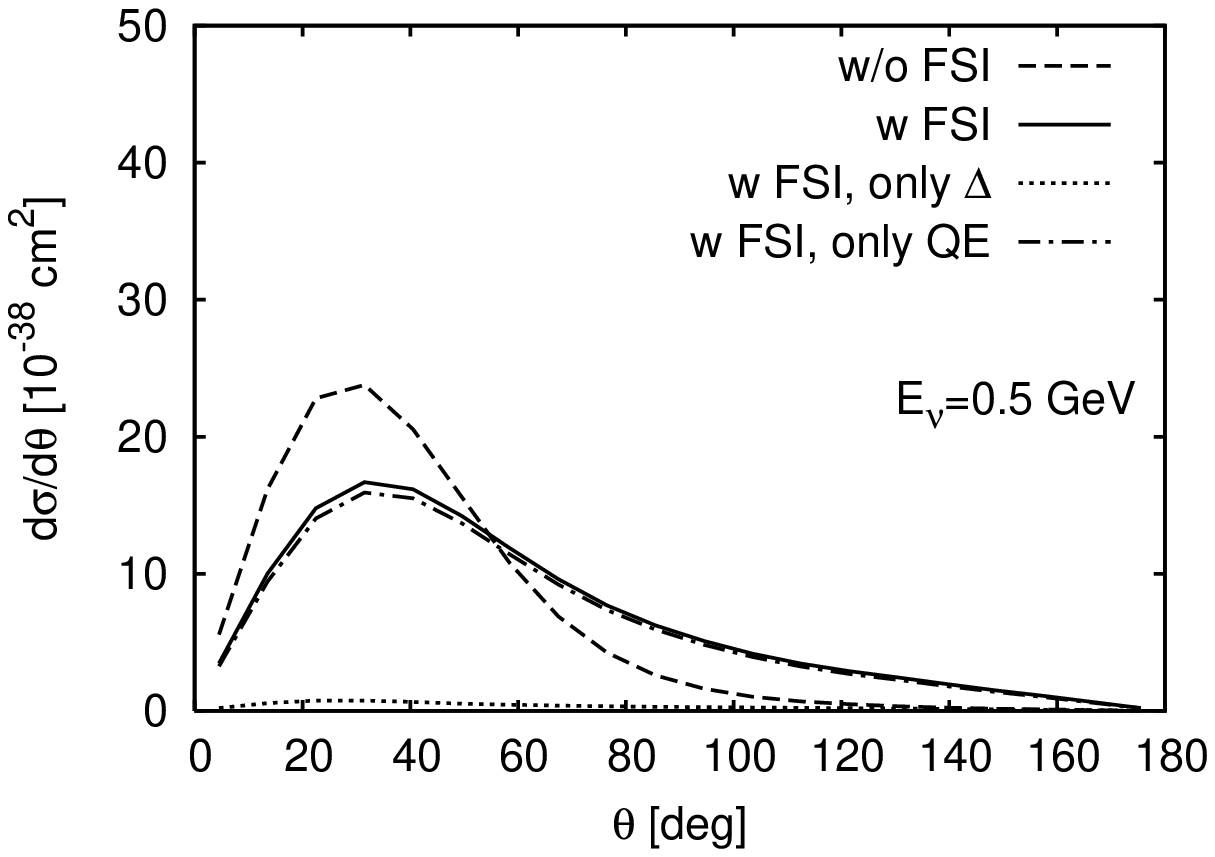}
       \end{center}
  \end{minipage}
  \begin{minipage}[t]{.48\textwidth}
      \begin{center}
        \includegraphics[height=5.1cm]{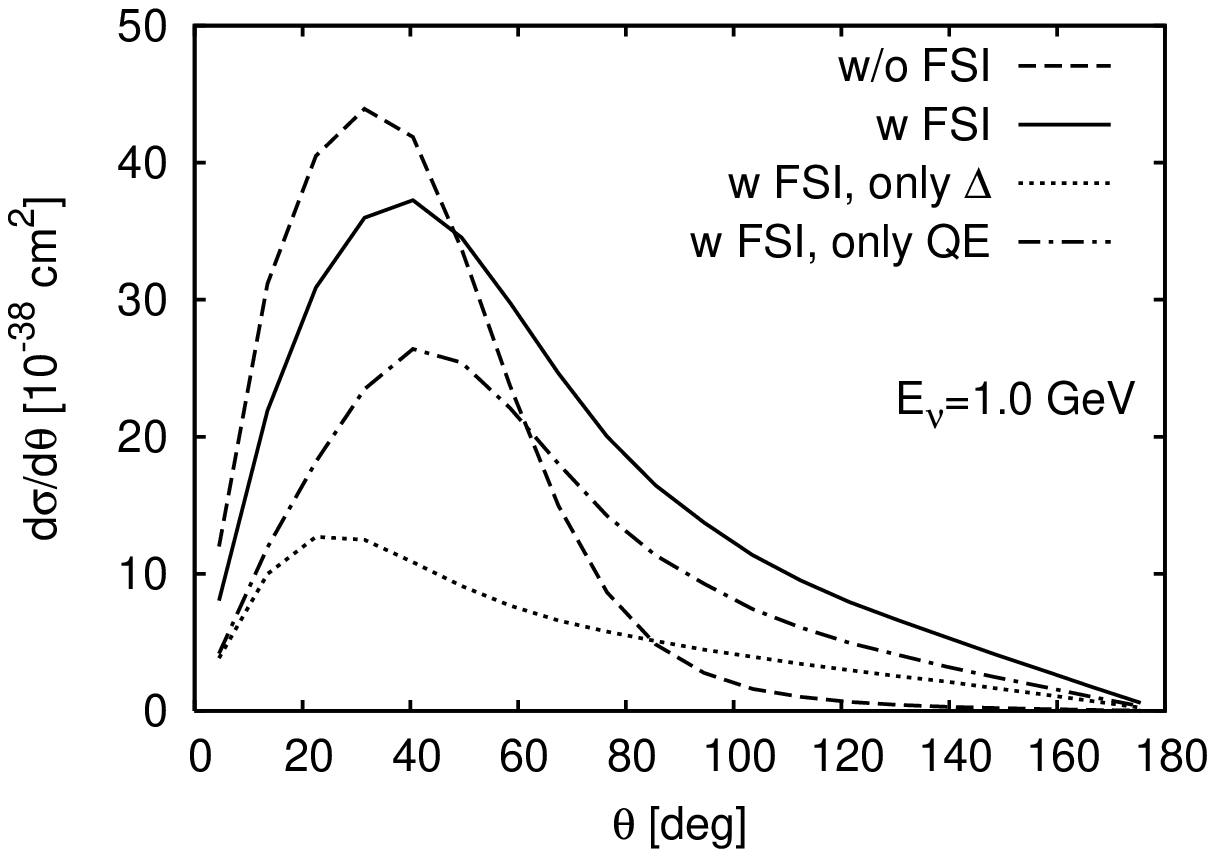}
      \end{center}
  \end{minipage}
 \\ \hfill \\
  \begin{minipage}[t]{.48\textwidth}
      \begin{center}
        \includegraphics[height=5.1cm]{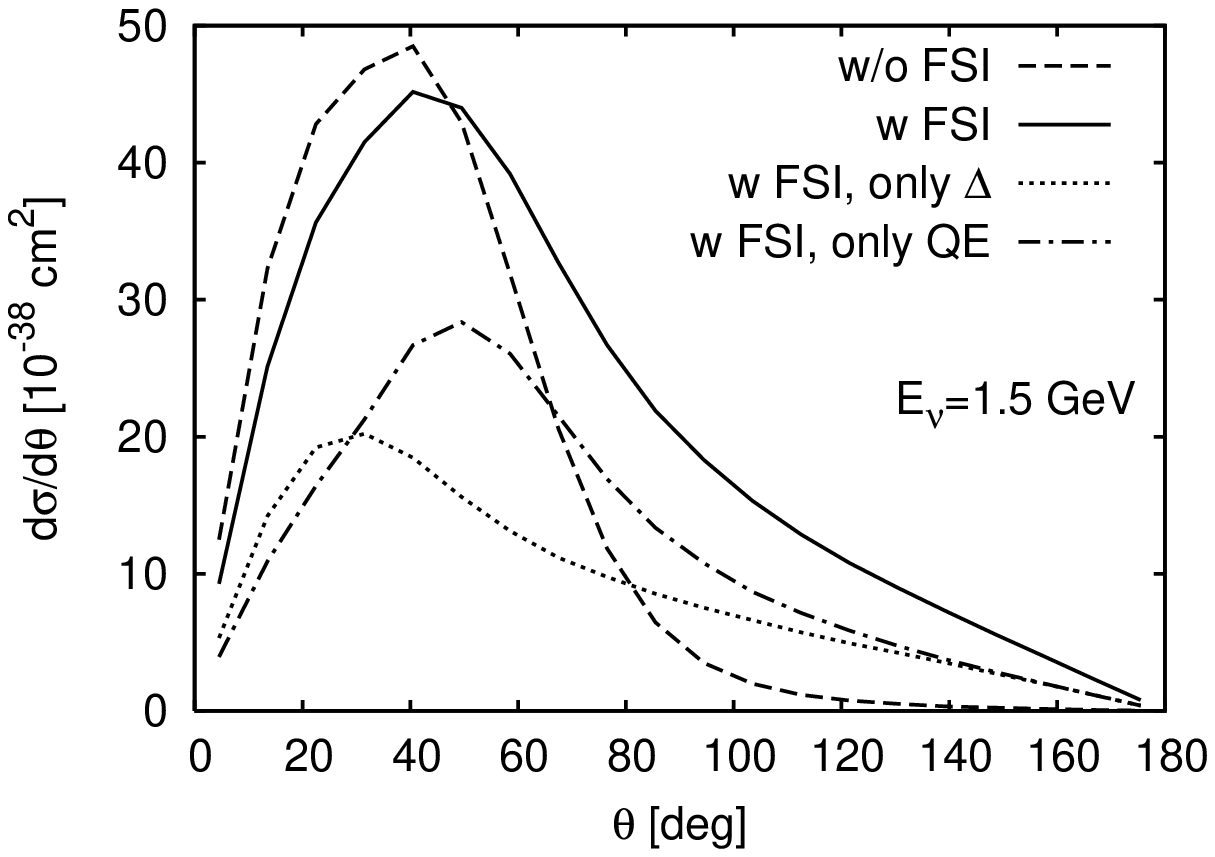}
      \end{center}
  \end{minipage}
  \begin{minipage}[t]{.48\textwidth}
      \begin{center}
        \includegraphics[height=5.1cm]{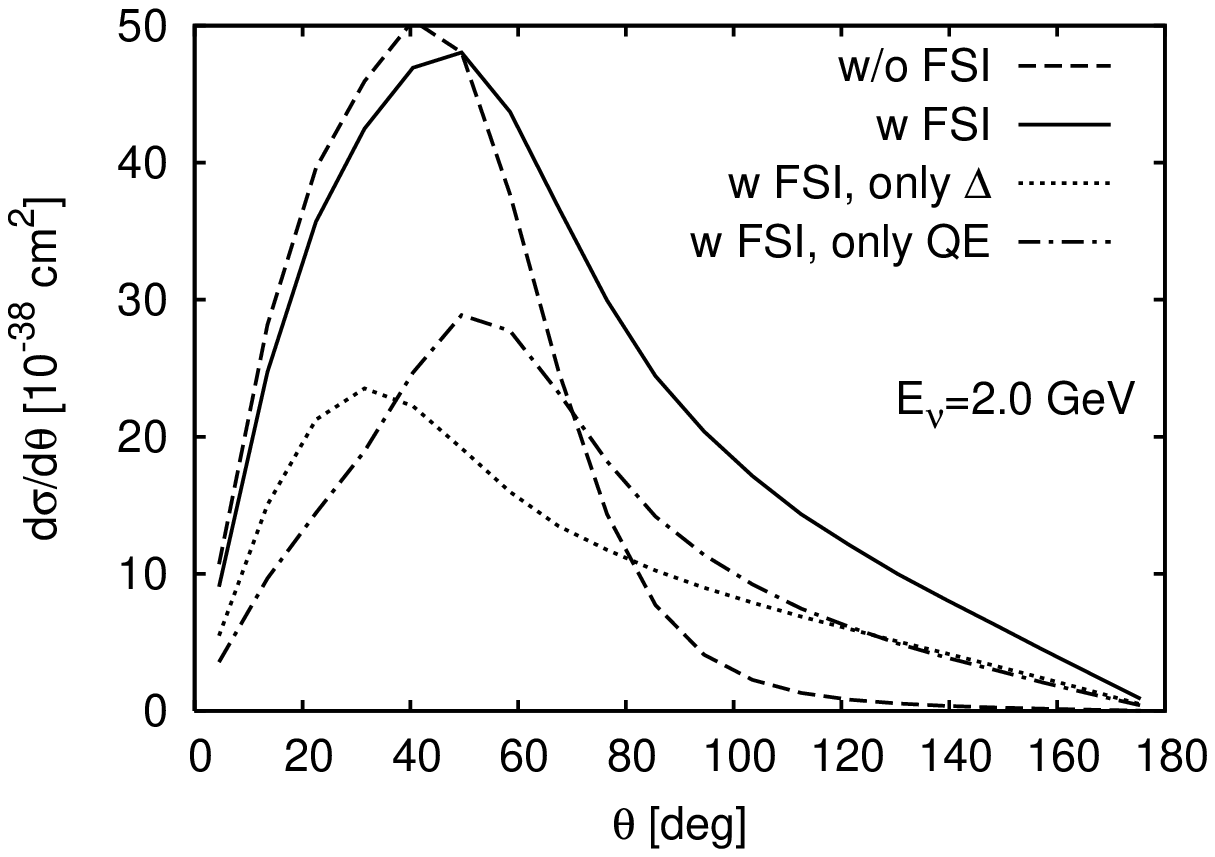}
      \end{center}
  \end{minipage}
\caption{Angular distribution for proton knockout on $^{56}\text{Fe}$ for various neutrino energies. The angle $\theta$ is measured with respect to the direction of the incoming neutrino. Labels are as in \reffig{fig:prot} and \reffig{fig:neut}. \label{fig:angularp}}  
\end{figure}

\begin{figure}[tb]
  \begin{minipage}[t]{.48\textwidth}
      \begin{center}
        \includegraphics[height=5.1cm]{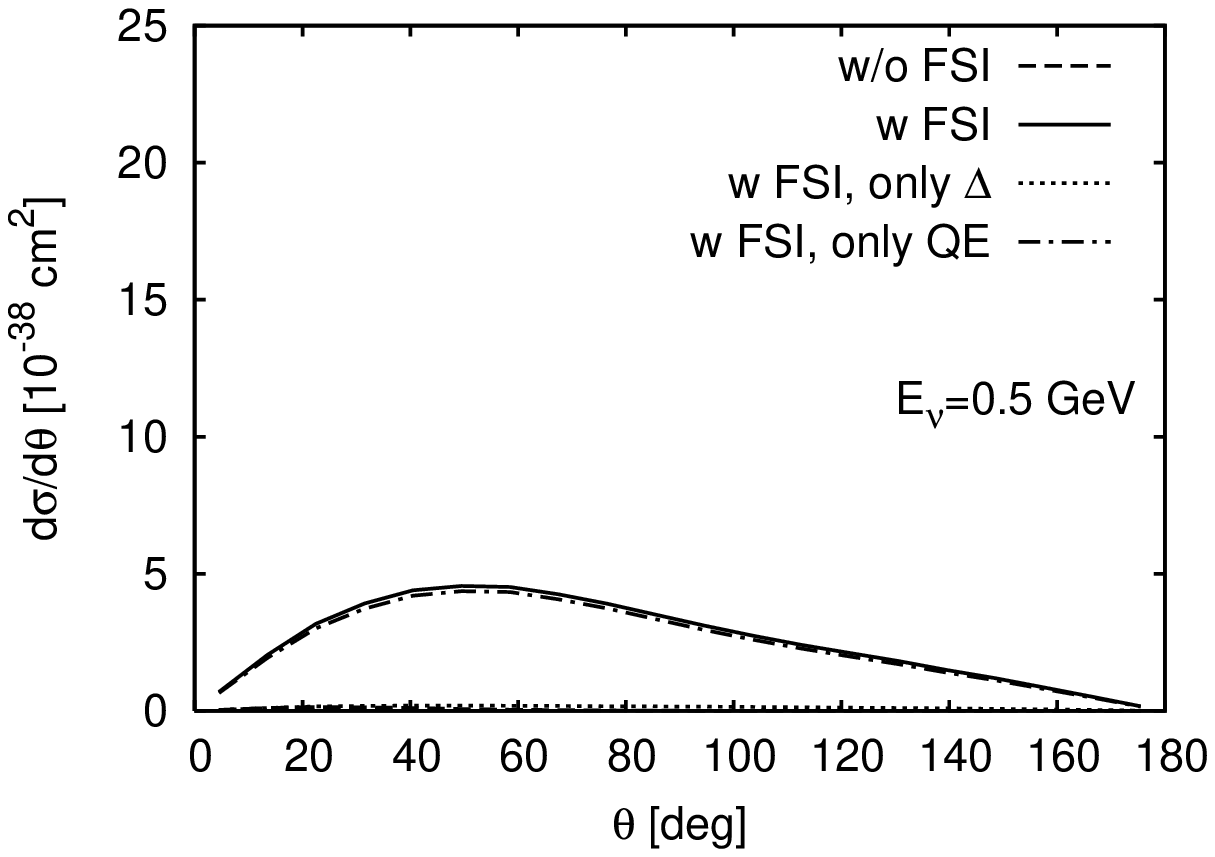}
       \end{center}
  \end{minipage}
  \begin{minipage}[t]{.48\textwidth}
      \begin{center}
        \includegraphics[height=5.1cm]{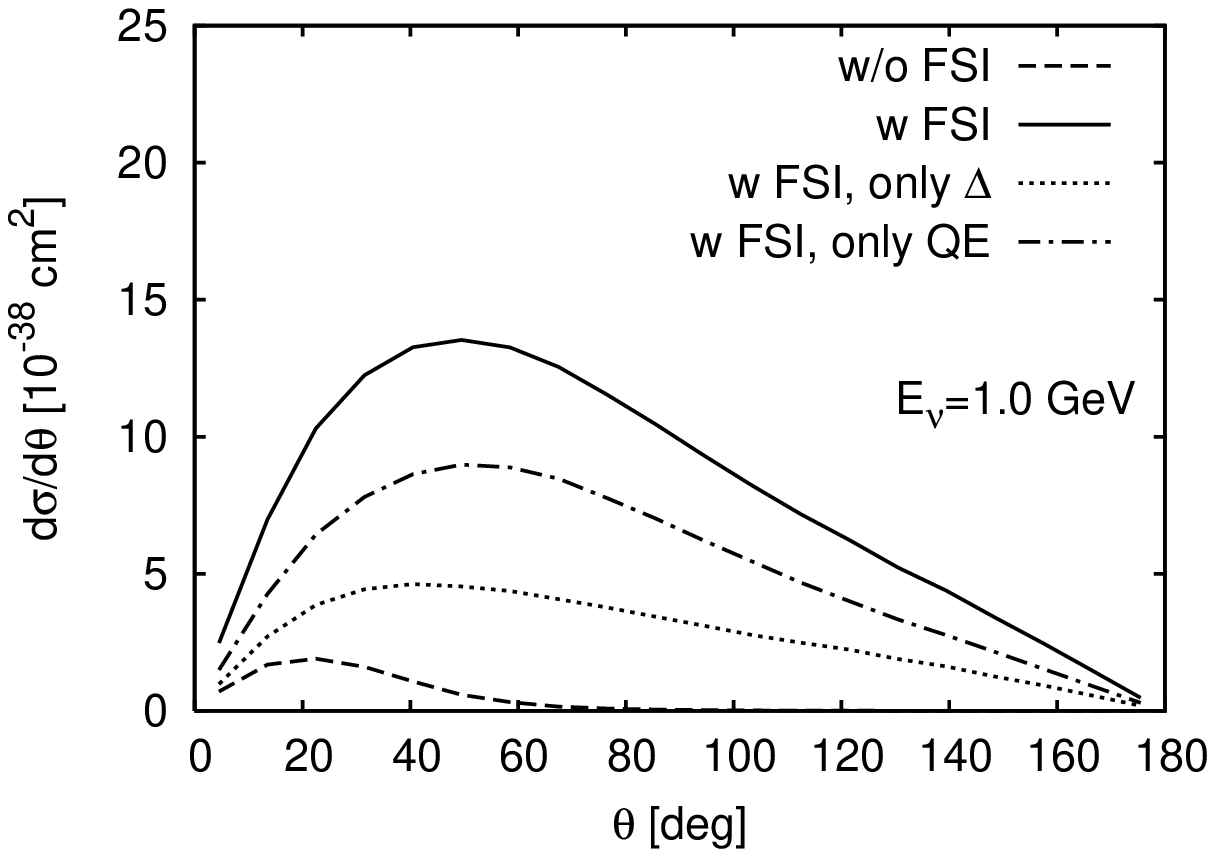}
      \end{center}
  \end{minipage}
 \\ \hfill \\
  \begin{minipage}[t]{.48\textwidth}
      \begin{center}
        \includegraphics[height=5.1cm]{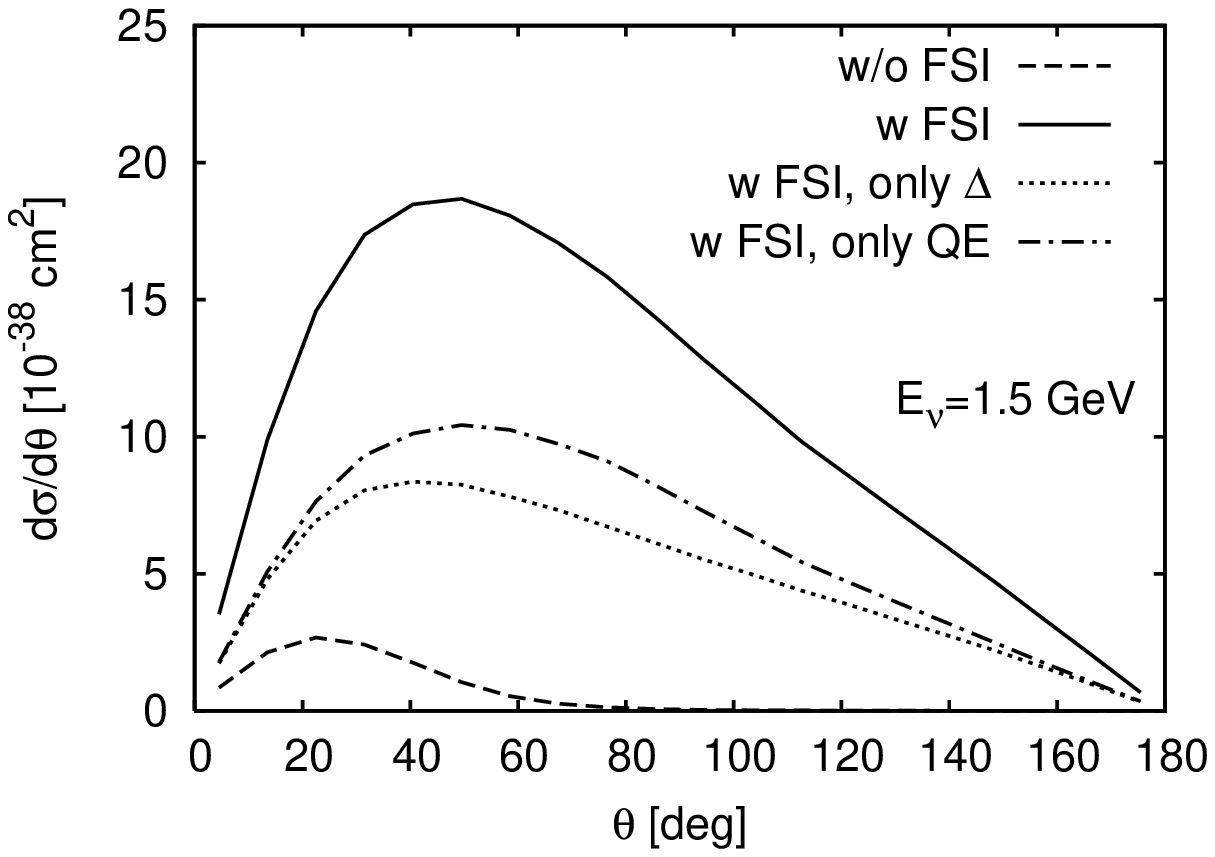}
      \end{center}
  \end{minipage}
  \begin{minipage}[t]{.48\textwidth}
      \begin{center}
        \includegraphics[height=5.1cm]{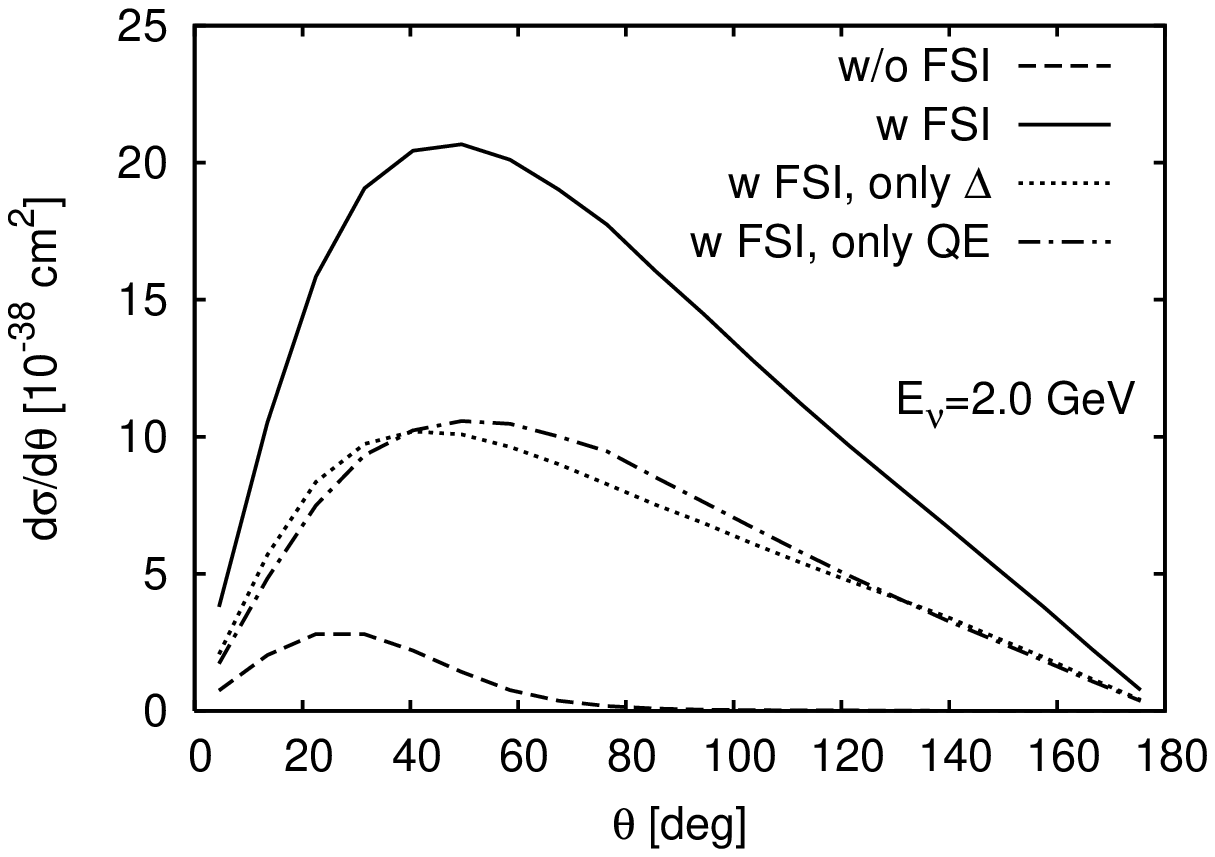}
      \end{center}
  \end{minipage}
\caption{Same as shown in \reffig{fig:angularp} for neutron knockout. \label{fig:angularn}}  
\end{figure}

The angular distribution is plotted in \reffig{fig:angularp} and \reffig{fig:angularn};
the line styles are as in the previous plots. In the case of the protons, it is clearly seen how FSI shift the strength from small ($\sim 30^{\circ}$) to large angles ($\gtrsim 60^{\circ}$)  as expected. The angular distribution for the neutrons shows an overall increase as a result of the described side-feeding. 

\begin{figure}
      \begin{center}
        \includegraphics[scale=.75]{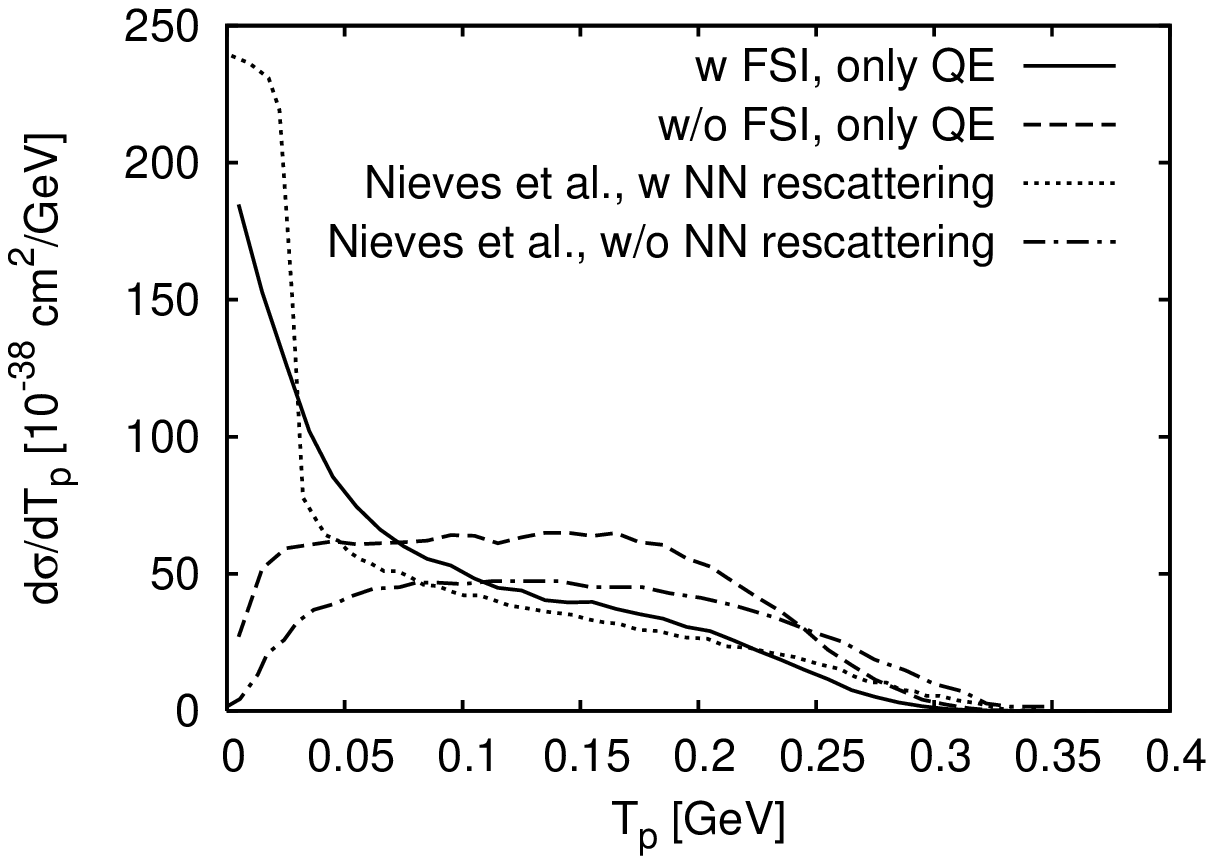}
        \caption{Muon neutrino-induced proton knockout on $^{40}$Ar at $E_{\nu}=0.5 \myunit{GeV}$ by QE scattering. The dotted and the dash-dotted line show the results of \refcite{Nieves:2005rq}, the solid and the dashed line are the results of our model without $\Delta$ excitation. \label{fig:nieves}}
      \end{center}
\end{figure}

Finally, we want to compare our results with others. \reffig{fig:nieves} shows our results together with those obtained by Nieves et al.~\cite{Nieves:2005rq}. 
Since in this calculation only QE scattering is included, we have "turned off" the $\Delta$ excitation. 
The difference between the results without FSI (dashed versus dash-dotted line) is due to the RPA correlations taken into account by Nieves et al.~which cause a reduction of the cross section and a spreading of the spectrum allowing, for larger energy transfers. 
However, when the rescattering of the outgoing nucleons inside the nucleus is considered, both calculations lead to very similar cross sections, namely a reduction of the flux for higher energetic protons and to a large number of secondary low energy protons (solid versus dotted line).

The flux reduction can also be achieved with optical potential models. However, by simple absorption of the nucleons they do not account for the rescattering in the medium which leads to the large number of secondary nucleons. Nucleons are not just absorbed but --- through rescattering --- ejected with a different energy, angle and/or charge. In particular, with a model not accounting for FSI in a realistic way, there would be no neutrons in the final state when the initial $\nu N$ collision was quasielastic.

\section{Summary and conclusions \label{sec:summary}}

In this work we have investigated neutrino interactions with nuclei. 
The model presented here is able to describe neutrino 
reactions on both nucleons and nuclei at intermediate energies on the same basis as photonuclear quasielastic and meson production processes.

We have focused on the region of the quasielastic and $\Delta (1232)$ peaks. For $\nu N$ collisions a fully relativistic formalism is used. The extension to finite nuclei has been done in the framework of a coupled-channel BUU transport theory where we have taken into account in-medium modifications due to Fermi motion, Pauli blocking, nuclear binding and collisional broadening of resonances. This gives us the possibility to study exclusive channels taking into account in-medium effects and final-state interactions.

To summarize our results on neutrino-induced pion production, we have found that
the in-medium effects and, especially, final-state interactions reduce the exclusive pion cross section and also give rise to a small fraction of $\pi^-$. Quasielastic scattering followed by $\pi$ production in $NN$ collisions contributes only weakly to the pion production cross section. In the kinematical region under investigation, the pions originate mainly from the initial $\Delta$ excitation. Furthermore, we have found an enhancement of the $\pi^0$ channel through side-feeding from the dominant $\pi^+$ channel.

For nucleon knockout, we have found that the influence of the final-state interactions is significant. High energy nucleons rescatter in the nucleus, which leads to a decrease of the flux at higher energies, but also to a large number of secondary nucleons at low nucleon energies. While the elementary quasielastic reaction cannot produce neutrons, but only protons, we found that, as a consequence of the final-state interactions, a large fraction of neutrons is nevertheless produced. Also in the $\Delta$ region we found a large enhancement of neutrons due to final-state interactions. 

We conclude that in-medium effects in $\nu A$ scattering, and in particular FSI, are essential. Their understanding --- within a well tested transport model --- is crucial for current and future experiments.

At the point of submission, we became aware of a article by Cassing et al.~\cite{Cassing:2006sg} studying the problem of neutrino-induced pion production.

\begin{acknowledgments}
This work was supported by DFG. One of us, L.~A.~R., has been supported in part by the Alexander-von-Humboldt-Foundation.
\end{acknowledgments}


\begin{thebibliography}{10}

\bibitem{Fukuda:1998mi}
Super-Kamiokande, Y.~Fukuda {\em et~al.},
\newblock Phys. Rev. Lett. {\bf 81}, 1562 (1998), [hep-ex/9807003].

\bibitem{Ahmad:2001an}
SNO, Q.~R. Ahmad {\em et~al.},
\newblock Phys. Rev. Lett. {\bf 87}, 071301 (2001), [nucl-ex/0106015].

\bibitem{minos}
MINOS,
\newblock \texttt{http://www-numi.fnal.gov/}.

\bibitem{k2k}
K2K,
\newblock \texttt{http://neutrino.kek.jp/}.

\bibitem{miniboone}
MiniBooNE,
\newblock \texttt{http://www-boone.fnal.gov/}.

\bibitem{minervaprop}
MINER$\nu$A, D.~Drakoulakos {\em et~al.},
\newblock hep-ex/0405002.

\bibitem{finesseprop}
FINeSSE, L.~Bugel {\em et~al.},
\newblock hep-ex/0402007.

\bibitem{Alberico:2001sd}
W.~M. Alberico, S.~M. Bilenky and C.~Maieron,
\newblock Phys. Rept. {\bf 358}, 227 (2002), [hep-ph/0102269].

\bibitem{Cavanna:2005yp}
F.~Cavanna, C.~Keppel, P.~Lipari and M.~Sakuda, editors,
\newblock {\em Neutrino Nucleus Interactions in the Few GeV Region.
  Proceedings, 3rd International Workshop, NuInt04, Assergi, Italy, March
  17-21, 2004}, , Nucl. Phys. Proc. Suppl. Vol. 139, 2005.

\bibitem{Haxton:1987kc}
W.~C. Haxton,
\newblock Phys. Rev. {\bf D36}, 2283 (1987).

\bibitem{Kolbe:1994xb}
E.~Kolbe, K.~Langanke and S.~Krewald,
\newblock Phys. Rev. {\bf C49}, 1122 (1994).

\bibitem{Volpe:2000zn}
C.~Volpe, N.~Auerbach, G.~Colo, T.~Suzuki and N.~Van~Giai,
\newblock Phys. Rev. {\bf C62}, 015501 (2000), [nucl-th/0001050].

\bibitem{Kolbe:2003ys}
E.~Kolbe, K.~Langanke, G.~Martinez-Pinedo and P.~Vogel,
\newblock J. Phys. {\bf G29}, 2569 (2003), [nucl-th/0311022].

\bibitem{Qian:1996db}
Y.~Z. Qian, W.~C. Haxton, K.~Langanke and P.~Vogel,
\newblock Phys. Rev. {\bf C55}, 1532 (1997), [nucl-th/9611010].

\bibitem{Smith:1972xh}
R.~A. Smith and E.~J. Moniz,
\newblock Nucl. Phys. {\bf B43}, 605 (1972).

\bibitem{Donnelly:1978tz}
T.~W. Donnelly and R.~D. Peccei,
\newblock Phys. Rept. {\bf 50}, 1 (1979).

\bibitem{Horowitz:1993rj}
C.~J. Horowitz, H.-C. Kim, D.~P. Murdock and S.~Pollock,
\newblock Phys. Rev. {\bf C48}, 3078 (1993).

\bibitem{Singh:1992dc}
S.~K. Singh and E.~Oset,
\newblock Nucl. Phys. {\bf A542}, 587 (1992).

\bibitem{Nieves:2004wx}
J.~Nieves, J.~E. Amaro and M.~Valverde,
\newblock Phys. Rev. {\bf C70}, 055503 (2004), [nucl-th/0408005].

\bibitem{Marteau:1999kt}
J.~Marteau,
\newblock Eur. Phys. J. {\bf A5}, 183 (1999), [hep-ph/9902210].

\bibitem{Benhar:2005dj}
O.~Benhar, N.~Farina, H.~Nakamura, M.~Sakuda and R.~Seki,
\newblock Phys. Rev. {\bf D72}, 053005 (2005), [hep-ph/0506116].

\bibitem{Alberico:1997vh}
W.~M. Alberico, M.~B. Barbaro, S.~M. Bilenky, J.~A. Caballero, C.~Giunti,
  C.~Maieron, E.~Moya~de Guerra and J.~M. Udias,
\newblock Nucl. Phys. {\bf A623}, 471 (1997), [hep-ph/9703415].

\bibitem{Meucci:2006ir}
A.~Meucci, C.~Giusti and F.~D. Pacati,
\newblock nucl-th/0601052.

\bibitem{vanderVentel:2003km}
B.~I.~S. van~der Ventel and J.~Piekarewicz,
\newblock Phys. Rev. {\bf C69}, 035501 (2004), [nucl-th/0310047].

\bibitem{vanderVentel:2005ke}
B.~I.~S. van~der Ventel and J.~Piekarewicz,
\newblock Phys. Rev. {\bf C73}, 025501 (2006), [nucl-th/0506071].

\bibitem{Martinez:2005xe}
M.~C. Martinez, P.~Lava, N.~Jachowicz, J.~Ryckebusch, K.~Vantournhout and J.~M.
  Udias,
\newblock Phys. Rev. {\bf C73}, 024607 (2006), [nucl-th/0505008].

\bibitem{Meucci:2004ip}
A.~Meucci, C.~Giusti and F.~D. Pacati,
\newblock Nucl. Phys. {\bf A744}, 307 (2004), [nucl-th/0405004].

\bibitem{Nieves:2005rq}
J.~Nieves, M.~Valverde and M.~J. Vicente~Vacas,
\newblock Phys. Rev. {\bf C73}, 025504 (2006), [hep-ph/0511204].

\bibitem{Singh:1998ha}
S.~K. Singh, M.~J. Vicente-Vacas and E.~Oset,
\newblock Phys. Lett. {\bf B416}, 23 (1998).

\bibitem{Paschos:2000be}
E.~A. Paschos, L.~Pasquali and J.-Y. Yu,
\newblock Nucl. Phys. {\bf B588}, 263 (2000), [hep-ph/0005255].

\bibitem{Kim:1996bt}
H.-C. Kim, S.~Schramm and C.~J. Horowitz,
\newblock Phys. Rev. {\bf C53}, 2468 (1996), [nucl-th/9507006].

\bibitem{Fogli:1979cz}
G.~L. Fogli and G.~Nardulli,
\newblock Nucl. Phys. {\bf B160}, 116 (1979).

\bibitem{Paschos:2003qr}
E.~A. Paschos, J.-Y. Yu and M.~Sakuda,
\newblock Phys. Rev. {\bf D69}, 014013 (2004), [hep-ph/0308130].

\bibitem{Adler:1974qu}
S.~L. Adler, S.~Nussinov and E.~A. Paschos,
\newblock Phys. Rev. {\bf D9}, 2125 (1974).

\bibitem{Ericson:1988gk}
T.~E.~O. Ericson and W.~Weise,
\newblock {\em Pions and Nuclei} (Clarendon Press, Oxford, 1988).

\bibitem{Ashie:2005ik}
Super-Kamiokande, Y.~Ashie {\em et~al.},
\newblock Phys. Rev. {\bf D71}, 112005 (2005), [hep-ex/0501064].

\bibitem{Zeller:2001hh}
NuTeV, G.~P. Zeller {\em et~al.},
\newblock Phys. Rev. Lett. {\bf 88}, 091802 (2002), [hep-ex/0110059].

\bibitem{Casper:2002sd}
D.~Casper,
\newblock Nucl. Phys. Proc. Suppl. {\bf 112}, 161 (2002), [hep-ph/0208030].

\bibitem{Gallagher:2002sf}
H.~Gallagher,
\newblock Nucl. Phys. Proc. Suppl. {\bf 112}, 188 (2002).

\bibitem{Hayato:2002sd}
Y.~Hayato,
\newblock Nucl. Phys. Proc. Suppl. {\bf 112}, 171 (2002).

\bibitem{Rein:1980wg}
D.~Rein and L.~M. Sehgal,
\newblock Ann. Phys. {\bf 133}, 79 (1981).

\bibitem{Leitner:2005jg}
T.~Leitner, L.~Alvarez-Ruso and U.~Mosel,
\newblock Prog. Part. Nucl. Phys. {\bf 57}, 395 (2006), [nucl-th/0511058].

\bibitem{Alvarez-Ruso:2006tv}
L.~Alvarez-Ruso, T.~Leitner and U.~Mosel,
\newblock to be published in the proceedings of PANIC 05, Santa Fe, USA, Oct
  2005  (2006), [nucl-th/0601021].

\bibitem{leitner_diplom}
T.~Leitner,
\newblock {\em Neutrino Interactions with Nucleons and Nuclei},
\newblock Diploma thesis, Justus-Liebig-Universit\"at Giessen, 2005,
\newblock ~\\
  \texttt{http://theorie.physik.uni-giessen.de/documents/diplom/leitner.pdf}.

\bibitem{Alvarez-Ruso:1998hi}
L.~Alvarez-Ruso, S.~K. Singh and M.~J. Vicente~Vacas,
\newblock Phys. Rev. {\bf C59}, 3386 (1999), [nucl-th/9804007].

\bibitem{Lalakulich:2005cs}
O.~Lalakulich and E.~A. Paschos,
\newblock Phys. Rev. {\bf D71}, 074003 (2005), [hep-ph/0501109].

\bibitem{Nowakowski:2004cv}
M.~Nowakowski, E.~A. Paschos and J.~M. Rodriguez,
\newblock Eur. J. Phys. {\bf 26}, 545 (2005), [physics/0402058].

\bibitem{Itzykson:1980rhy}
C.~Itzykson and J.~B. Zuber,
\newblock {\em Quantum Field Theory} (McGraw-Hill Book Company, New York,
  1980).

\bibitem{LlewellynSmith:1971zm}
C.~H. Llewellyn~Smith,
\newblock Phys. Rept. {\bf 3}, 261 (1972).

\bibitem{Budd:2003wb}
H.~Budd, A.~Bodek and J.~Arrington,
\newblock hep-ex/0308005.

\bibitem{Barish:1977qk}
S.~J. Barish {\em et~al.},
\newblock Phys. Rev. {\bf D16}, 3103 (1977).

\bibitem{Mann:1973pr}
W.~A. Mann {\em et~al.},
\newblock Phys. Rev. Lett. {\bf 31}, 844 (1973).

\bibitem{Baker:1981su}
N.~J. Baker {\em et~al.},
\newblock Phys. Rev. {\bf D23}, 2499 (1981).

\bibitem{Schreiner:1973mj}
P.~A. Schreiner and F.~von Hippel,
\newblock Nucl. Phys. {\bf B58}, 333 (1973).

\bibitem{Alvarez-Ruso:1997jr}
L.~Alvarez-Ruso, S.~K. Singh and M.~J. Vicente~Vacas,
\newblock Phys. Rev. {\bf C57}, 2693 (1998), [nucl-th/9712058].

\bibitem{Liu:1995bu}
J.~Liu, N.~C. Mukhopadhyay and L.-s. Zhang,
\newblock Phys. Rev. {\bf C52}, 1630 (1995), [hep-ph/9506389].

\bibitem{Feynman:1971wr}
R.~P. Feynman, M.~Kislinger and F.~Ravndal,
\newblock Phys. Rev. {\bf D3}, 2706 (1971).

\bibitem{Sato:2003rq}
T.~Sato, D.~Uno and T.~S.~H. Lee,
\newblock Phys. Rev. {\bf C67}, 065201 (2003), [nucl-th/0303050].

\bibitem{Barish:1978pj}
S.~J. Barish {\em et~al.},
\newblock Phys. Rev. {\bf D19}, 2521 (1979).

\bibitem{Radecky:1981fn}
G.~M. Radecky {\em et~al.},
\newblock Phys. Rev. {\bf D25}, 1161 (1982).

\bibitem{Kitagaki:1990vs}
T.~Kitagaki {\em et~al.},
\newblock Phys. Rev. {\bf D42}, 1331 (1990).

\bibitem{Adler:1968tw}
S.~L. Adler,
\newblock Ann. Phys. {\bf 50}, 189 (1968).

\bibitem{manley}
D.~M. Manley and E.~M. Saleski,
\newblock Phys. Rev. {\bf D45}, 4002 (1992).

\bibitem{Kitagaki:1983px}
T.~Kitagaki {\em et~al.},
\newblock Phys. Rev. {\bf D28}, 436 (1983).

\bibitem{Campbell:1973wg}
J.~Campbell {\em et~al.},
\newblock Phys. Rev. Lett. {\bf 30}, 335 (1973).

\bibitem{lenske}
H.~Lenske,
\newblock private communication.

\bibitem{welke}
G.~M. Welke, M.~Prakash, T.~T.~S. Kuo, S.~D. Gupta and C.~Gale,
\newblock Phys. Rev. {\bf C38}, 2101 (1988).

\bibitem{Effenberger:1999ay}
M.~Effenberger, E.~L. Bratkovskaya and U.~Mosel,
\newblock Phys. Rev. {\bf C60}, 044614 (1999), [nucl-th/9903026].

\bibitem{Oset:1987re}
E.~Oset and L.~L. Salcedo,
\newblock Nucl. Phys. {\bf A468}, 631 (1987).

\bibitem{Buss:2006vh}
O.~Buss, L.~Alvarez-Ruso, P.~Muehlich and U.~Mosel,
\newblock nucl-th/0603003.

\bibitem{Athar:2005hu}
M.~S. Athar, S.~Ahmad and S.~K. Singh,
\newblock Eur. Phys. J. {\bf A24}, 459 (2005), [nucl-th/0506057].

\bibitem{Teis:1996kx}
S.~Teis, W.~Cassing, M.~Effenberger, A.~Hombach, U.~Mosel and G.~Wolf,
\newblock Z. Phys. {\bf A356}, 421 (1997), [nucl-th/9609009].

\bibitem{Hombach:1998wr}
A.~Hombach, W.~Cassing, S.~Teis and U.~Mosel,
\newblock Eur. Phys. J. {\bf A5}, 157 (1999), [nucl-th/9812050].

\bibitem{Wagner:2004ee}
M.~Wagner, A.~B. Larionov and U.~Mosel,
\newblock Phys. Rev. {\bf C71}, 034910 (2005), [nucl-th/0411010].

\bibitem{Weidmann:1997vj}
T.~Weidmann, E.~L. Bratkovskaya, W.~Cassing and U.~Mosel,
\newblock Phys. Rev. {\bf C59}, 919 (1999), [nucl-th/9711004].

\bibitem{Lehr:1999zr}
J.~Lehr, M.~Effenberger and U.~Mosel,
\newblock Nucl. Phys. {\bf A671}, 503 (2000), [nucl-th/9907091].

\bibitem{Falter:2004uc}
T.~Falter, W.~Cassing, K.~Gallmeister and U.~Mosel,
\newblock Phys. Rev. {\bf C70}, 054609 (2004), [nucl-th/0406023].

\bibitem{Falter:2003uy}
T.~Falter, J.~Lehr, U.~Mosel, P.~Muehlich and M.~Post,
\newblock Prog. Part. Nucl. Phys. {\bf 53}, 25 (2004), [nucl-th/0312093].

\bibitem{Alvarez-Ruso:2004ji}
L.~Alvarez-Ruso, T.~Falter, U.~Mosel and P.~Muehlich,
\newblock Prog. Part. Nucl. Phys. {\bf 55}, 71 (2005), [nucl-th/0412084].

\bibitem{Andersson:1992iq}
B.~Andersson, G.~Gustafson and H.~Pi,
\newblock Z. Phys. {\bf C57}, 485 (1993).

\bibitem{Cugnon:1996kh}
J.~Cugnon, J.~Vandermeulen and D.~L'Hote,
\newblock Nucl. Instrum. Meth. {\bf B111}, 215 (1996).

\bibitem{Engel:1993jh}
A.~Engel, W.~Cassing, U.~Mosel, M.~Schafer and G.~Wolf,
\newblock Nucl. Phys. {\bf A572}, 657 (1994), [nucl-th/9307008].

\bibitem{Krusche:2004uw}
B.~Krusche {\em et~al.},
\newblock Eur. Phys. J. {\bf A22}, 277 (2004), [nucl-ex/0406002].

\bibitem{Paschos:2001np}
E.~A. Paschos and J.-Y. Yu,
\newblock Phys. Rev. {\bf D65}, 033002 (2002), [hep-ph/0107261].

\bibitem{Juszczak:2005wk}
C.~Juszczak, J.~A. Nowak and J.~T. Sobczyk,
\newblock Eur. Phys. J. {\bf C39}, 195 (2005).

\bibitem{Pohl:1979zm}
M.~Pohl {\em et~al.},
\newblock Lett. Nuovo Cim. {\bf 26}, 332 (1979).

\bibitem{Pohl:1979fw}
M.~Pohl {\em et~al.},
\newblock Lett. Nuovo Cim. {\bf 24}, 540 (1979).

\bibitem{Belikov:1983kg}
S.~V. Belikov {\em et~al.},
\newblock Z. Phys. {\bf A320}, 625 (1985).

\bibitem{Brunner:1989kw}
J.~Brunner {\em et~al.},
\newblock Z. Phys. {\bf C45}, 551 (1990).

\bibitem{Fluka}
G.~Battistoni, A.~Ferrari, A.~Rubbia and P.~R. Sala,
\newblock to appear in the proceedings of 2nd International Workshop on
  Neutrino Nucleus Interactions in the Few GeV Region (NuInt02), Irvine,
  California, 12-15 Dec 2002  (2002),
\newblock \texttt{http://www.ps.uci.edu/nuint/proceedings/sala.pdf}.

\bibitem{yuphd}
J.-Y. Yu,
\newblock {\em Neutrino Interactions and Nuclear Effects in Oscillation
  Experiments and the Nonpertubative Dispersive Sector in Strong
  (Quasi-)Abelian Fields},
\newblock PhD thesis, Universit\"at Dortmund, 2002,
\newblock \texttt{http://hdl.handle.net/2003/2396}.

\bibitem{Cassing:2006sg}
W.~Cassing, M.~Kant, K.~Langanke and P.~Vogel,
\newblock nucl-th/0601090.

\end{thebibliography}
\end{document}